\documentclass[a4paper,11pt]{article}
\pdfoutput=1 
\usepackage{jheppub,comment} 
\usepackage[T1]{fontenc} 

\usepackage{subfigure}
\renewcommand\parallel{\mathrel{/\mskip-2.5mu/}}
\usepackage{bbm}

\title{\boldmath Holographic Weak Measurement}

\author[a,b]{Xinyu Sun,}
\author[a]{ Shao-Kai Jian}

\affiliation[a]{Department of Physics and Engineering Physics, Tulane University, \\New Orleans, Louisiana, 70118, USA}
\affiliation[b]{Institute for Advanced Study, Tsinghua University, \\Beijing 100084, China}

\emailAdd{sunxy21@mails.tsinghua.edu.cn}
\emailAdd{sjian@tulane.edu}

\abstract{In this paper, we study a holographic description of weak measurements in conformal field theories (CFTs). 
Weak measurements can be viewed as a soft projection that interpolates between an identity operator and a projection operator, and can induce an effective central charge distinct from the unmeasured CFT.
We model the weak measurement by an interface brane, separating different geometries dual to the post-measurement state and the unmeasured CFT, respectively. 
In an infinite system, the weak measurement is related to ICFT via a spacetime rotation. 
We find that the holographic entanglement entropy with twist operators located on the defect is consistent in both calculations for ICFT and weak measurements. 
We additionally calculate the boundary entropy via holographic entanglement as well as partition function.
In a finite system, the weak measurement can lead to a rich phase diagram: for marginal measurements the emergent brane separates two AdS geometries, while for irrelevant measurements the post-measurement geometry features an AdS spacetime and a black hole spacetime that are separated by the brane.
Although the measurement is irrelevant in the later phase, the post-measurement geometry can realize a Python's lunch.
Finally, we discuss the thick brane construction for measurement and higher-dimension extensions of our model. 
For these general cases, our results above are still shown to be valid.}

\begin{document} 
\maketitle
\flushbottom

\newpage

\section{Introduction}
\label{sec:introduction}
In the holographic principle, dual geometries emerge from the entanglement structure of boundary quantum systems~\cite{maldacena1999large,gubser1998gauge,witten1998anti,maldacena2003eternal,ryu2006holographic,vanraamsdonk2010building}. 
It will be interesting to explore the effect of quantum information theoretic operations on the dual spacetime. 
For instance, local projection measurements are operations that radically change the entanglement of the wavefunction: the region being measured becomes unentangled after the measurements. 
It is, in general, very difficult to describe the effect of measurements on a many-body wavefunction because the measurement outcome is stochastic in nature, and one would not expect a universal description for all of them. 
Nevertheless, in conformal field theory (CFT), there is a big class of local projection measurements that can be described efficiently: when the measurement outcome is conditioned on the conformal invariant boundary states of the CFT~\cite{cardy1984conformal,cardy1986effect,cardy1989boundary,verlinde1988fusion,affleck1991universal}.
In 2d CFTs, these states are also called Cardy states~\cite{cardy2004boundary}.
After projecting (a subregion of) the wavefunction onto a Cardy state, the remaining parts still respect half of the conformal symmetry and are described by boundary conformal field theory (BCFT)~\cite{stephan2014emptiness,rajabpour2015post,rajabpour2016entanglement,najafi2016entanglement}.

\begin{figure}
    \centering
    \includegraphics[width=0.4\textwidth]{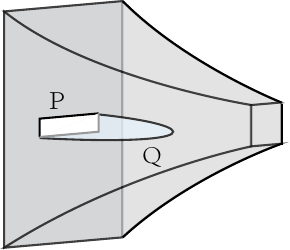}
    \caption{Local projection onto a Cardy state in a subregion. 
    $P$ denotes the slit induced by the projection and $Q$ denotes the ETW brane anchoring on the slit.
    The gray region is the spacetime dual to the remaining state after measurements.
    }
    \label{fig:slit}
\end{figure}

More concretely, the prescription to describe this class of measurements is to cut a slit, which represents the measurement, in the complex plane that represents the Euclidean path integral, and then map this plane with the slit, using conformal transformations, to a BCFT living in an upper half plane. 
This, according to AdS/BCFT~\cite{takayanagi2011holographic,Fujita_2011}, motivates the description of local projection measurements on the holography side.
As the holographic BCFT is dual to an end-of-the-world (ETW) brane ending at the boundary, the local projection measurements are also described by the ETW brane~\cite{numasawa2016epr}. 
When we consider the remaining state after measurements on a contiguous interval, as shown in figure~\ref{fig:slit}, it is dual to a spacetime with an ETW brane that ends on the slit, and cuts off a spacetime region that is homologous to the slit. 
We can also consider the bulk dual of the boundary state.
Because the boundary state has vanishing spatial entanglement, it is holographically dual to a trivial spacetime~\cite{miyaji2015boundary}.
This is consistent with the ETW brane description of the remaining parts, since the brane separates the dual spacetime of the remaining state, which still holds a good amount of entanglement to support the spacetime, and the trivial spacetime dual to the boundary state.
The informational aspect of holographic projective measurements has also been considered~\cite{antonini2022holographic,antonini2023holographic,antonini2023holographicb,goto2023entanglement}.

While the projection measurements onto a boundary state have a holographic description in terms of the ETW brane, to the extent of our understanding, much less is known about general measurements.
In this paper, we consider weak measurements, and study its holographic description.
Weak measurements are distinct from local projection measurements: the subregion being measured still holds finite entanglement.
A way to describe a weak measurement is by introducing a parameter to describe the measurement strength. 
Such a parameter ranges from zero to infinity, corresponding to no measurement and projection measurement, respectively.
We are interested in a generic measurement strength other than zero or infinity.
The weak measurement has been studied in CFTs recently~\cite{garratt2023measurements, sun2023new,yang2023entanglement,weinstein2023nonlocality}.
In reference~\cite{sun2023new}, we consider weak measurements performed on a Luttinger liquid described by a compactified free boson CFT with central charge $c=1$.
The compactified free boson CFT can be realized in the XXZ model, $H = -\sum_i ( X_i X_{i+1} + Y_i Y_{i+1} + \Delta Z_i Z_{i+1}) $, where $X_i, Y_i, Z_i$ denote Pauli matrices and $|\Delta| < 1$. 
The weak measurement operator is~\cite{sun2023new}
\begin{equation} \label{eq:measurement_example}
    M = e^{-W \sum_i (-1)^i Z_i}, 
\end{equation}
where $W \in [0, \infty)$ is the parameter that captures the measurement strength. 
The weak measurement is performed on all qubits, and its effect is irrelevant, marginal, or relevant when $\Delta<0$, $\Delta =0$, or $\Delta>0$, respectively. 
Compared to projection measurements, weak measurements have richer behaviors.
In particular, when it is marginal, the subsystem entanglement entropy of the resulted state exhibits a continuous effective central charge, $c_\text{eff}(W)$. 
More specifically, the entanglement entropy of a subregion with length $L_A$ (with lattice constant $a$)
\begin{equation}
    S_A = \frac{c_\text{eff}(W)}{3} \log \frac{L_A}{a}.
\end{equation}
The continuous central charge interpolates the original value (of central charge of the unmeasured system) and zero between the two limits of $W$, $W=0$ and $W=\infty$, that stand for no measurements and local projection measurements respectively.
Namely, \begin{equation}
    0 \le c_\text{eff}(W) \le c, \quad c_\text{eff}(0) = c, \quad  c_\text{eff}(\infty) = 0.
\end{equation}
While in the example of the XXZ model, we have the anisotropy parameter $\Delta$ to tune the relevance/irrelevance of the  measurement strength, here, in the holographic setup, we focus on the fixed point that is characterized by $c_{\rm eff}$, namely,~\footnote{We will also denote the central charge of the unmeasured CFT as $c_1$ in the following discussion.}
\begin{equation}
     \begin{split}
        c_\text{eff} = c & \qquad \text{irrelevant} \\
        0 < c_\text{eff} < c & \qquad \text{marginal} \\
        c_\text{eff} = 0 & \qquad \text{relevant}
    \end{split}
\end{equation}

\begin{figure}
    \centering
    \subfigure[]
    {\includegraphics[width=0.4 \textwidth]{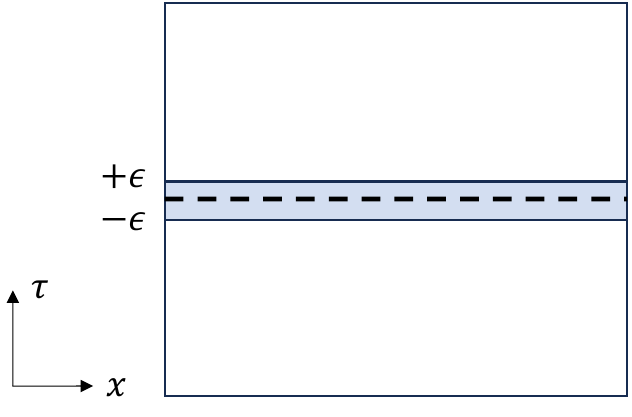}}
    \qquad \qquad \qquad
    \subfigure[]
    {\includegraphics[width=0.3\textwidth]{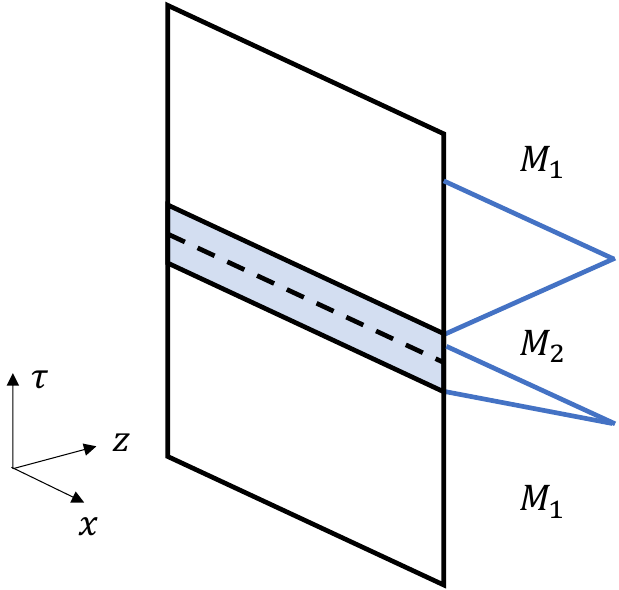}}
    \caption{(a) CFT with weak measurements on an infinite plane.
    A regularized measurement occurs in $\tau \in (-\epsilon, \epsilon)$. 
    The dashed line indicates the time reflection invariant line, $\tau = 0$.
    (b) Illustration of the bulk duality.
    $M_1$ and $M_2$ denote the bulk region dual to the boundary, the unmeasured CFT and the measurement.
    We also illustrate the branes separating $M_1$ and $M_2$.}
    \label{fig:illustration}
\end{figure}

From this example, we can deduce that weak measurements can support a state with nontrivial spatial entanglement, and thus, a nontrivial spacetime via AdS/CFT correspondence.
In the case of local projection measurements, the ETW brane separates a nontrivial spacetime of the remaining state and a trivial spacetime of the boundary state.
Now, different from the trivial spacetime dual to the boundary state after projection, because the state after weak measurements is dual to a nontrivial spacetime, the ETW brane is replaced by an interface brane separating two nontrivial spacetimes with different central charges.
To be concrete, on the CFT side, considering a weak measurement, given by $M = e^{-W H_M}$, acting on the ground state $|\Psi\rangle $,
we can use path integral to present the matrix element, $\langle \Psi| M^\dag M | \Psi \rangle = \lim_{\epsilon \rightarrow 0}  \langle \Psi| M_\epsilon^\dag M_\epsilon | \Psi \rangle $
\begin{equation}
\label{eq:measurement_CFT_action}
    \langle \Psi| M_\epsilon^\dag M_\epsilon | \Psi \rangle = \int D\phi \  e^{-S}, \quad S = \int d\tau \left[ L_{\text{CFT}} + f_\epsilon(\tau) W H_M \right],
\end{equation}
where $f_\epsilon(\tau)$ is a regularization function of the Dirac delta function, 
\begin{equation}
    f_\epsilon(\tau) = \begin{cases}
        \frac1{2\epsilon} \quad & |\tau| < \epsilon \\
        0  \quad & |\tau| > \epsilon
    \end{cases}.
\end{equation}
Such a Euclidean path integral is shown in figure~\ref{fig:illustration} (a), where $|\tau| = \epsilon$ separates two regions with different (effective) central charges.
On the gravity side, we need to fill in the spacetime that is consistent with the boundary quantum system.
Because we have different regions with different central charges at the boundary, the dual spacetime metrics are, in general, different, and separated by interface branes, as shown in figure~\ref{fig:illustration} (b).
$M_1$ and $M_2$ denote the bulk regions dual to the boundary of the unmeasured CFT and the measurement, and they are separated by interface branes.~\footnote{Here, $M_1$ is dual to the unmeasured CFT for both $\tau > \epsilon$ and $\tau < - \epsilon$ owing to the time reflection symmetry.}

The interface branes generalize the holographic local projection measurements to weak measurements.
This spacetime is similar to AdS/ICFT, where ICFT stands for interface CFT.
The ICFT and its holographic duality have been extensively studied \cite{kane1992transmission,ghoshal1994boundary,oshikawa1997boundary,leclair1999minimal,peschel2005entanglement,quella2007reflection,eisler2012entanglement,gaiotto2012domain,bachas2013fusion,gliozzi2015boundary,meineri2020colliders,karch2001open,bachas2002permeable,dewolfe2002holography,azeyanagi2007holographic,bachas2007fusion,sakai2008entanglement,gutperle2015entanglement,erdmenger2015bending,brehm2015enganglement,rozali2019information,bachas2020energy,Simidzija_2020,chen2020quantum,karch2021universal,bachas2021phases,karch2022universal,Anous_2022,bachas2022energy,karch2023universality,baig2023transport,tang2023universal,afrasiar2023island,cogburn2023cft}.
It describes joint spatial regions of possibly different CFTs.
Nevertheless, an obvious difference is that ICFT concerns the spatial interface, whereas weak measurements occur on an equal time slice. 
In this paper, we will investigate a few key features of the interface brane description of holographic weak measurements. 
Here is a brief summary of the main contribution in this paper:
\begin{itemize}
    \item We propose that the weak measurements that support an effective central charge $0< c_\text{eff} < c$ on an infinite line are described by interface branes as shown in figure~\ref{fig:illustration}.
    The dual spacetime is given by two AdS geometries with different radii that originate from the weak measurement and the unmeasured CFT, respectively, see~\eqref{eq:ads_radii_central_charge}.
    
    \item We establish the relation between the weak measurement and the spatial interface: we show that the weak measurement of CFTs defined on an infinite complex plane is related to ICFT via a spacetime rotation.
    In particular, the logarithmic entanglement entropy with a continuous effective central charge created by the spatial interface is the same as that from a weak measurement at an equal time slice.
    Figure~\ref{fig:infinite_EE} is a clear illustration of the spacetime rotation between the weak measurement and the spatial interface.
    
    \item We calculate the boundary entropy from holographic entanglement entropy of an ICFT, and show that it is the same as the measurement probability by evaluating the onshell action.

    \item We investigate the holographic weak measurement in CFTs defined on a circle, which leads to different phases (see figure~\ref{fig:phase_diagram}): 
    for marginal measurements, the dual geometry is two AdS spacetimes with different central charges similar to the case for infinite systems (this is called the no-bubble phase~\footnote{A similar calculation with different motivations has been studied in reference~\cite{Simidzija_2020}.
    We use their conventions for the names of different phases for convenience.}); 
    for irrelevant measurements, the dual geometry features an AdS spacetime and a black hole that are separated by the brane. 
    More concretely, when measurements are irrelevant, the AdS spacetime dual to the unmeasured CFT has a bubble, which is separated by an interface brane from the black hole dual to the weak measurement region, in the time reversal invariant slice. 
    If the black hole horizon is (not) contained in the solution, it is the bubble-inside (outside)-horizon phase.
    For the three phases, the system exhibits different behaviors of entanglement entropy, where the prefactor of $\log$-term are $c_1$ and $c_2$ for irrelevant and marginal phases.
    Interestingly, the post-measurement geometry of the bubble-inside-horizon phase realizes a Python's lunch~\cite{brown2019pythons}.

    \item While we primarily consider the thin brane description of the holographic weak measurement for ${\rm AdS_3/CFT_2}$, we extend our results to a thick-brane description in an infinite system and the higher-dimension generalization of our models (see section~\ref{sec:several extensions}).
    We find a consistent result as before.
\end{itemize}

The paper is organized as follows.
In section~\ref{sec:measurement/measurement geodesic in infinite system}, we consider the holographic dual of weak measurements on a CFT defined on an infinite complex plane.
The geometry corresponds to pure AdS spacetime with different AdS radii separated by the interface branes, as shown in figure~\ref{fig:illustration}.
We will also show that the weak measurement is related to interface CFT via a spacetime rotation.
In section~\ref{sec:CFT/CFT geodesic in infinite system for symmetric case}, we calculate the boundary entropy from holographic entanglement entropy and from the measurement partition function.
We find that they are equal.
In section~\ref{sec:phase transition for measurement and geodesic for two phases with black hole}, we consider holographic weak measurements in the CFTs defined on a circle. 
Depending on the AdS radii and the brane tension, there are three phases.
We discuss the entanglement properties of different phases. 
In section~\ref{sec:several extensions}, we present additional extensions of our results.
We introduce a thick brane description of weak measurements in an infinite system. 
Besides, we also discuss the generalization of higher-dimension cases, which are demonstrated to possess similar properties above.
Furthermore, we point out the bubble-inside-horizon phase in the case of weak measurements in finite systems has a structure of Python's lunch.
Finally, in section~\ref{sec:discussion}, we discuss a few future directions.
In appendix~\ref{sec:Geodesic with spatial point defect}, we consider the asymmetric entanglement entropy with a point defect, where the locations of twist operators are asymmetric.
Besides, we also consider the effect of ETW brane on the geodesic.
In appendix~\ref{sec:boundary entropy with path integral}, we introduce some basic notations for calculating the partition function.
In appendix~\ref{sec:different phases and phase transition for measurement (quantum quench)}, we provide a detailed derivation of geodesic solutions in general and in the bubble phase.
In appendix~\ref{sec:tensor network realization of Python's Lunch}, we construct a tensor network to realize Python's lunch, whose complexity is exponential in the difference between the maximal and outermost minimal areas.

\begin{figure}
    \centering
    \subfigure[]{\includegraphics[height=0.35\textwidth]{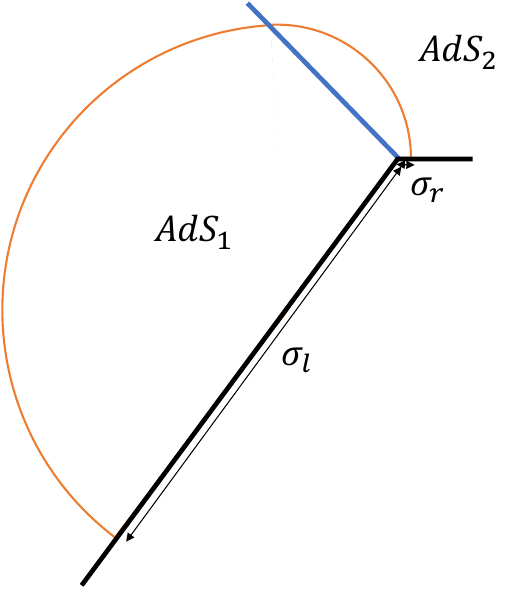}} \qquad \qquad
    \subfigure[]{\includegraphics[height=0.35\textwidth]{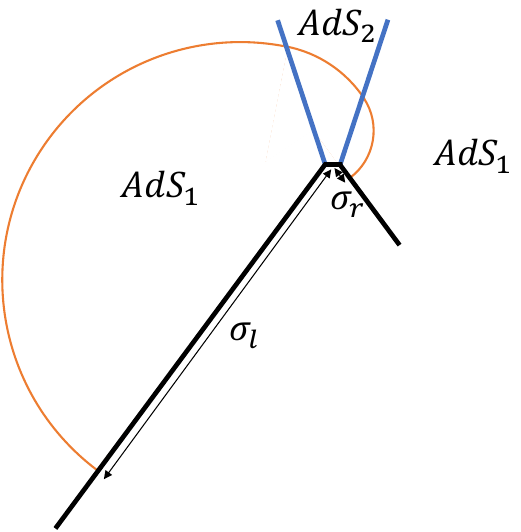}}
    \caption{Illustration of the RT surface dual to the entanglement entropy of an interval in CFTs with different spatial interface.
    We plot the limit where the left boundary of the interval is much longer than the right boundary $\sigma_l \gg \sigma_r$. 
    See the Appendix~\ref{sec:Geodesic with spatial point defect} for detailed discussions.
    (a) The boundary has two CFTs with different central charges, denoted by $c_1$ and $c_2$, respectively.
    They are separated by an interface.
    The dual geometry is given by two AdS spacetimes with different AdS radii separated by an interface brane colored in blue. 
    The entanglement entropy of an interval $(-\sigma_l, \sigma_r)$ is dual to the RT surface colored in orange.
    (b) The boundary is a CFT with a defect. 
    The central charge of the CFT is denoted by $c_1$.
    The defect supports a nontrivial AdS spacetime with a radius dual to an effective central charge $c_2$.
    These AdS spacetimes are separated by interface branes colored in blue. 
    The entanglement entropy of an interval $(-\sigma_l, \sigma_r)$ is dual to the RT surface colored in orange. }
    \label{fig:summarize_appendix_a}
\end{figure}

The calculation in Appendix~\ref{sec:Geodesic with spatial point defect} leads to some interesting results of holographic interface CFT that we would like to summarize in the following.
For the readers who are only interested in the results of weak measurements may skip this summary.
\begin{itemize}
\item We consider two semi-infinite CFTs with different central charges, denoted by $c_1$ and $c_2$. 
They are separated by a spatial interface. 
The gravity dual of such an interface CFT is given by two AdS spacetimes with radii determined by $c_1$ and $c_2$, respectively. 
The spacetimes are separated by an interface brane, as shown in figure~\ref{fig:summarize_appendix_a} (a). 
The interface is located at $0$, and we consider the entanglement entropy of an interval $(-\sigma_l, \sigma_r)$.
The general result for the holographic entanglement entropy of such a subsystem has been considered previously~\cite{Anous_2022}.
We obtain a simple result at the leading order in the limit for $\sigma_l \gg \sigma_r$ as follows~\footnote{We note that a similar result has also been reported in reference~\cite{tang2023universal}.},
\begin{equation}
    S_{-\sigma_l, \sigma_r} = \left( \frac{c_1}6 + \frac16 \min(c_1, c_2) \right) \log \sigma_l,
\end{equation}
where the left and right twist operators contribute to $\frac{c_1}6$ and $\frac16 \min(c_1, c_2)$, respectively.
It is interesting to see that while the right twist operator is located in the region of CFT with central charge $c_2$, the holographic entanglement entropy tries to ``minimize'' the central charge: $\frac16 \min(c_1, c_2)$.

\item We consider a CFT of central charge $c_1$ with a defect.
The defect supports dual AdS spacetime with an effective central charge $c_2$.~\footnote{There is no essential difference between a defect CFT and an interface CFT.}
The gravity dual of such a defect CFT is given by two AdS spacetimes with radii determined by $c_1$ and $c_2$, respectively. 
The spacetimes are separated by an interface brane, as shown in figure~\ref{fig:summarize_appendix_a} (b). 
Again, we consider the entanglement entropy of an interval $(-\sigma_l, \sigma_r)$.
We obtain a simple result at the leading order in the limit for $\sigma_l \gg \sigma_r$ as follows,
\begin{equation}
    S_{-\sigma_l, \sigma_r} = \left( \frac{c_1}6 + \frac16 \min(c_1, c_2) \right) \log \sigma_l,
\end{equation}
where the left and right twist operators contribute to $\frac{c_1}6$ and $\frac16 \min(c_1, c_2)$, respectively.
It is surprising to see that while the right twist operator is located in the region of the CFT with central charge $c_1$, the holographic entanglement entropy has the information about the effective central charge created by the defect when $c_2 < c_1$.
This can be intuitively understood as in the limit of $\sigma_l \gg \sigma_r$, there is a proxy effect to the defect for the right twist operator.

\end{itemize}

\section{Weak measurement in an infinite system}
\label{sec:measurement/measurement geodesic in infinite system}

In this section, we consider a semiclassical gravity in ${\rm AdS_3}$ describing interface CFTs, and use Ryu-Takayanagi formula to calculate the entanglement entropy of various subregions.
We will see that the entanglement structure of a CFT upon weak measurements is related to the interface CFTs via a spacetime rotation. 

\subsection{Review of AdS/ICFT}

The interface conformal field theory has been studied extensively both in condensed matter~\cite{kane1992transmission,ghoshal1994boundary,oshikawa1997boundary,leclair1999minimal,peschel2005entanglement,quella2007reflection,eisler2012entanglement,gaiotto2012domain,bachas2013fusion,gliozzi2015boundary,meineri2020colliders}, and in holography~\cite{karch2001open,bachas2002permeable,dewolfe2002holography,azeyanagi2007holographic,sakai2008entanglement,gutperle2015entanglement,erdmenger2015bending,bachas2007fusion,brehm2015enganglement,rozali2019information,bachas2020energy,Simidzija_2020,chen2020quantum,karch2021universal,bachas2021phases,karch2022universal,Anous_2022,bachas2022energy,karch2023universality,baig2023transport,tang2023universal,afrasiar2023island,cogburn2023cft}.
The basic structure of interface CFTs considered in this paper is given by two CFTs, each lives in a half plane and are separated by an interface.  
Here, we consider a thin brane construction.
In general, the two CFTs have different central charges. 
The gravity dual of this interface CFTs consists of the corresponding AdS geometry in bulk.
There are two different AdS radii in two half spaces, accounting for the different central charges.
The bulk AdS spacetimes with different radii are separated by a brane, connecting to the interface at the boundary. 
The Euclidean action of this gravity theory reads 
\begin{equation}
\begin{split}
\label{eq:bottom-up model}
    S_{{\rm EH}}=-\frac{1}{16\pi G_{(3)}} &\left[\int_{\mathcal{M}_1}{\rm d}^3 x\ \sqrt{g_1}\left(R_1+\frac{2}{L^2_1}\right)+\int_{\mathcal{M}_2}{\rm d}^3 x\ \sqrt{g_2}\left(R_2+\frac{2}{L^2_2}\right) \right.\\
    &+\left. 2\int_{\mathcal{S}} {\rm d}^2 y\ \sqrt{h}(K_1-K_2)-2T\int_{\mathcal{S}}{\rm d}^2y \ \sqrt{h}\right]+{\rm corner\ and\ counter\ terms},
\end{split}
\end{equation}
where $G_{(3)}$ is Newton constant in 3d. $\mathcal{M}_{1,2}$ denote the bulk spacetimes dual of the two CFTs, $L_{1,2}$ are the corresponding AdS radii, $g_{1,2}$ denote the metrics of two AdS bulks, and $R_{1,2}$ are the Ricci scalars.
$\mathcal S$ denotes the brane separating two AdS bulk spacetimes.
$K_{1,2}$ denote the extrinsic curvatures and $h_{ab}$ is the induced metric on the brane.
$T$ is the tension of the brane.
For simplicity, we omit the corner and counter terms.

The central charge of boundary CFT is related to the AdS radius via 
\begin{equation}
    c_{1,2}=\frac{3L_{1,2}}{2G_{(3)}}. 
\end{equation}
We work in the semiclassical limit, $c_{1,2} \gg 1$.
There are two junction conditions for induced metric $h_{ab}$ and extrinsic curvatures $K_{1,2}$,
\begin{subequations}
\label{eq:matching condition}
\begin{align}
    \label{eq:matching condition 1}
    &h_{ab}=\frac{\partial x_1^\mu}{\partial y^a} \frac{\partial x_1^\nu}{\partial y^b} g_{\mu\nu}^1=\frac{\partial x_2^\mu}{\partial y^a} \frac{\partial x_2^\nu}{\partial y^b} g_{\mu\nu}^2,\\
    \label{eq:matching condition 2}
    &\Delta K_{ab}-h_{ab}\Delta K=-T h_{ab},
\end{align}
\end{subequations}
where $\Delta K_{ab}=K_{1,ab}-K_{2,ab}$.
After tracing \eqref{eq:matching condition 2} we have $\Delta K=2T$, and then $\Delta K_{ab}=Th_{ab}$.

Before solving the junction conditions, we introduce a few coordinate systems that will be used in this paper.
A Euclidean ${\rm AdS_3}$ can be considered as a hyperboloid
\begin{equation}
\label{eq:embedded condition}
    -(X^0)^2+(X^3)^2+\sum_{i=1}^{2}(X^i)^2=-L^2,
\end{equation}
in four dimensional Minkowski spacetime with metric $G_{\mu\nu}={\rm Diag}(-1,1,1,1)$.
The first set of coordinates, which is useful for solving the junction conditions, is given by
\begin{subequations}
\begin{equation}
\label{eq:coordinates 1}
\begin{split}
    X^0=&\frac{L^2+\tau^2+y^2}{2y}\cosh{\frac{\rho}{L}},\qquad X^1=L\sinh{\frac{\rho}{L}},\\
    X^2=&\frac{L^2-\tau^2-y^2}{2y}\cosh{\frac{\rho}{L}},\qquad X^3=\frac{L\tau}{y}\cosh{\frac{\rho}{L}},
\end{split}
\end{equation}
\begin{equation}
\label{eq:metric of coordinate 1}
    {\rm d}s^2={\rm d}\rho^2+L^2\cosh^2{\left(\frac{\rho}{L}\right)}\left(\frac{{\rm d}\tau^2+{\rm d}y^2}{y^2}\right),
\end{equation}
\end{subequations}
where $-\infty<\rho<\infty$ and $y>0$.
Under a coordinate transformation,
\begin{equation}
\label{eq:coordinate transformation}
    z=\frac{y}{\cosh{\left(\frac{\rho}{L}\right)}},\qquad x=y \tanh{\left(\frac{\rho}{L}\right)},
\end{equation}
we get the second set of coordinates $(\tau,x,z)$ with the familiar metric
\begin{equation}
\label{eq:metric of coordinate 2}
    {\rm d}s^2=L^2\frac{{\rm d}\tau^2+{\rm d}x^2+{\rm d}z^2}{z^2}.
\end{equation}
To simplify the coordinate transformation, we introduce an angular variable $\sin{\chi}=\tanh{\left(\frac{\rho}{L}\right)}$, such that $z=y \cos{\chi},\ x=y \sin{\chi}$.
The asymptotic boundary is $z=0$ or equivalently $\chi=\pm \frac{\pi}{2}$.

\begin{figure}
    \centering
    \subfigure[]{
    \includegraphics[width=0.55\textwidth]{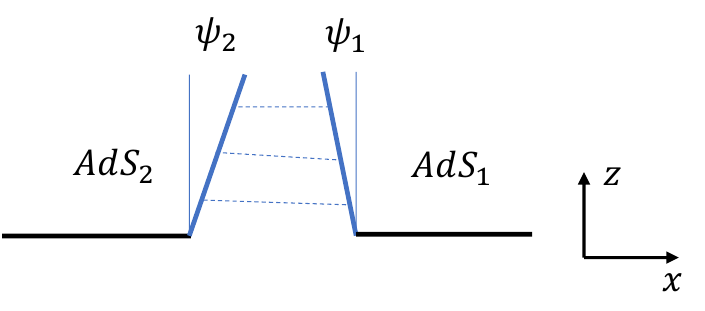}} \qquad\qquad
    \subfigure[]{
    \includegraphics[width=0.22\textwidth]{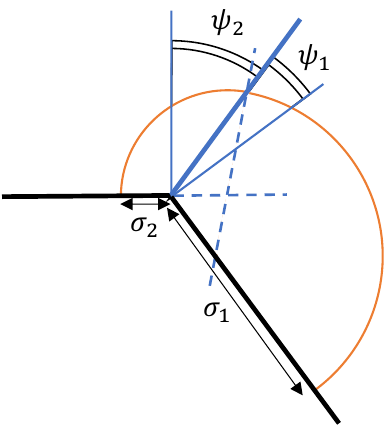}
    }
    \caption{(a) An interface brane between two AdS spacetimes. 
    The two branes are identified.
    (b) RT surface that connects $[-\sigma_2, \sigma_1]$.}
    \label{fig:ICFT_general}
\end{figure}

To solve the junction condition \eqref{eq:matching condition 1}, we use \eqref{eq:coordinates 1} and consider a brane located at $\rho_i=\rho_i^*$, $i=1,2$, or equivalently, $\sin \psi_{i} = \tanh \left( \frac{\rho_i^\ast}{L_i}\right)$. 
The convention for $\chi_i$ is that $\chi_i \rightarrow -\pi/2$ corresponds to the asymptotic boundary of the region $M_i$.
An example of the branes is shown in figure~\ref{fig:ICFT_general}.
Then the junction condition requires
\begin{subequations}
\begin{equation}
\label{eq:special form of matching condition 1}
    L_1^2\cosh^2{\left(\frac{\rho_1^*}{L_1}\right)}\left(\frac{{\rm d}\tau_1^2+{\rm d}y_1^2}{y_1^2}\right)=L_2^2\cosh^2{\left(\frac{\rho_2^*}{L_2}\right)}\left(\frac{{\rm d}\tau_2^2+{\rm d}y_2^2}{y_2^2}\right),
\end{equation}
\begin{equation}
\label{eq:special form of matching condition 2}
    \frac{1}{L_1}\tanh{\frac{\rho_1^*}{L_1}}+\frac{1}{L_2}\tanh{\frac{\rho_2^*}{L_2}}=T.
\end{equation}
\end{subequations}
The first equation leads to $y_1=y_2,\ \tau_1=\tau_2$ and $L_1\cosh{\left(\frac{\rho_1^*}{L_1}\right)}=L_2\cosh{\left(\frac{\rho_2^*}{L_2}\right)}$.
Combining with the second equation, we have
\begin{equation}
\label{eq:solution of matching condition initial}
    \tanh{\frac{\rho_1^*}{L_1}}=\frac{L_1}{2T}\left(T^2+\frac{1}{L_1^2}-\frac{1}{L_2^2}\right),\quad
    \tanh{\frac{\rho_2^*}{L_2}}=\frac{L_2}{2T}\left(T^2+\frac{1}{L_2^2}-\frac{1}{L_1^2}\right).
\end{equation}
Here $T\in(T_{\rm min},T_{\rm max})$ where $T_{\rm max,min}=\left|\frac{1}{L_1}\pm\frac{1}{L_2}\right|$.
In terms of the angular variable, we have 
\begin{equation}
    \sin{\psi_{1,2}}=\tanh{\left(\frac{\rho^\ast_{1,2}}{L_{1,2}}\right)}. 
\end{equation}
Therefore, the brane is determined by the AdS radii and the tension.

In 3d gravity, the entanglement entropy of an interval $[\sigma_{1}, \sigma_2]$ in the boundary theory is given by the geodesic length $d(\sigma_1,\sigma_2)$ via the RT formula,
\begin{equation}
\label{eq:entanglement entropy with RT formula initial}
    S_{\sigma_1,\sigma_2}=\frac{d(\sigma_1,\sigma_2)}{4G_{(3)}}.
\end{equation}

In general, we have to solve the geodesic equation subjected to the connection conditions to get the RT surface.
In the simple metric~\eqref{eq:metric of coordinate 2}, the relevant geodesic is given by an arc with the arc center located at the boundary (or the extension line of the boundary which will be clear later), and the connection condition is that the tangents of two arcs at the junction are the same.
With these simple rules, we are able to determine the RT surface and calculate the geodesic length.

In the following, we will investigate a special case with three AdS bulk regions separated by two interfaces.
The full theory has a reflection symmetry, as indicated in figure~\ref{fig:infinite_interface_brane}.

\subsection{Weak measurement and effective central charge}

Considering a weak measurement on the ground state $|\Psi\rangle $ defined in an infinite line,
we can use path integral to present (unnormalized) measurement amplitude $\langle \Psi| M^\dag M | \Psi \rangle$ with \eqref{eq:measurement_CFT_action}.
Such a Euclidean path integral is shown in figure~\ref{fig:illustration} (a).
We assume that the state after weak measurement, $M|\Psi\rangle$, has an effective central charge $c_\text{eff}$.
Because there are different regions with different central charges at the boundary, the dual spacetime metrics are separated by interface branes, as in figure~\ref{fig:illustration} (b).

\begin{figure}
    \centering
    \includegraphics[width=0.7\textwidth]{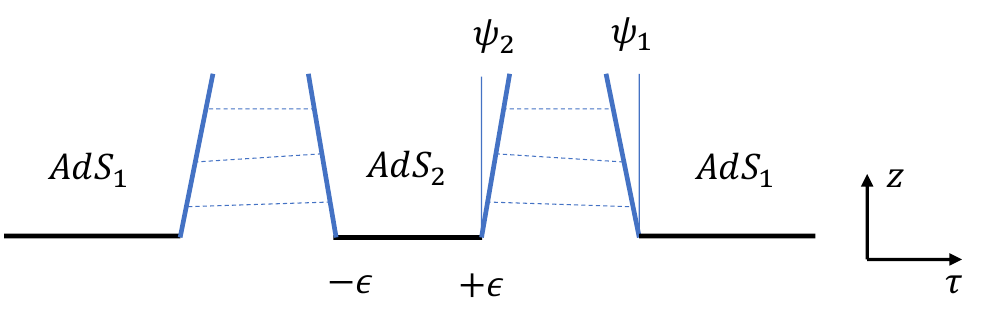}
    \caption{The gravity dual of weak measurements. 
    $\tau = \pm \epsilon$ denote the regularized weak measurement.
    Note that the geometry is symmetric w.r.t. $\tau = 0$.
    Two branes connected by the dashed line are identified as the same one.}
    \label{fig:infinite_interface_brane}
\end{figure}

As shown in figure~\ref{fig:infinite_interface_brane}, $AdS_1$ and $AdS_2$ are dual to the original CFT and the weak measurement.
\footnote{Here we use ${\rm AdS}_n$ to represent an AdS space with dimension $n$, and ${AdS_m}$ to represent an AdS space labeled by $m$.}
The action is the same as~\eqref{eq:bottom-up model}, except that the brane ends at $ \tau = \pm \epsilon$, $z=0$.
The metric is
\begin{equation} \label{eq:metric_measurement_infinite}
    {\rm d}s^2=L_i^2\frac{{\rm d}\tau^2+{\rm d}x^2+{\rm d}z^2}{z^2},
\end{equation}
with the AdS radii given respectively by
\begin{equation} \label{eq:ads_radii_central_charge}
    c_1 = \frac{3L_1}{2G_\text{(3)}}, \quad c_\text{eff} = \frac{3L_2}{2G_\text{(3)}}. 
\end{equation} 
Because $c_\text{eff} \le c_1$ for weak measurements, we have $L_1 \ge L_2$. 

The two AdS spacetimes are jointed via an interface brane, which is determined by the junction condition.
Because the metric is symmetric under the exchange of $x$ and $\tau$, we can directly get the solution from AdS/ICFT in the appendix via an exchange of $x$ and $\tau$.
Because the weak measurements respect translational symmetry in the spatial direction, the interface brane can be described by $(z,\tau_i(z),x)$. 
Here $\tau_{1,2}$ denotes the interface bane for $AdS_{1,2}$.
The solutions of the interface branes ending at $\tau = \epsilon$~\footnote{The solution of the interface branes ending at $\tau = - \epsilon$ can be obtained by time reflection transformation, $\tau \rightarrow - \tau$.} are given by,
\begin{equation}\label{eq:solution of matching condition}
    \begin{split}
    &\tau_1(z) = - \tan \psi_1 z + \epsilon, \quad \sin \psi_1 =\frac{L_1}{2T}\left(T^2+\frac{1}{L_1^2}-\frac{1}{L_2^2}\right),  \\
    &\tau_2(z) = \tan \psi_2 z + \epsilon, \quad
    \sin \psi_2 =\frac{L_2}{2T}\left(T^2+\frac{1}{L_2^2}-\frac{1}{L_1^2}\right).
    \end{split}
\end{equation}
For the solution to exist, it is easy to see that the tension should satisfy $T\in(T_{\rm min},T_{\rm max})$, where $T_{\rm max} = \frac{1}{L_1}+\frac{1}{L_2}$, $T_{\rm min}=\frac{1}{L_2}-\frac{1}{L_1}$.
The minimal tension is non-negative, and $c_\text{eff} \le c$ implies $\psi_2 \ge 0$ with \eqref{eq:solution of matching condition}. 
The two branes in $AdS_2$ will not have a crossing even at the limit $\epsilon \rightarrow 0$ as seen from figure~\ref{fig:infinite_interface_brane}.
\footnote{Here we start from the CFT side with an assumption $c_{\rm eff}<c$, so we have $\psi_2>\psi_1$ and two brane will not cross.
On the other hand, if we start from the gravity side and require the two branes not to cross, then we can actually allow $L_2>L_1$, which means $c_{\rm eff}>c$.
It is because, for $L_2>L_1$, there are the two branes which cross at some $T$ and don't cross at other $T$.
Therefore, non-crossing condition from the gravity side is not enough to require $c_{\rm eff} < c$ in the CFT side.
In the following, we will find stricter conditions on the gravity side for the marginal (relevant) phase, which turns out to imply $c_{\rm eff}<c$ in the CFT side.}

In the following, we consider entanglement entropy of the post-measurement state $|\Psi_M \rangle = M|\Psi\rangle$. 
To calculate holographic entanglement entropy of a subregion $\{ x: x \in (\sigma_1, \sigma_2) \}$ of $|\Psi_M \rangle $, we will use the RT formula,
\begin{equation}
\label{eq:entanglement entropy with RT formula}
    S_{\sigma_1,\sigma_2}=\frac{d(\sigma_1,\sigma_2)}{4G_{(3)}}.
\end{equation}
In 3d, the RT surface is nothing but geodesics. 
$d(\sigma_1, \sigma_2)$ denotes the length of a geodesic ending at $\sigma_1$ and $\sigma_2$.

Generically, for a geodesic with two endpoints $(\tau,x,z)$ and $(\tau',x',z')$ in one uniform AdS spacetime, its length reads
\begin{equation}
\label{eq:length of arc geodesic}
    d=L\cosh^{-1}{\left(\frac{(\tau-\tau')^2+(x-x')^2+z^2+z'^2}{2zz'}\right)}.
\end{equation} 
To this end, we consider two endpoints at the time reversal invariant plane.
And they are located on $(x=0,z=\varepsilon,\tau=0)$ and $(x'=l,z'=\varepsilon,\tau'=0)$~\footnote{$\varepsilon$ is the UV cutoff in the $z$ direction near the boundary. Do not confuse with $\epsilon$ that is introduced to regularize the measurements.}. 
Because of the time reflection symmetry, the geodesic will be located at $\tau=0$ slice, which is contained in $AdS_2$.
A schematic illustration is shown in figure~\ref{fig:infinite_EE} (a).
The orange curve denotes the geodesic at the time reversal invariant slice.
Therefore, the geodesic length is
\begin{equation}
\label{eq:length of geodesic for measurement/CFT geodesic on infinite system}
    d=L_2\cosh^{-1}{\left(\frac{l^2}{2\varepsilon^2}\right)}\approx L_2\log{\left(\frac{l^2}{\varepsilon^2}\right)},
\end{equation}
which leads to
\begin{equation}
\label{eq:entanglement entropy for measurement/measurement geodesic on infinite system}
    S_{0,l}=\frac{c_\text{eff}}{3}\log{\frac{l}{\varepsilon}}.
\end{equation}
The prefactor of $\log$-term is $\frac{c_\text{eff}}{3}$, and it is because both endpoints are located in the region $M_2$.

From this result, we can find that without measurement, the entanglement entropy is $S=\frac{c_1}{3}\log{\frac{l}{\varepsilon}}$;
while in the presence of weak measurements, the prefactor will depend on the AdS radius of $AdS_2$, which can be continuously tuned.
This is consistent with the marginal weak measurements in~\cite{sun2023new}.
If we take the limit $c_{\text{eff}}\rightarrow0$, there is no $\log$-term, which means we also retain the relevant phase with an area law.
This is consistent with local projection measurements, which cannot support nontrivial spatial entanglement.
On the gravity side, this corresponds to $L_2=0$, and the interface brane reduces to the ETW brane.
Therefore, we can recover the AdS/BCFT description of local projection measurements.

\subsection{Spacetime rotation}

\begin{figure}
    \centering
    \includegraphics[width=0.9 \textwidth]{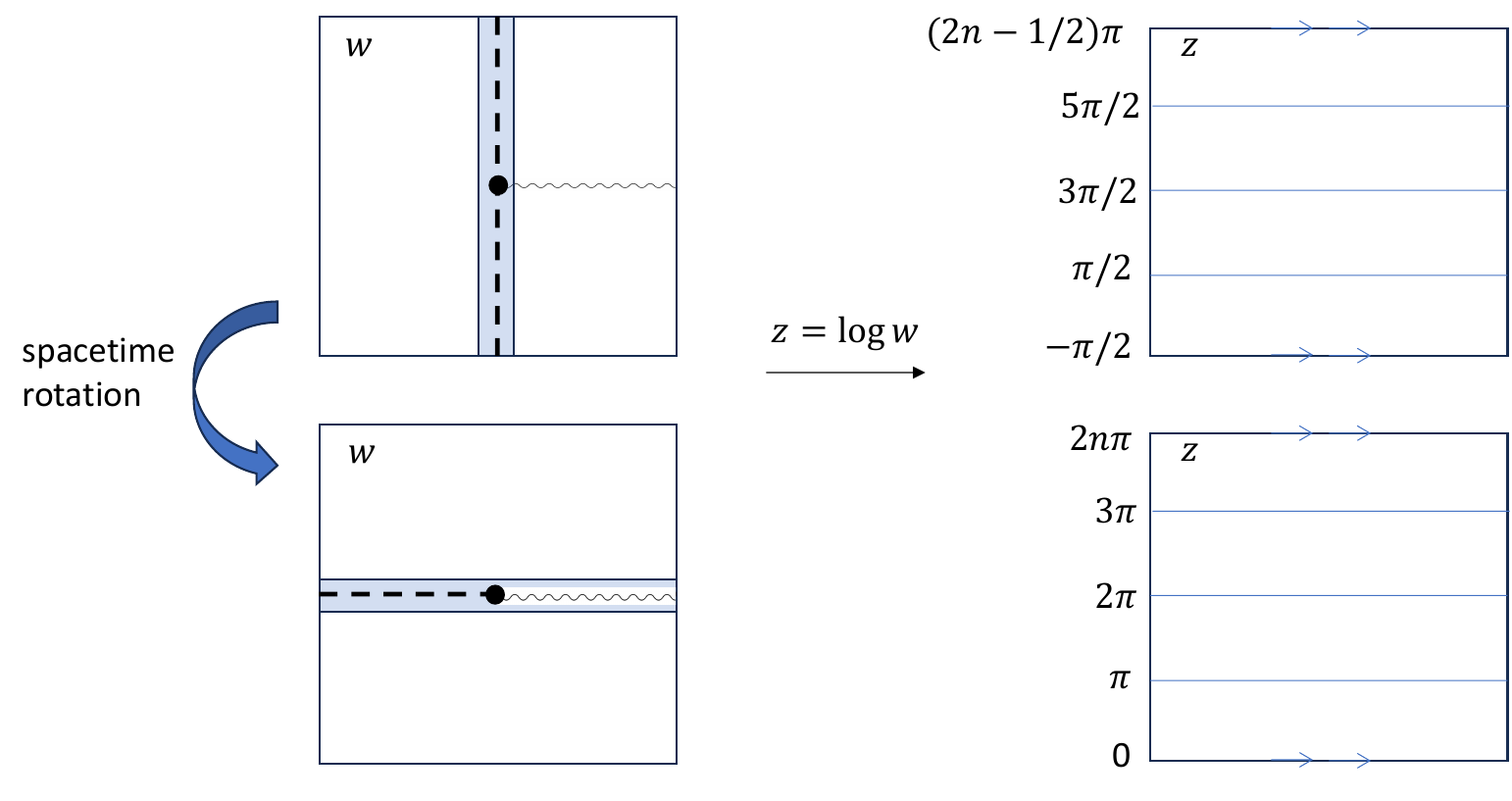}
    \caption{Illustration of spacetime rotation between ICFT and weak measurements. 
    The blue region denotes an interface with regularization.
    The wavy line denotes the branch cut created by the twist operator at the defect.
    Top penal: ICFT defined in the $w$ plane. The defect is located at $(x=0, - \infty < \tau < \infty)$.
    Bottom penal: Weak measurements in the $w$ plane. 
    The defect is located at $(- \infty < x < \infty, \tau=0)$.
    $z= \log w$ is the coordinate transformation to map the Riemann sheet with defects to a cylinder.
    The defect lines are mapped onto $(k-1/2)\pi$ and $k \pi$ for ICFT and for weak measurements, respectively.}
    \label{fig:spactime_rotation}
\end{figure}

In this subsection, we show that a spacetime rotation can relate the spatial ICFT and the weak measurement.
Consider an ICFT on a disk with a defect at $x=0$.
We are interested in the entanglement entropy of $\{x: x \in (0, l)\}$, where $l>0$ is the disk radius.
To calculate this entanglement entropy, a twist operator is inserted at the defect, and it creates a branch cut $\{x: x \in (0, l)\}$.
This is shown in the top panel of figure~\ref{fig:spactime_rotation}.
We denote the partition function with this branch cut by $Z_n$.
The entanglement entropy reads
\begin{equation}
    S = \lim_{n \rightarrow 1} S_n, \quad S_n = \frac1{1-n} \log \frac{Z_n}{(Z_1)^n} 
\end{equation}
where $Z_1$ is the original partition function without twist operator, and $S_n$ is the Renyi entanglement entropy.
To evaluate $Z_n$, we can make a coordinate transformation $z = \log w$, as shown in the top panel of figure~\ref{fig:spactime_rotation}. 
The defect line is mapped to $ \text{Im}(z) = (k- \frac12)\pi $, $k=0, 1, ..., 2n-1$, and in the $z$ plane, it has the periodic boundary condition $z = z + i 2n \pi $. 
Therefore, after this map, $Z_n$ is a partition function on a cylinder with defect lines $ \text{Im}(z) = (k- \frac12)\pi $, $k=0, 1, ..., 2n-1$.

It has been shown that, the entanglement entropy of one region with endpoints on the defect and the system boundary is characterized by a logarithm with a continuous coefficient.
And the coefficient depends on the strength of the defect, namely,
\begin{equation}
    S = \frac{c_{\rm eff}}{6} \log \frac{l}{\varepsilon},
\end{equation}
where ${c_{\rm eff}}$ denotes the effective central charge. 
$\varepsilon$ stands for the lattice constant, serving as a UV cutoff.
We have assumed that the other boundary of the system (as it is defined in a disk) has a normal boundary condition such that it does not contribute to the entanglement entropy.

Consider a spacetime rotation of the defect from $(x=0,-\infty < \tau < \infty )$ to $(-\infty < x < \infty , \tau = 0 )$~\footnote{On a disk, this should be from $(x=0,-l < \tau < l )$ to $(-l < x < l , \tau = 0 )$}. 
This corresponds to the weak measurement given by~\eqref{eq:measurement_CFT_action}. 
The entanglement entropy of the subregion $\{x: x \in (0, l)\}$ is again calculated by inserting a twist operator at $x = 0$, as shown in the bottom panel of figure~\ref{fig:spactime_rotation}.
We can use the same coordinate transformation $z= \log w$ to evaluate $\tilde Z_n$ in this case.
Note that we use $\tilde Z_n$ to denote the partition function with branch cut in the measurement case. 
In the bottom panel of figure~\ref{fig:spactime_rotation}, the defect line is mapped to $ \text{Im}(z) = k\pi $, $k=0, 1, ..., 2n-1$, with $z = z + i 2n \pi$. 
After this map, $\tilde Z_n$ is a partition function on a cylinder with defect lines $ \text{Im}(z) = k\pi $, $k=0, 1, ..., 2n-1$.
However, since the $\text{Im}(z)$ is periodic, we can shift the coordinate by $z \rightarrow z - i \pi/2$, which means $\tilde Z_n$ is equivalent to $Z_n$ in the spatial ICFT. 

Therefore, under the spacetime rotation, we can deduce that after weak measurements, the entanglement entropy is the same as that in the case of ICFT when the twist operator is located at the defect, 
\begin{equation}
    \tilde S = \frac{c_{\rm eff}}6 \log \frac{l}{\varepsilon}.
\end{equation}
Indeed, in reference~\cite{sun2023new}, the effective central charge from weak measurement reads
\begin{equation}
    c_{\rm eff} = - \frac6{\pi^2} \left( [(1+s) \log (1+s) + (1-s) \log (1-s)] \log s + (1+s) {\rm Li}_2 (-s) + (1-s) {\rm Li}_2(s) \right), 
\end{equation}
where $s$ is a function of measurement parameter and $s = 1/\cosh(2W)$ in~\eqref{eq:measurement_example}.
The same effective central charge~\footnote{Actually, $c_{\rm eff}/2$ appears in the entanglement entropy across a defect of Ising CFT because reference~\cite{sun2023new} investigated a Luttinger liquid, which can be decoupled into two Ising CFTs at the noninteracting limit.} also appears in the entanglement entropy across a defect of Ising CFT.

\begin{figure}
    \centering
    \subfigure[]{
    \includegraphics[width=0.4\textwidth]{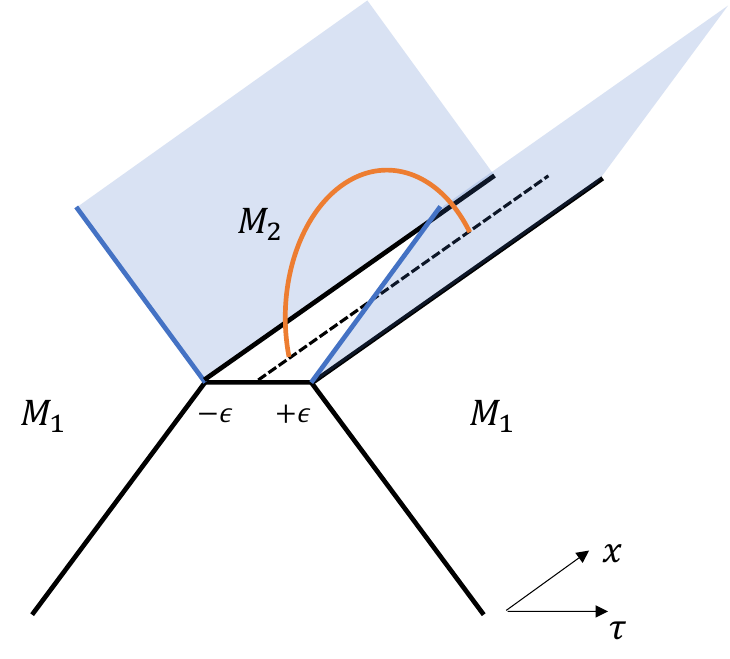}} 
    \qquad
    \subfigure[]{
    \includegraphics[width=0.4\textwidth]{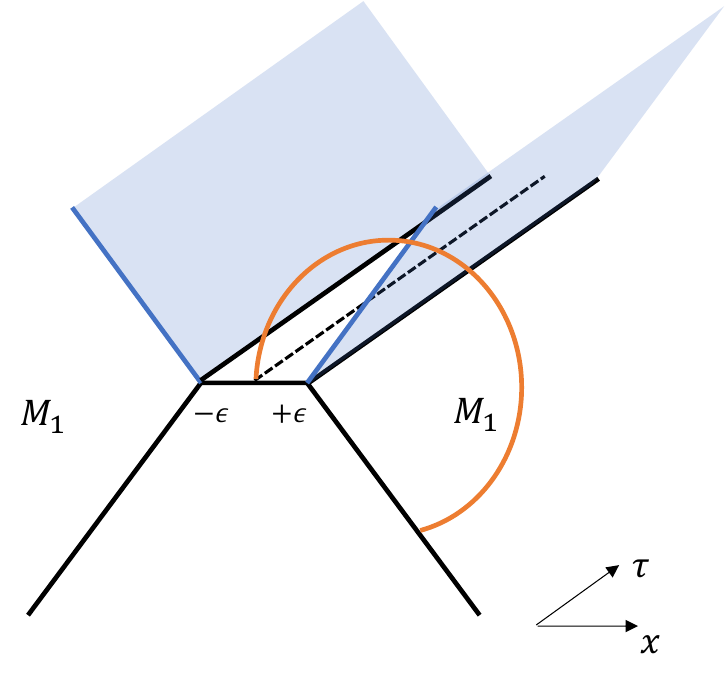}}
    \caption{(a) The spacetime dual to weak measurements. 
    $M_{1,2}$ denote the bulk dual to the unmeasured CFT and the weak measurements, respectively.
    $\epsilon$ denotes the regularization of the weak measurement.
    The blue surface indicates the interface brane.
    (b) The spacetime dual to the spatial ICFT upon a spacetime rotation in $(x ,\tau)$ for the measurement case.
    The orange curve denotes the RT surface of a subregion.}
    \label{fig:infinite_EE}
\end{figure}

On the gravity side, because the metric is symmetric~\eqref{eq:metric_measurement_infinite} under spacetime rotation, the solution of brane can also be obtained via a spacetime rotation.
To explore the entanglement structure in both cases, we consider the entanglement entropy of a subregion with length $l$.
For the ICFT, we consider that one of the twist operators is located at the defect.
The RT surface in the ICFT case and the weak measurement case is illustrated in figure~\ref{fig:infinite_EE}.
It is given by~\eqref{eq:entanglement entropy for measurement/measurement geodesic on infinite system} for weak measurements.
For ICFT, we expect the entanglement entropy to be 
\begin{equation}
    S_{0,l} = \frac{c + c_{\rm eff}}{6} \log \frac{l}{\varepsilon},
\end{equation}
because the two ends of the RT surface are located in $M_1$ and $M_2$ regions, respectively.
We will calculate the entanglement entropy for the ICFT case in the next subsection, and find that this is indeed the case.

\subsection{Entanglement entropy in holographic ICFTs}

\begin{figure}
    \centering
    \subfigure[]{\includegraphics[width=0.6 \textwidth]{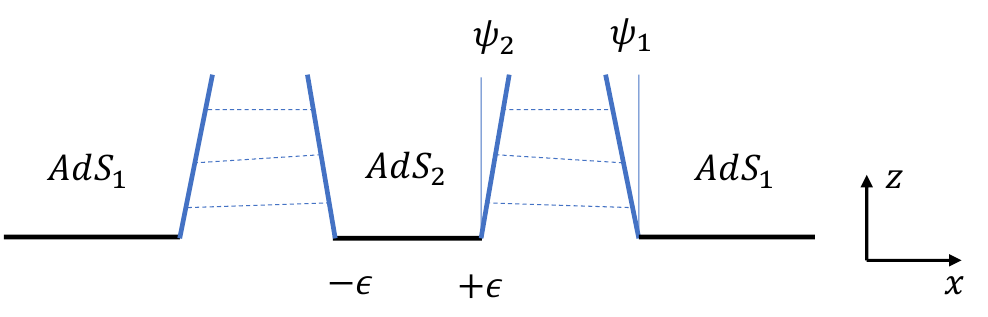}}
    \qquad  \qquad
    \subfigure[]{\includegraphics[width=0.25 \textwidth]{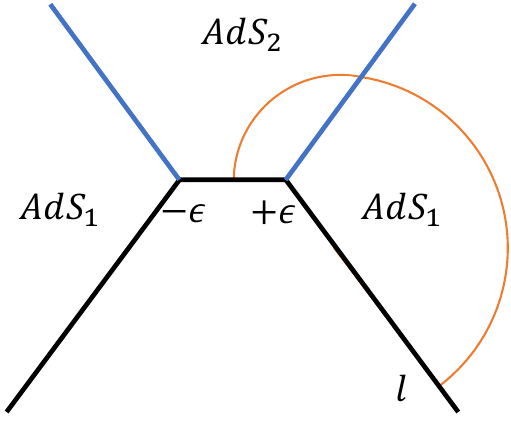}}
    \caption{(a) Interface branes of the spatial ICFT.
    It is related to the weak measurements via a spacetime rotation.
    (b) RT surface ending at $[0,l]$}
    \label{fig:infinite_ICFT}
\end{figure}

Upon a spacetime rotation, the interface brane dual to the weak measurement becomes a spatial interface brane of ICFT.
In the spatial ICFT, we arrive at branes $(z, \tau, x_i(z))$ with
\begin{equation} \label{eq:brane_solution_ICFT}
    \begin{split}
    &x_1(z) = - \tan \psi_1 z + \epsilon, \quad \sin \psi_1 =\frac{L_1}{2T}\left(T^2+\frac{1}{L_1^2}-\frac{1}{L_2^2}\right),  \\
    &x_2(z) = \tan \psi_2 z + \epsilon, \quad
    \sin \psi_2 =\frac{L_2}{2T}\left(T^2+\frac{1}{L_2^2}-\frac{1}{L_1^2}\right),
    \end{split}
\end{equation}
as shown in figure~\ref{fig:infinite_ICFT} (a).

\begin{figure}[tbp]
\centering 
\includegraphics[width=.45\textwidth]{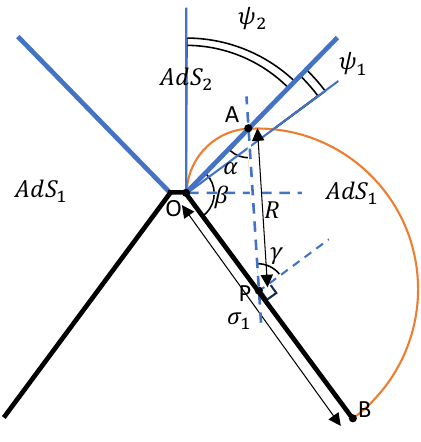}
\caption{\label{fig:defect_CFT_geodesic_infinite_system} AdS dual of CFT with interface defect for infinite system and its geodesic.}
\end{figure}

We consider the entanglement entropy of a subregion $\{x: x \in [0, l] \}$.
The RT surface that ends on these two points is illustrated in figure~\ref{fig:infinite_ICFT}. 
Because we take the limit $\sigma_2\rightarrow0$, we may refer to figure~\ref{fig:defect_CFT_geodesic_infinite_system}.
To solve the geodesic, we just need to compute $|{\rm OA}|$.
For $\triangle {\rm OAP}$, we have $\alpha=\frac{\pi}{2}-\psi_2$ and $\beta=\frac{\pi}{2}+\psi_1$.
Then the law of sines gives
\begin{equation}
\label{eq:the law of sines for radius}
    \frac{\sigma_1-R}{\sin{\alpha}}=\frac{R}{\sin{\beta}}=\frac{x}{\sin{(\pi-\alpha-\beta)}},
\end{equation}
where $|{\rm OA}|=x$.
The solutions are $R=\frac{\cos{\psi_1}}{\cos{\psi_1}+\cos{\psi_2}}\sigma_1$ and $x=\frac{\sin{(\psi_2-\psi_1)}}{\cos{\psi_1}}R$.
Now we can calculate the length of geodesic.
For ${\rm OA}$ we have ${\rm O}=(0,\varepsilon)$ and ${\rm A}=(x\sin{\psi_2},x\cos{\psi_2})$, and 
\begin{equation}
\label{eq:geodesic length of OA}
\begin{split}
    d_{\rm OA}=&L_2 \cosh^{-1}{\frac{\varepsilon^2+(x\cos{\psi_2})^2+(x\sin{\psi_2})^2}{2\varepsilon x\cos{\psi_2}}}\approx L_2 \cosh^{-1}{\left(\frac{x}{2\varepsilon\cos{\psi_2}}\right)}\\
    \approx&L_2\log{\left(\frac{\sigma_1}{\varepsilon}\frac{\sin{(\psi_2-\psi_1)}}{\cos{\psi_2}(\cos{\psi_1}+\cos{\psi_2})}\right)}.
\end{split}
\end{equation}
For ${\rm AB}$, with origin ${\rm P}$ and $\gamma=\frac{\pi}{2}+\psi_1-\psi_2$ we have ${\rm A}=(-R\sin{\gamma},R\cos{\gamma})$ and ${\rm B}=(R,\varepsilon)$, so 
\begin{equation}
\label{eq:geodesic length of AB}
\begin{split}
    d_{\rm AB}=&L_1 \cosh^{-1}{\left(\frac{\varepsilon^2+(R\cos{\gamma})^2+(R+R\sin{\gamma})^2}{2\varepsilon R\cos{\gamma}}\right)}\approx L_1 \cosh^{-1}{\left(\frac{R(2+2\sin{\gamma})}{2\varepsilon\cos{\gamma}}\right)}\\
    \approx&L_1\log{\left(\frac{\sigma_1}{\varepsilon}\frac{2\cos{\psi_1}}{\tan{(\frac{\psi_2-\psi_1}{2})}(\cos{\psi_1}+\cos{\psi_2})}\right)}.
\end{split}
\end{equation}
Thus, the total length of geodesic with $\sigma_1=l$ is 
\begin{equation}
    d(0,l)=d_{\rm OA}+d_{\rm AB}=(L_1+L_2)\log{\frac{l}{\varepsilon}}+d_{\rm sub},
\end{equation}
where $d_{\rm sub}$ is given by
\begin{equation}
\label{eq:boundary geodesic for defect/CFT geodesic on infinite system without reference}
\begin{split}
    d_{\rm bdy}
    =&L_2\log{\left(\frac{\sin{(\psi_2-\psi_1)}}{\cos{\psi_2}(\cos{\psi_1}+\cos{\psi_2})}\right)}+L_1\log{\left(\frac{2\cos{\psi_1}}{\tan{(\frac{\psi_2-\psi_1}{2})}(\cos{\psi_1}+\cos{\psi_2})}\right)}.
\end{split}
\end{equation}

Therefore, with \eqref{eq:entanglement entropy with RT formula} we have \footnote{For the geodesic with two endpoints in CFTs with different central charges, we assume the same UV-cutoff $\varepsilon$ for two endpoints.}
\begin{equation}
\label{eq:entanglement entropy for defect/CFT geodesic on infinite system}
    S_{0,\sigma_2}=\frac{c_1+c_{\rm eff}}{6}\log{\frac{l}{\varepsilon}}+S_{\rm sub},
\end{equation}
where $S_{\rm sub}=\frac{d_{\rm sub}}{4G_{(3)}}$.
The subleading entropy $S_{\rm }$ is UV dependent because the leading term is a logarithm with the UV dependence, and there is no way to subtract the leading term.~\footnote{We will see in the symmetric case discussed later, the subleading term is independent of the UV cutoff because there is a well-defined procedure to subtract the leading term.}
Here the prefactor of $\log$-term is $\frac{c_1+c_{\rm eff}}{6}$, which is because the location of two endpoints are in region $1$ and $2$.

Now we discuss the condition of the existence of a nontrivial geodesic solution.
Because we need $\triangle{\rm OAP}$, it means there must be a crossing point for lines ${\rm AP}$ and ${\rm OB}$.
Therefore, $\alpha+\beta<\pi$, which is equivalent to $\psi_1<\psi_2$.
With \eqref{eq:solution of matching condition}, we have
\begin{equation}
\begin{split}
\label{eq:relation between psi and L}
    \sin{\psi_2}-\sin{\psi_1}=\frac{L_2}{2T}\left(T^2+\frac{1}{L_2^2}-\frac{1}{L_1^2}\right)-\frac{L_1}{2T}\left(T^2+\frac{1}{L_1^2}-\frac{1}{L_2^2}\right)
    =\frac{(\frac{1}{L_1}+\frac{1}{L_2})^2-T^2}{2T}(L_1-L_2).
\end{split}
\end{equation}
Because $T<T_{\rm max}=\frac{1}{L_1}+\frac{1}{L_2}$, $\psi_1<\psi_2$ corresponds to $L_1>L_2$, which means $c_1>c_{\rm eff}$.
Therefore, the existence of a nontrivial geodesic corresponds to marginal (relevant) phase, which is consistent with the CFT results.

This motives us to derive $c_{\rm eff} < c_1$ for marginal (relevant) measurement in the CFT side from the gravity perspective.
In the two regions as shown in figure~\ref{fig:ICFT_general}, if one region is dual to the measurement or defect after the spacetime rotation,~\footnote{In general, we have three regions with a reflection symmetry, where two are dual to the unmeasured CFT and one is dual to the measurement. In the discussion here, only two regions are relevant, so we refer to figure~\ref{fig:ICFT_general} for simplicity.} the boundary of that region has the length scale in the order of regularization parameter $\epsilon$.
For the defect case, the limit of $\sigma\rightarrow0$ must be well-defined. 
It means for a geodesic ending at a general $\sigma_l$ and $\sigma_r$, including the limit case, it is supposed to cross the interface brane.
Therefore, for the case of measurements in figure~\ref{fig:infinite_ICFT} (b), we expect that the geodesic continues to be well-defined in the limit $\epsilon \rightarrow 0$.
We find that this is only true when $\psi_1 < \psi_2$, otherwise, the condition that $\alpha + \beta < \pi$ is not satisfied. 
Therefore, the existence of a limit $\epsilon \rightarrow 0$ with a well-defined geodesic that crosses the interface brane implies $L_2 < L_1$.

\section{Boundary entropy of holographic measurement in an infinite system}
\label{sec:CFT/CFT geodesic in infinite system for symmetric case}

As mentioned above, the entanglement of a spatial ICFT is related to the weak measurement via spacetime rotation. 
In particular, the same effective central charge appears in both cases.
It is given by the AdS radius that is dual to the weak measurement regions, $c_\text{eff} = \frac{3L_2}{2G_{\text{(3)}}}$.
To determine the interface brane solution, we also need the tension $T$ of the brane.
How is the tension related to the weak measurement?
In this section, we will show that the tension $T$ is related to the boundary entropy of weak measurements. 
To this end, we first calculate the entanglement entropy of a region symmetric w.r.t. the defect in the spatial ICFT.
The boundary entropy, $S_{\rm bdy}$, appears as the excess of entanglement entropy induced by the defect.
Then we calculate the amplitude in the weak measurement case, 
\begin{equation} \label{eq:partition_function_measurement}
    \begin{split}
        Z &= \frac{\langle \Psi | M^\dag M | \Psi \rangle}{\langle \Psi |  \Psi \rangle} \\
        &= e^{-(I_M - I_0) }
    \end{split}
\end{equation}
where in the second line, we use the saddle point approximation, and $I_M$ and $I_0$ denote the on-shall action with and without weak measurements, respectively.
We find that the same boundary entropy~\footnote{We call it a boundary entropy because the ICFT can be mapped to a BCFT by folding trick. } is given by the partition function $S_{\rm bdy} = \log Z$.
\begin{figure}[tbp]
\centering 
\includegraphics[width=.45\textwidth]{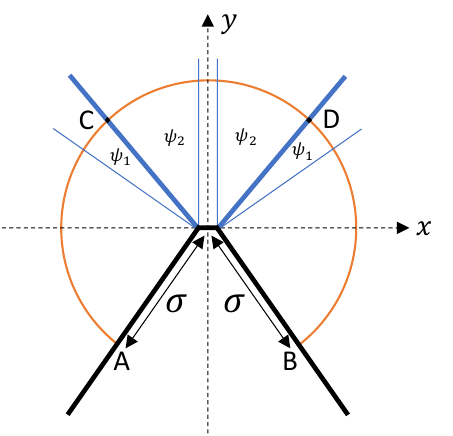}
\caption{\label{fig:CFT_CFT_geodesic_infinite_system_symmetric_with_brane} AdS dual of CFT with interface defect for infinite system with symmetry and its geodesic.}
\end{figure}

\subsection{Entanglement entropy of a symmetric region in the spatial ICFT}

We consider a symmetric configuration in the spatial ICFT, as shown in figure~\ref{fig:CFT_CFT_geodesic_infinite_system_symmetric_with_brane}.
Assuming $\sigma_l=\sigma_r=\sigma$, we can get analytical result for entanglement entropy and corresponding boundary entropy in the following.
By symmetry, the center of the circle must be on the defect line, and its radius is $\sigma$.
Therefore, for ${\rm AC}$ with ${\rm A}=(-\sigma,\varepsilon),\ {\rm C}=(\sigma\sin{\psi_1},\sigma\cos{\psi_1})$, the length of the geodesic connecting them is 
\begin{subequations}
\begin{equation}
\label{eq:CFT/CFT geodesic in infinite system symmetric solution of length eq1}
    d_{\rm AC}=L_1\cosh^{-1}{\left(\frac{(-\sigma-\sigma\sin{\psi_1})^2+\varepsilon^2+(\sigma\cos{\psi_1})^2}{2\varepsilon\sigma\cos{\psi_1}}\right)} \approx L_1\log{\left(\frac{2\sigma}{\varepsilon}\frac{1+\sin{\psi_1}}{\cos{\psi_1}}\right)}.
\end{equation}
For ${\rm CD}$ with ${\rm C}=(-\sigma\sin{\psi_2},\ \sigma\cos{\psi_2})$ and ${\rm D}=(\sigma\sin{\psi_2},\ \sigma\cos{\psi_2})$, the length of the geodesic connecting them is
\begin{equation}
\label{eq:CFT/CFT geodesic in infinite system symmetric solution of length eq2}
    d_{\rm CD}=L_2\cosh^{-1}{\left(\frac{(-2\sigma\sin{\psi_2})^2+2(\sigma\cos{\psi_2})^2}{2(\sigma\cos{\psi_2})^2}\right)}=L_2\cosh^{-1}{\left(\frac{1+\sin^2{\psi_2}}{\cos^2{\psi_2}}\right)}.
\end{equation}
\end{subequations}
By symmetry, we also have $d_{\rm AC}=d_{\rm DB}$, so the total entanglement entropy for the symmetric case in figure~\ref{fig:CFT_CFT_geodesic_infinite_system_symmetric_with_brane} is 
\begin{equation}
\label{eq:entanglement entropy for symmetric CFT/CFT geodesic on infinite system}
\begin{split} 
    S_{-\sigma,\sigma}=\frac{d_{\rm AC}+d_{\rm CD}+d_{\rm DB}}{4G_{(3)}}=
    \frac{c_1}{3}\log{\left(\frac{2\sigma}{\varepsilon}\right)}+S_{\rm bdy},
\end{split}
\end{equation}
where the boundary entropy $S_{\rm bdy}$ is 
\begin{equation}
\label{eq:boundary entropy for symmetric CFT/CFT geodesic on infinite system}
    S_{\rm bdy}=\frac{c_1}{3}\log{\left(\frac{1+\sin{\psi_1}}{\cos{\psi_1}}\right)}+\frac{c_\text{eff}}{6}\cosh^{-1}{\left(\frac{1+\sin^2{\psi_2}}{\cos^2{\psi_2}}\right)}.
\end{equation}

Actually, the two terms in the boundary entropy have the same forms after some manipulation.
For the first term, we have $\frac{1+\sin{\psi_1}}{\cos{\psi_1}}=\tan{\left(\frac{\psi_1}{2}+\frac{\pi}{4}\right)}$.
To simplify the second term, (i) we first prove that 
\begin{subequations}
\begin{equation}
\label{eq:helpful equation 1 for proof}
    \cosh^{-1}{\left(\frac{1+\sin^{2}{\psi}}{\cos^2{\psi}}\right)}=2\tanh^{-1}{\left(\sin{\psi}\right)}.
\end{equation}
Define ${\rm l.h.s.}=2p$, then $\frac{2}{\cos^2{\psi}}-1=\cosh{2p}$, so $\cos^2{\psi}=\left[\frac{1}{2}\left(\cosh{2p}+1\right)\right]^{-1}=\frac{1}{\cosh^{2}{p}}$.
Therefore, $\sin^2{\psi}=1-\frac{1}{\cosh^{2}{p}}=\tanh^2{p}$.
Finally, we get $p=\tanh^{-1}{\left(\sin{\psi}\right)}$, and finish the proof.
(ii) Now we prove
\begin{equation}
\label{eq:helpful equation 2 for proof}
    \tanh^{-1}{\left(\sin{\psi}\right)}=\log{\left(\tan{\left(\frac{\psi}{2}+\frac{\pi}{4}\right)}\right)}.
\end{equation}
Similar to the proof above, we define ${\rm r.h.s.}=p$, then $e^p=\tan{\left(\frac{\psi}{2}+\frac{\pi}{4}\right)}$.
Therefore, $\sin{\psi}=-\cos{\left(\psi+\frac{\pi}{2}\right)}=(\tan^2{\frac{\psi+\frac{\pi}{2}}{2}}-1)/(\tan^2{\frac{\psi+\frac{\pi}{2}{2}}+1})=\tanh{p}$.
It means $p=\tanh^{-1}{\left(\sin{\psi}\right)}$ and we finish the proof.
\end{subequations}
With \eqref{eq:helpful equation 1 for proof} and \eqref{eq:helpful equation 2 for proof}, the second term in \eqref{eq:boundary entropy for symmetric CFT/CFT geodesic on infinite system} can be expressed as $\cosh^{-1}{\left(\frac{1+\sin^2{\psi_2}}{\cos^2{\psi_2}}\right)}=2\tanh^{-1}{\left(\sin{\psi_2}\right)}=2\log{\left(\tan{\left(\frac{\psi_2}{2}+\frac{\pi}{4}\right)}\right)}$.
Therefore, the boundary entropy \eqref{eq:boundary entropy for symmetric CFT/CFT geodesic on infinite system} can be simplified as 
\begin{equation}
\label{eq:simplified boundary entropy for symmetric CFT/CFT geodesic on infinite system}
    S_{\rm bdy}=\frac{c_1}{3}\log{\left(\tan{\left(\frac{\psi_1}{2}+\frac{\pi}{4}\right)}\right)}+\frac{c_\text{eff}}{3}\log{\left(\tan{\left(\frac{\psi_2}{2}+\frac{\pi}{4}\right)}\right)}.
\end{equation}

Here are some remarks about the results:
(i) One may wonder if the boundary entropy is UV dependent from~\eqref{eq:entanglement entropy for symmetric CFT/CFT geodesic on infinite system} by rescaling the UV cutoff $a$.
However, we can properly regularize this dependence by defining the boundary entropy as the excess of entanglement entropy due to the defect
\begin{equation}
    S_{\rm bdy} = S_{-\sigma, \sigma} - S^{(0)}_{-\sigma, \sigma},
\end{equation}
where the second term $S^{(0)}_{-\sigma, \sigma}$ denotes the entanglement entropy of the same region but without a defect. 
(ii) While~\eqref{eq:simplified boundary entropy for symmetric CFT/CFT geodesic on infinite system} seemingly has two independent terms from two regions, this is not the case.
Recall that the solutions of the branes $\psi_{1,2}$ in~\eqref{eq:brane_solution_ICFT} depend on both radii $L_{1,2}$ and the tension $T$.
It is clear that the parameters in the holographic description are uniquely determined by the central charge of the unmeasured CFT $c_1$, the effective central charge after measurement $c_\text{eff}$, and the boundary entropy $S_\text{bdy}$. 
(iii) In the symmetric case considered here, the prefactor of the logarithm in~\eqref{eq:entanglement entropy for symmetric CFT/CFT geodesic on infinite system} is $\frac{c_1}{3}$. 
This can be intuitively understood: the two twist operators are located in region 1, and when we take the long-wave length limit, they approach deep in the bulk of the system symmetrically.
However, when the interval is not symmetric and the limit is not taken symmetrically, the prefactor of the logarithm can also be related to $c_{\rm eff}$ as we will show in appendix~\ref{sec:Geodesic with spatial point defect}. 

\subsection{Boundary entropy from weak measurements partition function}

\begin{figure}[tbp]
\centering 
\subfigure[]{\includegraphics[width=0.9\textwidth]{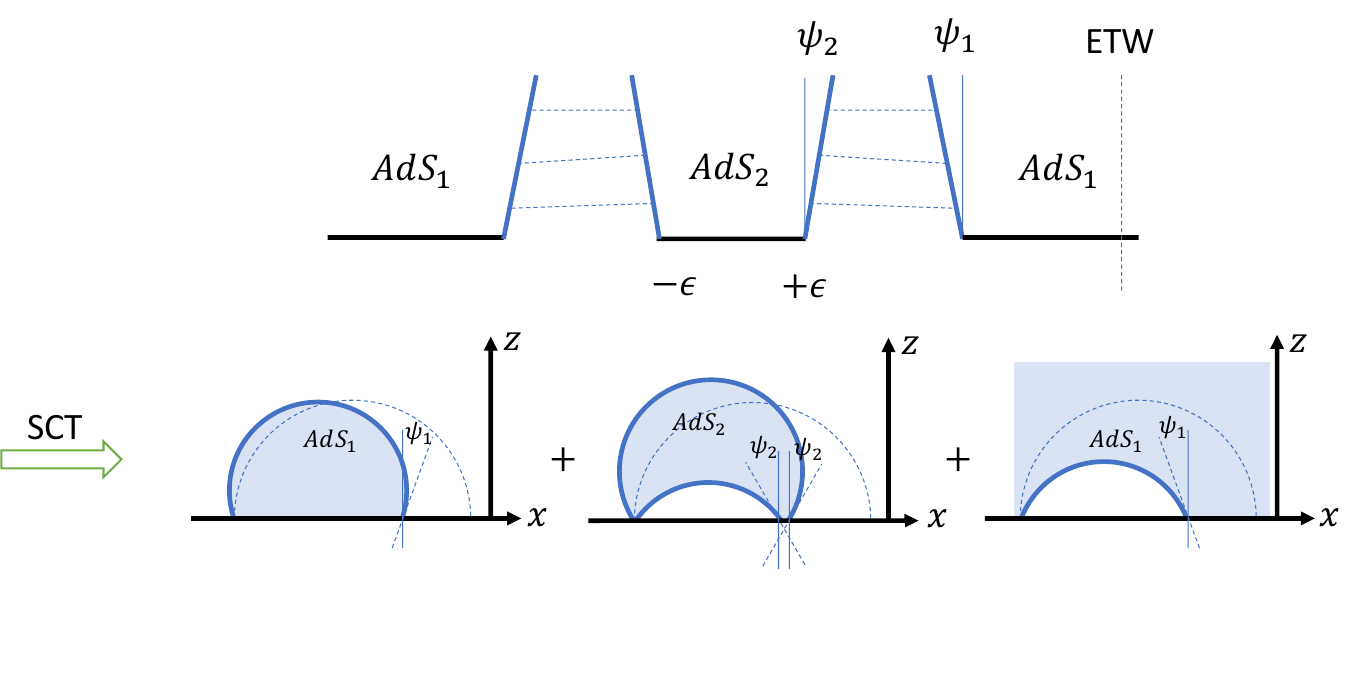}}\qquad
\subfigure[]{\includegraphics[width=.75\textwidth]{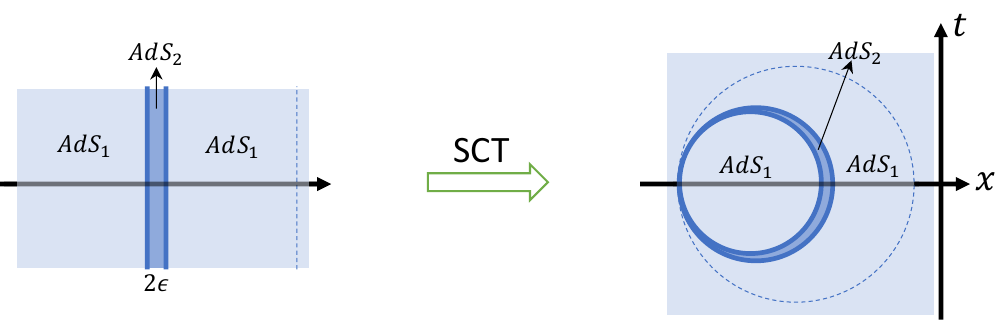}}
\caption{\label{fig:diagram_three_bulk_two_brane} Spacial conformal transformation that maps the brane into part of a sphere. 
(a) Illustration of the map in the xz-plane. 
(b) Illustration of the map in the xt-plane.}
\end{figure}

In a BCFT, the boundary entropy \eqref{eq:simplified boundary entropy for symmetric CFT/CFT geodesic on infinite system} can be related to g-theorem with the definition of boundary entropy $S_{\rm bdy}=\log{g}$ where $g=\left<0|B\right>$, and $\left|B\right>$ is the boundary state of the BCFT.
We consider a similar quantity in this subsection: the partition function with the measurement, as is given by~\eqref{eq:partition_function_measurement}.
In reference~\cite{Fujita_2011}, the authors have given the boundary entropy of one AdS region with a single ETW brane, for completeness, we review the calculation in appendix~\ref{sec:boundary entropy with path integral}.
For our interested case shown in figure~\ref{fig:CFT_CFT_geodesic_infinite_system_symmetric_with_brane}, we apply a similar procedure and obtain the boundary entropy in the following.
As mentioned in appendix~\ref{sec:boundary entropy with path integral}, we first apply a special conformal transformation, which will lead to the geometry in figure~\ref{fig:diagram_three_bulk_two_brane}.
There are three bulk regions and two interface branes.
Although it seems that there is overlap between two connected regions, we should calculate the action separately and consider that the different regions are connected by some magic glue.

To this end, we define the corresponding action
\begin{equation}
\label{eq:three bluk action for boundary entropy}
\begin{split}
    &I=-\frac{1}{16\pi G_{(3)}}\left[\int_{\mathcal{M}_1}{\rm d}^3 x\ \sqrt{g_1}\left(R_1+\frac{2}{L^2_1}\right)+\int_{\mathcal{M}_2}{\rm d}^3 x\ \sqrt{g_2}\left(R_2+\frac{2}{L^2_2}\right)+\int_{\mathcal{M}_3}{\rm d}^3 x\ \sqrt{g_3}\left(R_3+\frac{2}{L^2_3}\right) \right.\\
    &+\left. 2\int_{\mathcal{S}_{12}} {\rm d}^2 y\ \sqrt{h_{12}}(K_1-K_2)-2T_{12}\int_{\mathcal{S}_{12}}{\rm d}^2y \ \sqrt{h_{12}}+2\int_{\mathcal{S}_{23}} {\rm d}^2 y\ \sqrt{h_{23}}(K_2-K_3)-2T_{23}\int_{\mathcal{S}_{23}}{\rm d}^2y \ \sqrt{h_{23}}\right].
\end{split}
\end{equation}
By the symmetry, we have $L_1=L_3$ and $T_{12}=T_{23}=T$.
The connection conditions \eqref{eq:matching condition 2} lead to $K_1-K_2=K_2-K_3=2T$.
With a similar method in appendix~\ref{sec:boundary entropy with path integral}, we have the onshell action
\begin{equation}
\label{eq:three bluk action for boundary entropy after transformation}
\begin{split}
    I=&-\frac{1}{16\pi G_{(3)}}\left[\int_{\mathcal{M}_1}{\rm d}^3 x\ \sqrt{g_1}\left(R_1+\frac{2}{L^2_1}\right)+\int_{\mathcal{M}_2}{\rm d}^3 x\ \sqrt{g_2}\left(R_2+\frac{2}{L^2_2}\right)+\int_{\mathcal{M}_3}{\rm d}^3 x\ \sqrt{g_3}\left(R_3+\frac{2}{L^2_3}\right) \right.\\
    &+\left. \int_{\mathcal{S}_{12}} {\rm d}^2 y\ \sqrt{h_{12}}(K_1-K_2)+\int_{\mathcal{S}_{23}} {\rm d}^2 y\ \sqrt{h_{23}}(K_2-K_3)\right]\\
    =&-\frac{1}{16\pi G_{(3)}} \sum_{\alpha=1,2,3}\left[\int_{\mathcal{M}_\alpha} {\rm d}^3 x\ \sqrt{g_\alpha}\left(R_\alpha+\frac{2}{L^2_\alpha}\right)+\sum_{\beta={\rm L,R}}\left(\int_{\mathcal{S}_{\alpha,\beta}} {\rm d}^2 y\ \sqrt{h_{\alpha,\beta}}K_{\alpha,\beta}\right)\right] \\
    =& \sum_{\alpha=1,2,3} I_\alpha,
\end{split}
\end{equation}
where we define $\beta$ as the label of boundary of the region $\alpha$.
Therefore, $K_{\alpha,\beta}$ corresponds to the left and right branes of the region $\alpha$, and the corresponding signs are absorbed by redefining that, the directions for $K_{\alpha,\beta}$ point outside from their bulk regions.
Here for $AdS_1$ we only have one boundary brane and one boundary term in action, while for region $AdS_2$ we have two boundary terms.
However, from the discussion in appendix~\ref{sec:boundary entropy with path integral}, for the largest region after SCT (here it is the right $AdS_1$), we must apply UV-cutoff, which can be an ETW brane on the right of the right $AdS_1$ in figure~\ref{fig:diagram_three_bulk_two_brane}.
The ETW brane is perpendicular to the boundary CFT and has no contribution to the boundary entropy.
Besides, we can also add another ETW$'$ brane on the left of the left $AdS_1$ without contribution.
Then for each region we can calculate the action with the same method in appendix~\ref{sec:boundary entropy with path integral} with \eqref{aeq:integral result of AdS/BCFT action in ref}, and the result of each part is 
\begin{equation}
\label{eq:integral for each region with two brane}
\begin{split}
    I_\alpha=&\frac{L}{4 G_N}\left[\left(\frac{r_D^2}{2\varepsilon^2}+\log{\frac{\varepsilon}{r_D}}-\frac{\rho^*}{L}-\frac{1}{2}+\frac{r_D}{\varepsilon}\sinh{\frac{\rho^*}{L}}\right)-\left(\frac{{r_D'}^{2}}{2\varepsilon^2}+\log{\frac{\varepsilon}{r_D'}}-\frac{\rho^{*}{}'}{L}-\frac{1}{2}+\frac{r_D'}{\varepsilon}\sinh{\frac{\rho^{*}{}'}{L}}\right)\right],
\end{split}
\end{equation}
where $r_D, \rho^*$ correspond to one brane and $r_D', \rho^*{}'$ correspond to another brane.
\footnote{For example, for $AdS_2$ we can consider its region after special conformal transformation is bounded by two branes.
We can first calculate the contribution without the left brane, which is the first term.
Then the region bounded by the left brane which we should subtract corresponds to the second term.
So actually the direction of ${\rho^*}'$ in the second term is opposite to the direction of the brane of $AdS_2$.}
Therefore, the total boundary entropy is 
\begin{equation}
\label{eq:total boundary entropy from action integral}
\begin{split}
    S_{\rm bdy}=&\frac{\rho_1^*-(-\rho_0^*)}{4G_{(3)}}+\frac{\rho_2^*-(-\rho_2^*)}{4G_{(3)}}+\frac{\rho_0^*-(-\rho_1^*)}{4G_{(3)}}=\frac{2\rho_1^*+2\rho_2^*}{4G_{(3)}}\\
    =&\frac{c_1}{3}\log{\left(\tan{(\frac{\psi_1}{2}+\frac{\pi}{4})}\right)}+\frac{c_{\rm eff}}{3}\log{\left(\tan{(\frac{\psi_2}{2}+\frac{\pi}{4})}\right)},
\end{split}
\end{equation}
where $\rho_0^*=0$.
As mentioned before, here we change the sign of second $\rho^*$ in the numerator because we use the proper definition in reference~\cite{Anous_2022}.
Besides, although here we add ETW brane which is perpendicular to the CFT surface artificially, we can remove it as discussed in appendix~\ref{sec:boundary entropy with path integral}.
Without ETW brane, we can consider it as the case with ETW brane with location $x\rightarrow\infty$.
Under corresponding conformal transformation, we can map the ETW brane to a semi-sphere at the infinity, and it covers the whole space with $r_D\rightarrow\infty$.
As shown in \eqref{eq:total boundary entropy from action integral}, the final boundary entropy doesn't rely on $r_D$, so we can safely apply the limit above and get the same result.
And this result is also consistent with \eqref{eq:simplified boundary entropy for symmetric CFT/CFT geodesic on infinite system} without the ETW brane.

\section{Holographic weak measurements in a finite system: phase transition and entanglement entropy}
\label{sec:phase transition for measurement and geodesic for two phases with black hole}

In the previous section about weak measurements in an infinite system, the holographic description is given by interface branes.
The effective central charge $c_\text{eff} \le c_1$ distinguishes the different cases of the weak measurements:
\begin{equation}
     \begin{split}
        c_\text{eff} = c_1 & \qquad \text{irrelevant} \\
        0 < c_\text{eff} < c_1 & \qquad \text{marginal} \\
        c_\text{eff} = 0 & \qquad \text{relevant}
    \end{split}
\end{equation}
The interface brane solution exists for the irrelevant and marginal cases.
In the relevant case, the interface brane becomes an ETW brane~\cite{Anous_2022}.

In this section, we consider weak measurements in a finite system. 
We will see that in a finite system, the transition between the irrelevant and marginal weak measurements is dual to a transition of interface branes with different topologies.
After discussing the phase diagram, we calculate the subregion entanglement entropy in different phases, and identify the fate of weak measurements.

\subsection{Phase diagram}

\begin{figure}
    \centering
    \includegraphics[width=0.5\textwidth]{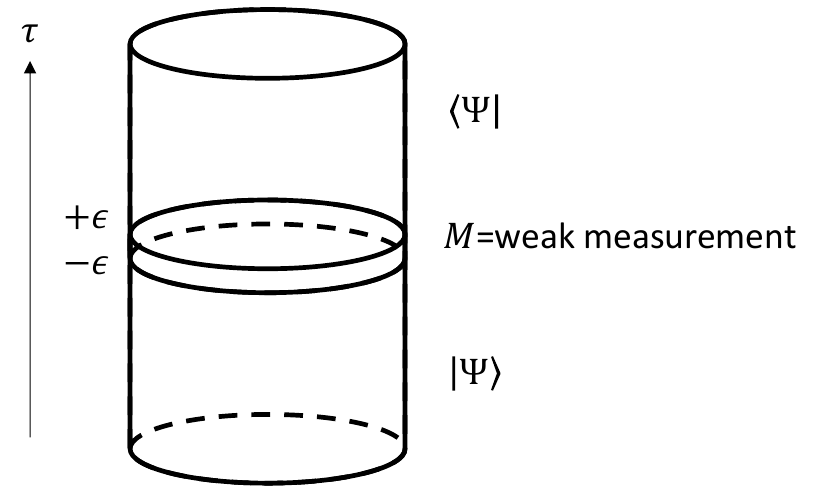}
    \caption{Euclidean path integral of a CFT, defined in a circle, with weak measurements regularized at $\tau = \pm \epsilon$. 
    $|\Psi \rangle$ ($\langle \Psi |$) indicates the ket (bra) of the ground state prepared from Euclidean path integral.}
    \label{fig:measurement_finite_CFT}
\end{figure}

The Euclidean path integral of a CFT,  defined on a circle $x=x +2\pi $, with weak measurements is shown by the cylinder in figure~\ref{fig:measurement_finite_CFT}. 
$|\Psi \rangle $ denotes the ground state of the CFT and $M |\Psi \rangle$ denotes the state after the weak measurements.
The measurements are performed in the full system, and preserve translational symmetry.
Again, we denote the central charge of the unmeasured CFT as $c_1$ and the effective central charge of the post-measurement state $M | \Psi \rangle$ as $c_{\rm eff} \le c_1$.

The gravity dual is to ``fill in'' the bulk of the cylinder with the boundary condition given by the Euclidean path integral defined on the surface of the cylinder.
In reference~\cite{Simidzija_2020}, the authors have discussed the phase diagram of the dual gravity solution, though they considered a different problem.
In the following, we summarize the results in reference~\cite{Simidzija_2020}.

To this end, we start from a similar action in~\eqref{eq:bottom-up model}.
This time, $\mathcal M_{1,2}$ denotes the regions dual to the original unmeasured CFT on a circle and the post-measurement state, respectively.
We use a different convention for the metric: 
\begin{equation}
\label{eq:general metric for AdS 3D}
    {\rm d}s^2=f_i(r){\rm d}\tau^2+\frac{{\rm d}r^2}{f_i(r)}+r^2{\rm d}x^2,
\end{equation}
where $f_i(r)=(1-\mu_i)+\lambda_i r^2$ with $\lambda_i=L_i^{-2}$.
In this metric, $\tau$ denotes the imaginary time, and $x = x + 2\pi$ denotes the spatial coordinate.
$r$ is the radius, with $r \rightarrow \infty$ corresponding to the boundary.
$\mu_i$ determines whether the metric is AdS ($\mu_i<1$) or black hole ($\mu_i>1$).
Again, we have the relation
\begin{equation}
    c_1 = \frac{3L_1}{2G_\text{(3)}}, \quad c_\text{eff} = \frac{3L_2}{2G_\text{(3)}}.
\end{equation}

The parameter $\mu_i$ and the brane solution are determined by the radius $L_{i}$ and the tension $T$.
Notice that there is a regularization parameter $\epsilon$ for the measurement, at which the interface brane is located.
To determine the phase diagram of weak measurements, we take $\epsilon \rightarrow 0$.

\begin{figure}[tbp]
\centering 
\subfigure[]{\includegraphics[width=.6\textwidth]{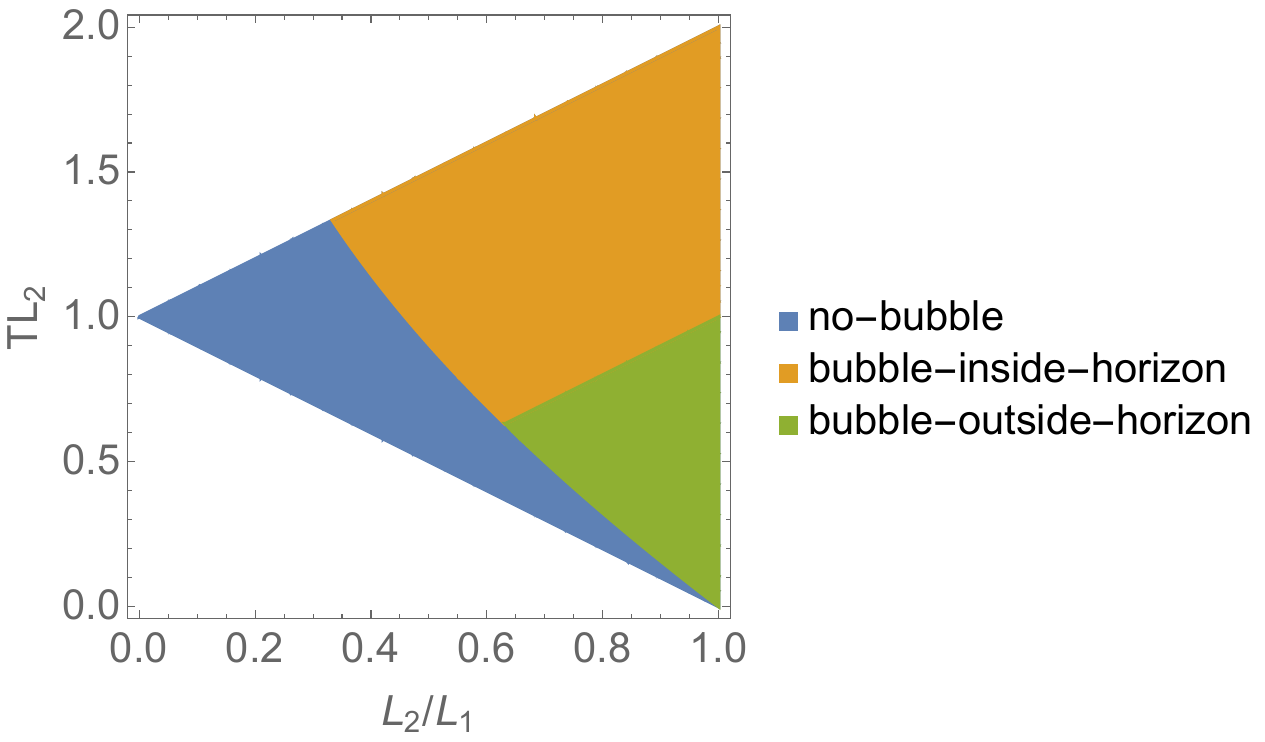}} \\
\subfigure[]{\includegraphics[width=0.4\textwidth]{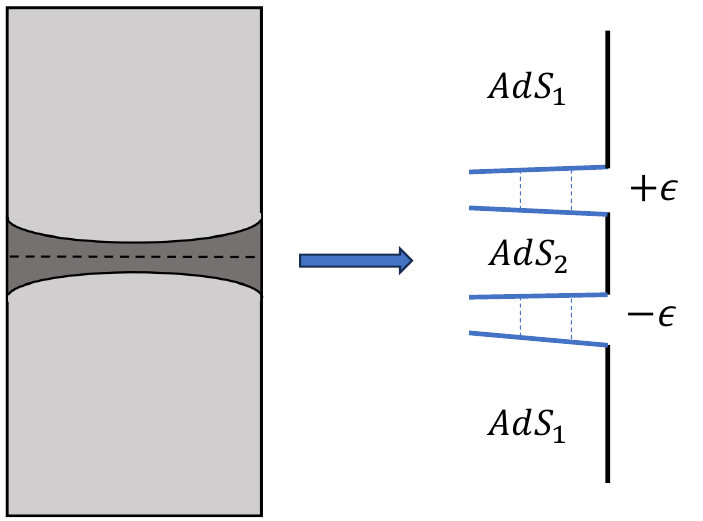}} \qquad \qquad
\subfigure[]{\includegraphics[width=0.4\textwidth]{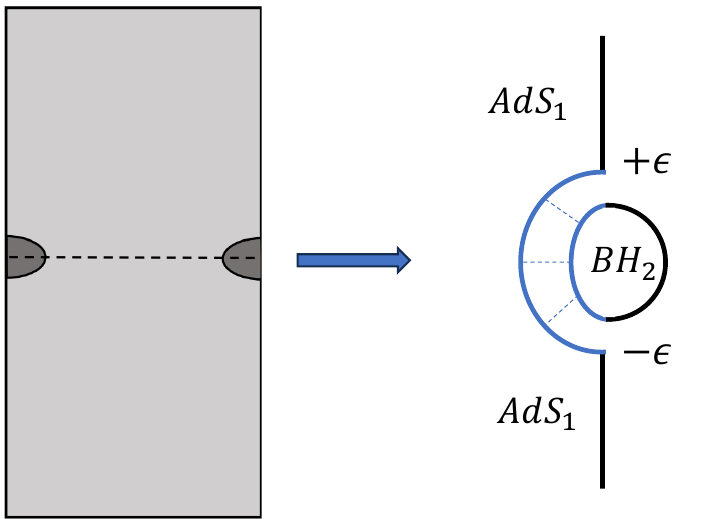}}
\caption{\label{fig:phase_diagram} (a) The phase diagram of the weak measurement in a finite system. 
The vertical cross-section of the cylinder in the no-bubble phase (b) and in the bubble phase (c). 
The light (dark) region is dual to the unmeasured CFT (the weak measurement).
(b) The dual geometry consists of two AdS spacetimes with different radii.
(c) The dual geometry of the unmeasured CFT (the weak measurement) is given by the AdS (black hole) spacetime.}
\end{figure}

Analogous to the previous case, the  solution exists when the tension falls in the following range:
\begin{equation}
\label{eq:constraint of tension}
    \frac1{L_2} -\frac1{L_1} \le T \le \frac1{L_2} + \frac1{L_1}.
\end{equation}
Note that we focus on $L_2 \le L_1$.
In this range of parameters, there are three different phases, which are called ``no-bubble'', ``bubble-inside-horizon'' and  `bubble-outside-horizon'' in reference~\cite{Simidzija_2020}.
We show the phase diagram in figure~\ref{fig:phase_diagram}. 
In all three phases, for the region dual to the unmeasured CFT, the metric is given by AdS space with $\mu_1=0$.
For simplicity, we define $\mu_2 = \mu$ without loss of generality.
Depending on whether $\mu < 1$ or $\mu > 1 $, the spacetime dual to the weak measurement region is given by AdS or black hole, respectively. 
In particular, there are two classes of solutions, corresponding to $\mu=0$ and $\mu>1$.

The no-bubble phase is given by $\mu = 0$.
The dual gravity theories of the unmeasured CFT and the weak measurement are given by AdS spacetimes, $AdS_{1,2}$, with different radii, similar to the weak measurement in the infinite system. 
The two AdS spacetimes are separated by interface branes, as is illustrated in figure~\ref{fig:phase_diagram} (b).
In this phase, the time reversal invariant slice $\tau = 0$ is included in $AdS_2$ that is dual to the weak measurement region.
Hence, it is referred to as the no-bubble phase.

On the other hand, in the bubble-inside-horizon and bubble-outside-horizon phases, $\mu > 1$. 
The dual gravity theory of the weak measurement region is given by a BTZ black hole. 
The interface brane separates the AdS spacetime and the black hole spacetime, as is illustrated in figure~\ref{fig:phase_diagram} (c).
In this phase, the time reversal invariant slice will cut through both the AdS spacetime and the black hole spacetime.
The AdS spacetime emerges as a bubble, surrounded by the interface brane, in the black hole spacetime. 
Furthermore, depending on whether the horizon of the black hole is included or not, the bubble solution splits into two cases. 
When the horizon is not included in the solution, we have the bubble-outside-horizon phase.
Namely, the interface brane (the boundary of the bubble) is located outside the black hole horizon, while the interior of the bubble is the AdS spacetime, so the solution does not include the horizon.
When the horizon is included in the solution, we have the bubble-inside-horizon phase.
The interface brane (the boundary of the bubble) is located inside the black hole horizon, so the full solution does include the horizon.

The phase boundary between the no-bubble phase ($\mu=0$) and the bubble phase ($\mu>1$) including both the bubble-outside-horizon phase and the bubble-inside-horizon phase is given by $f(\hat R, \hat T) = 0$~\cite{Simidzija_2020}, with
\begin{equation} \label{eq:phase_boundary}
    \begin{split}
        f(\hat R, \hat T) =& 2\pi \Theta\left( \frac{\hat T}{\hat R} - 1 \right) - \frac1{\sqrt{\hat R\hat T}} \Pi\left(0, \frac12 \sqrt{\frac{1-(\hat R-\hat T)^2}{\hat R\hat T}}   \right) \\ 
        & + \frac1{\sqrt{\hat R\hat T}} \frac{\hat R+\hat T}{\hat R-\hat T} \Pi \left( 1- \frac1{(\hat R-\hat T)^2}, \frac12 \sqrt{\frac{1-(\hat R-\hat T)^2}{\hat R\hat T}} \right),
    \end{split}
\end{equation}
where $\hat T = T L_2$, $\hat R = L_2/ L_1$, and the function $\Pi(\nu,z)$ is defined by
\begin{equation}
\label{eq:special function Pi and K}
    \Pi(\nu,z)=\int_0^1 \frac{{\rm d}t}{(1-\nu t^2)\sqrt{(1-t^2)(1-z^2t^2)}}.
\end{equation}
The phase boundary between the bubble-outside-horizon phase and the bubble-inside-horizon phase is given by $T L_1 = 1$.
The phase diagram is given in figure~\ref{fig:phase_diagram} (a). 
Notice that these phase boundaries are obtained in the limit $\epsilon \rightarrow 0$.

We further review a few quantities for the bubble phase from reference~\cite{Simidzija_2020} that will be used later in the discussion of the bubble phases. 
In the black hole solution, there is a relation between the measurement regularization $\epsilon$ and $\mu$,
\begin{equation}
\label{eq:relation of mu and epsilon}
    2 \epsilon = \frac1{\sqrt \mu} f(\hat R, \hat T).
\end{equation}
where $f(\hat R, \hat T)$ is given in~\eqref{eq:phase_boundary}.
Apparently, the measurement solution requires $\epsilon > 0 $.
The onshell action difference between the black hole phase and the AdS phase dual to the weak measurement is given by $S^{\rm BH} - S^{\rm AdS} = - \frac{\mu \epsilon}{\sqrt {\lambda_2}} + \mathcal O (\mu^{-1/2})$, which means that as long as the black hole solution exists, it dominates over the AdS solution.
Therefore, this gives the phase boundary between the no-bubble phase and the bubble phase $f(\hat R, \hat T)=0$ described before.
Specifically, when $f(\hat R, \hat T)<0$, the black hole solution does not exist, while when $f(\hat R, \hat T)>0$ the black hole solution exists and dominates over the AdS solution.
Besides, in the bubble phase with $f(\hat R,\hat T) > 0$, if we take $\epsilon \ll 1$, \eqref{eq:relation of mu and epsilon} will lead to $\mu \propto \epsilon^{-2} \gg 1$. 

\begin{figure}
    \centering
    \includegraphics[width=0.5\textwidth]{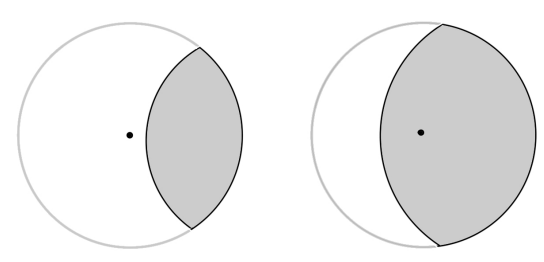}
    \caption{A zooming-in illustration of the interface brane trajectory in the black hole geometry for the bubble-outside-horizon (left) and bubble-inside-horizon phase (right).
    The shaded region is dual to the post-measurement state. 
    The horizon is present in the shaded region in the bubble-inside-horizon phase, while absent in the shaded region in the bubble-outside-horizon phase. 
    Figure from reference~\cite{Simidzija_2020}.}
    \label{fig:bubble_transition}
\end{figure}

The black hole metric dual to the weak measurement has a horizon at
\begin{equation} \label{eq:horizon_position}
    r_H = \sqrt{\frac{\mu-1}{\lambda_2}}.
\end{equation}
The interface brane, on the other hand, on the time reflection invariant slide is located at
\begin{equation}
\label{eq:zero point for Veff}
    r_0=\frac{\mu}{\sqrt{2\sqrt{\mu^2 T^2\lambda_1+\mu T^2(\lambda_2-\lambda_1- T^2)+ T^4}+\mu(\lambda_2-\lambda_1- T^2)+2 T^2}}.
\end{equation}
The interface brane trajectory in the bubble-outside-horizon phase and the bubble-inside-horizon phase is illustrated in figure~\ref{fig:bubble_transition}.
The brane trajectory in the black hole geometry starts from the boundary $r = \infty$, moves into the bulk, and returns to the boundary.
It is symmetric under time reflection transformation.
The minimal radius $r$ of the brane trajectory is given by~\eqref{eq:zero point for Veff}.
In figure~\ref{fig:bubble_transition}, the shaded region is dual to the post-measurement state. 
The horizon is present in the shaded region in the bubble-inside-horizon phase, while absent in the shaded region in the bubble-outside-horizon phase.
In both phases, we have the relation $r_0 > r_H$, and the transition point is when these two are degenerate, $r_0 = r_H$.
In the limit $\epsilon \ll 1$, i.e., $\mu \propto \epsilon^{-2} \gg 1$, we have  
\begin{equation}
    r_H \approx \sqrt{\frac{\mu}{\lambda_2}}, \qquad r_0 \approx \sqrt{\frac{\mu}{\lambda_2 - (\sqrt{\lambda_1} - T)^2 }}. 
\end{equation}
which gives the transition between the bubble-outside-horizon phase and the bubble-inside-horizon phase at $T = \sqrt{\lambda_1} $ as described before.

\subsection{Brief summary of geodesics and their lengths}

Later we will discuss the lengths of geodesics in different phases, which lead to different entanglement behaviors.
Here we briefly summarize the geodesic solutions and their lengths for different metrics.
The details of the derivation are given in appendix~\ref{convention of metric and corresponding geodesic}.

We consider the geodesics at the time reversal invariant slice $\tau = 0$ of the spacetime with the metric
\begin{equation}
\label{eq:general_metric}
    {\rm d}s^2=f(r){\rm d}\tau^2+\frac{{\rm d}r^2}{f(r)}+r^2{\rm d}x^2,
\end{equation}
where $f(r)=(1-\mu)+(r/L)^2$.  
As mentioned before, when $\mu>1$ ($\mu<1$), it describes a black hole (AdS) spacetime. 
Parametrized as $(\tau =0 , r(x) , x)$, the geodesics are given by the following relation
\begin{subequations}
\label{eq:general solution of final differential equation for geodesic for used metric sum}
\begin{equation}
\label{eq:general solution of final differential equation for geodesic for used metric black hole sum}
    r^{-2}+\left(\frac{1}{L^2(\mu-1)}-c_1 \right)\sinh^2{[\sqrt{\mu-1}(x+c_2)]} = c_1 \qquad \mu>1,
\end{equation}
\begin{equation}
\label{eq:general solution of final differential equation for geodesic for used metric thermal AdS sum}
    r^{-2}+ \left(\frac{1}{L^2(1-\mu)}+c_1 \right)\sin^2{[\sqrt{1-\mu}(x+c_2)]}=c_1 \qquad \mu<1,
\end{equation}
\end{subequations}
in the black hole geometry and the AdS geometry, respectively.
$c_1,c_2$ are two parameters to be determined.~\footnote{Do not confuse $c_{1,2}$ with the central charges.} 
The parameter $c_1$ can be considered as the size of the geodesic, which is directly related to the subsystem size.
The parameter $-c_2$ represents the location of the center of the subsystem: this is evident as the equation \eqref{eq:general solution of final differential equation for geodesic for used metric sum} is symmetric w.r.t. $x=-c_2$.
The geodesics above are called usual geodesics for black hole and AdS metrics.
While for the black hole metric, $\mu>1$, we have another geodesic solution,
\begin{equation}
\label{eq:second solution for general solution of final differential equation for geodesic for used metric black hole sum}
    r^{-2}+ \left(\frac{1}{L^2(\mu-1)}-c_1 \right)\sinh^2{[\sqrt{\mu-1}(x+c_2)]}=\frac{1}{L^2(\mu-1)}.
\end{equation}
The key difference between it and \eqref{eq:general solution of final differential equation for geodesic for used metric black hole sum} is that, \eqref{eq:second solution for general solution of final differential equation for geodesic for used metric black hole sum} has a fixed point $(r=L\sqrt{\mu-1},x=-c_2)$ on the horizon.
We call it an unusual geodesic because it is always longer than \eqref{eq:general solution of final differential equation for geodesic for used metric black hole sum} in a pure black hole spacetime.~\footnote{The comparison of the lengths of the two geodesics can be found in appendix~\ref{convention of metric and corresponding geodesic}.}
Intuitively, when we construct a geodesic with two endpoints on the boundary, it is not necessary for it to touch the horizon and then return to the boundary. 
This implies that the geodesic described by \eqref{eq:general solution of final differential equation for geodesic for used metric black hole sum} will be shorter.
Nevertheless, this difference will play an important role when we try to construct a geodesic in the bubble phase, because the geodesic will cross the interface brane.

To evaluate the length of geodesics, we start from $d=\int {\rm d}s=\int \sqrt{(r'(x))^2/f(r)+r^2}\ {\rm d}x$, where $f(r)$ is the function defined in metric \eqref{eq:general metric for AdS 3D}, and $r'(x)= \frac{{\rm d} r(x)}{{\rm d} x}$ is the derivative.
For $r(x)$ implicitly given by~\eqref{eq:general solution of final differential equation for geodesic for used metric sum}, the integral can be evaluated to give 
\begin{subequations}
\label{eq:solution of integral of geodesic length for used metric sum}
\begin{equation}
\label{eq:solution of integral of geodesic length for used metric for black hole sum}
    d=L\tanh^{-1}{\left[\frac{\tanh{(\sqrt{\mu-1}(x+c_2))}}{L\sqrt{c_1(\mu-1)}}\right]} \qquad \mu>1,
\end{equation}
\begin{equation}
\label{eq:solution of integral of geodesic length for used metric for thermal AdS sum}
    d=L\tanh^{-1}{\left[\frac{\tan{(\sqrt{1-\mu}(x+c_2))}}{L\sqrt{c_1(1-\mu)}}\right]} \qquad \mu<1.
\end{equation}
\end{subequations}
They are the lengths of the usual geodesics with two endpoints $(r^{-1}=\sqrt{c_1},x=-c_2)$ and $(r^{-1},x)$ in the black hole geometry and the AdS geometry, respectively.
Namely, it is the geodesic length starting from the symmetric point and ending at an arbitrary point $(r^{-1},x)$.
Notice that in \eqref{eq:general solution of final differential equation for geodesic for used metric sum} and \eqref{eq:solution of integral of geodesic length for used metric sum} the solutions of two cases are related: for example, starting from $\mu>1$ in \eqref{eq:solution of integral of geodesic length for used metric for black hole sum}, it can be continued to $\mu<1$, then $\sqrt{\mu-1}={\rm i}\sqrt{1-\mu}$ leads to~\eqref{eq:solution of integral of geodesic length for used metric for thermal AdS sum}.

Let's discuss the asymptotic behavior of the geodesic length.
We assume two endpoints are located at $(r=\frac{1}{\varepsilon},x=\pm\frac{1}{2}x_0)$, which correspond to a subregion $[-\frac{x_0}2, \frac{x_0}2]$ at the boundary. 
In the AdS spacetime for $\mu<1$, we can further determine $c_{1,2}$, and thus the geodesic length,
\begin{subequations}
\begin{equation}
\label{eq:total length of geodesic with unused metric for sued metric for thermal AdS sum}
\begin{split}
    \Delta d
    \approx2L\log{\left(\frac{2\sin{(\sqrt{1-\mu}\frac{x_0}{2})}}{L\sqrt{1-\mu}\varepsilon}\right)},
\end{split}
\end{equation}
where the prefactor $2$ is because the geodesic is symmetric and \eqref{eq:solution of integral of geodesic length for used metric for thermal AdS sum} computes only half of it.
Therefore, for $x_0\ll \frac{1}{\sqrt{1-\mu}}$, we obtain a logarithm $\Delta d=2L\log{\frac{x_0/L}{\varepsilon}}$, as expected.
On the other hand, in the black hole spacetime for $\mu>1$, the total length of the geodesic is 
\begin{equation}
\label{eq:total length of geodesic with unused metric for sued metric for black hole sum}
\begin{split}
    \Delta d
    \approx2L\log{\left(\frac{2\sinh{(\sqrt{\mu-1}\frac{x_0}{2})}}{L\sqrt{\mu-1}\varepsilon}\right)}.
\end{split}
\end{equation}
\end{subequations}
Therefore, for $x_0\ll \frac{1}{\sqrt{\mu-1}}$, we obtain $\Delta d=2L\log{\frac{x_0/L}{\varepsilon}}$.
But, for $x_0\gg \frac{1}{\sqrt{\mu-1}}$, we obtain $\Delta d=2L\log{(\frac{1}{\varepsilon L\sqrt{\mu-1}})}+L\sqrt{\mu-1}x_0$, which corresponds to volume-law entanglement.

For the unusual geodesic \eqref{eq:second solution for general solution of final differential equation for geodesic for used metric black hole sum} with the fixed point $(r^{-1}=\sqrt{\frac{1}{L^2(\mu-1)}}, x=-c_2)$, the length of the unusual geodesic \eqref{eq:second solution for general solution of final differential equation for geodesic for used metric black hole sum} is
\begin{equation}
\label{eq:nonphysical solution of integral of geodesic length for used metric for black hole sum}
    d=L\tanh^{-1}{\left[\sqrt{2-c_1 L^2(\mu-1)}{\tanh{(\sqrt{\mu-1}(x+c_2))}}\right]},
\end{equation}
where the other end point is at $(r^{-1},x)$. 
Again, if we assume two endpoints $(r=\frac1\varepsilon,\pm\frac{x_0}{2})$ on the geodesic, then we can solve $c_{1,2}$ and get the total length 
\begin{equation}
\label{eq:total length of nonphysical geodesic for black hole sum}
\begin{split}
    \Delta d=&2L\tanh^{-1}{\left[\left(1-\frac{L^2(\mu-1)\varepsilon^{2}}{\cosh^2{[\sqrt{\mu-1}\frac{x_0}{2}]}}\right)^{\frac{1}{2}}\right]}
    \approx2L\log{\left(\frac{2\cosh{(\sqrt{\mu-1}\frac{x_0}{2})}}{L\sqrt{\mu-1}\varepsilon}\right)}.
\end{split}
\end{equation}
However, different from \eqref{eq:total length of geodesic with unused metric for sued metric for black hole sum}, for $x_0\ll \frac{1}{\sqrt{\mu-1}}$, it does not lead to a logarithmic function.
While for $x_0\gg \frac{1}{\sqrt{\mu-1}}$, we have $\Delta d=2L\log{(\frac{1}{\varepsilon L\sqrt{\mu-1}})}+L\sqrt{\mu-1}x_0$, which is the same as the usual geodesic.
We see that the length of geodesic \eqref{eq:total length of nonphysical geodesic for black hole sum} will always be greater than \eqref{eq:total length of geodesic with unused metric for sued metric for black hole sum}.
The difference is significant for a small subsystem size $x_0$, while there is no difference at the first two orders in the large subsystem size.

Having discussed the geodesics and the corresponding lengths, we are ready to present the results of the geodesics in different phases in the following.

\subsection{No-bubble phase: marginal weak measurements}

Recall that the no-bubble phase is given by $\mu_1 = 0 $ and $\mu_2 = \mu = 0$. 
The dual gravity theories of the unmeasured CFT and the weak measurement are given by two AdS spacetimes, $AdS_{1,2}$ with different radii $L_{1,2}$, separated by interface branes.
See figure~\ref{fig:phase_diagram} (b). 
In this phase, the time reversal invariant slice is included in the $AdS_2$ spacetime with radius $L_2$.
Due to the time reversal symmetry, the RT surface of a subregion is located within the time reversal invariant slice. 
Therefore, the entanglement entropy is given accordingly by 
\begin{equation}
    S = \frac{c_\text{eff}}3 \log \frac{l}\varepsilon,
\end{equation}
where $l$ is the length of the subregion. 
We have taken the limit $\varepsilon/l \ll 1$ and $l/(2\pi) \ll 1 $.
It means that the subregion is significantly large  compared to the microscopic scale $\varepsilon$, but still much smaller than the total size of the system, which is set to be $2\pi$.
Notice that $c_\text{eff} = \frac{3L_2}{2G_\text{(3)}}$.
This corresponds to the marginal phase with a continuous effective central charge.

If we take the limit $L_2\rightarrow0$, then the entanglement will change to an area law.
This corresponds to the case where the measurement is relevant and flows towards a projection measurement.
In this case, the interface brane becomes the ETW brane.
The region dual to the weak measurement is a trivial spacetime, consistent with the gravity dual of a boundary state resulted from projection measurements.

\subsection{Geodesic in the bubble-outside-horizon phase}
\label{geodesic of bubble outside horizon}

Unlike the no-bubble phase, in the bubble phase including both the bubble-outside-horizon phase and the bubble-inside-horizon phase, the time reflection slice consists of both the black hole geometry and the AdS geometry. 
See figure~\ref{fig:phase_diagram} (c).
Accordingly, the geodesic will have different sections located within these two geometries.  
To determine the parameters in the geodesic solutions, we use not only the condition of endpoints at the boundary, but also the gluing condition at the brane that separates different geometries. 
We simplify the parameters in certain physical limits, and evaluate the corresponding geodesic length. 

\begin{figure}[tbp]
\centering 
\subfigure[]{\includegraphics[width=.45\textwidth]{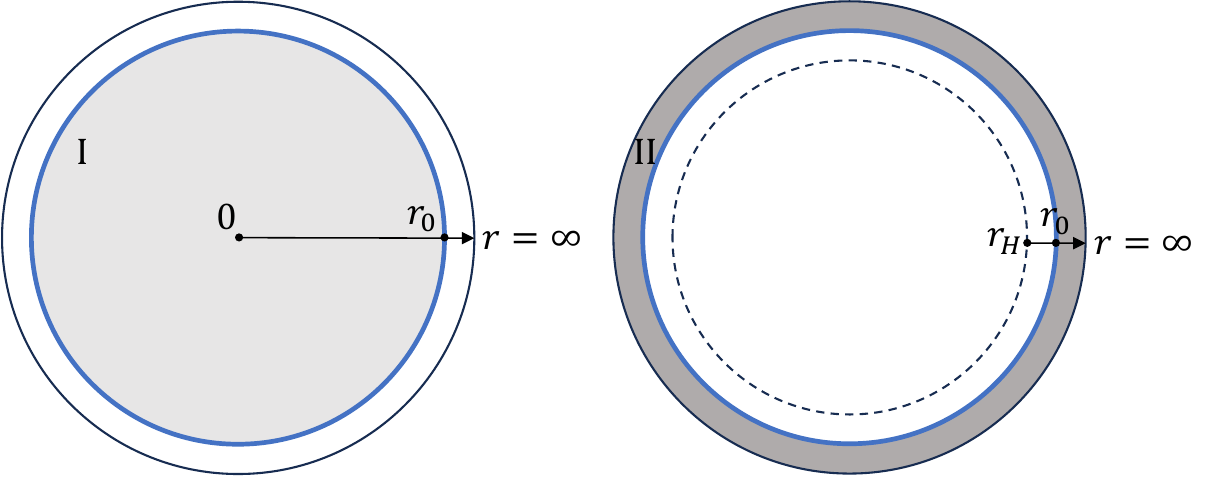}}\qquad
\subfigure[]{\includegraphics[width=.45\textwidth]{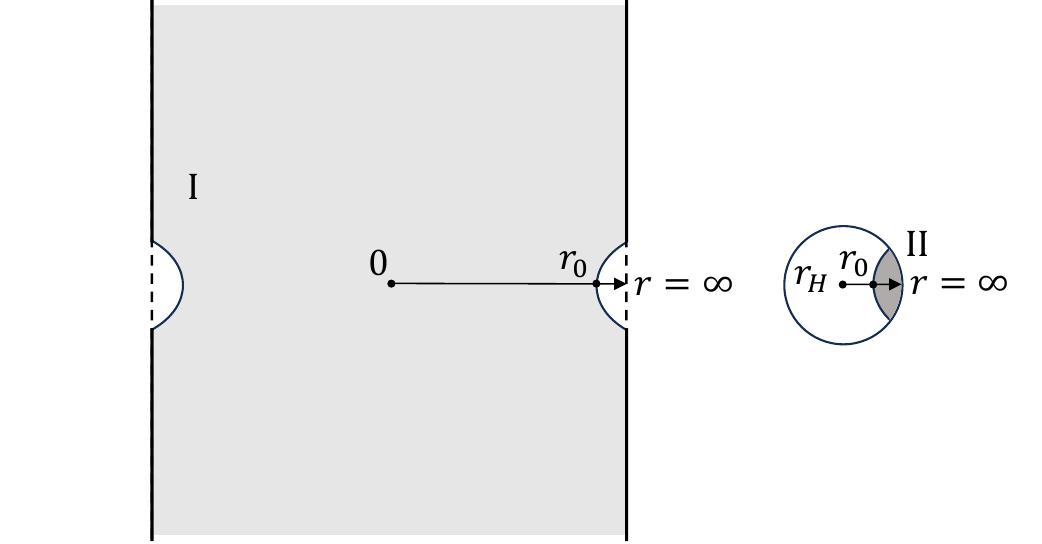}}
\caption{\label{fig:xy_yz_plane} The light (dark) shaded region denotes the AdS (black hole) spacetime. 
(a) The time reversal invariant slice in the bubble-outside-horizon phase. 
The blue circle indicates the interface brane at $r_0$ and the dashed circle indicates the horizon $r_H$. 
(b) The vertical cross-section of the cylinder indicates the AdS spacetime, and the circle indicates the black hole spacetime.
}
\end{figure}

We first consider the bubble-outside-horizon phase.
The geometry of the bubble-outside-horizon phase is shown in figure~\ref{fig:xy_yz_plane}.
The blue (dashed) circle represents the interface brane (horizon). 
It is clear that the interface brane is located outside the horizon.
We expect the entanglement entropy to satisfy a $\log$-law with prefactor $\frac{c_1}{3}$, because the black hole horizon is cut off by the interface brane and the RT surface can cross the interface brane to enter the AdS spacetime.

\begin{figure}[tbp]
\centering 
\subfigure[]{\includegraphics[width=.45\textwidth]{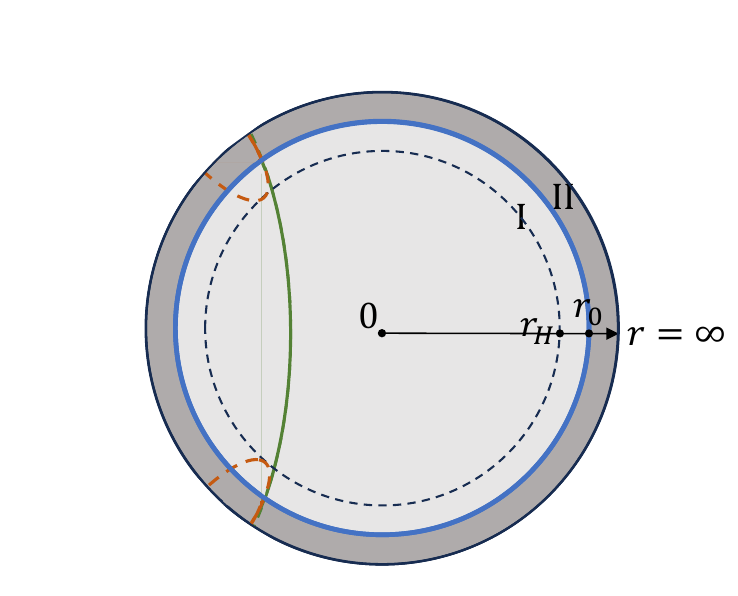}}\qquad
\subfigure[]{\includegraphics[width=.45\textwidth]{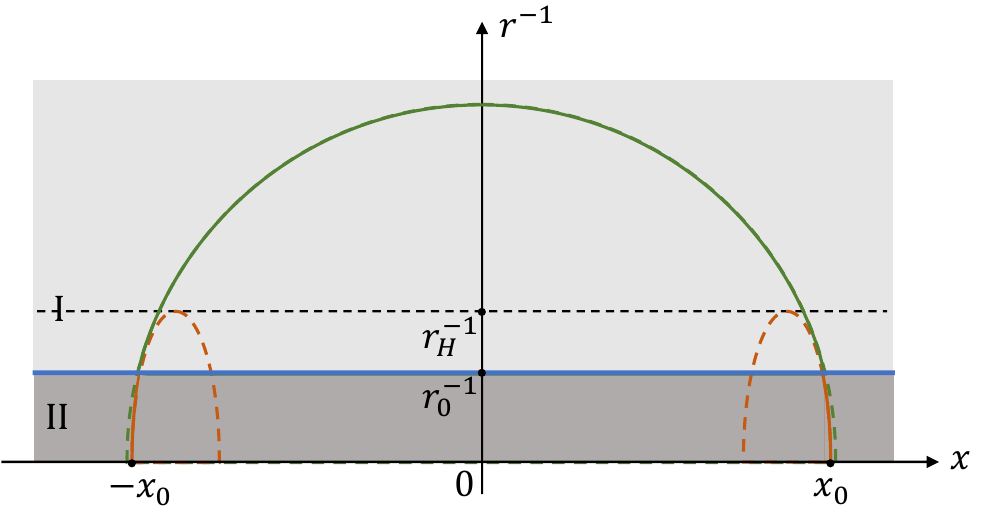}}
\caption{\label{fig:naive_geodesic} (a) Illustration of the geodesic in the bubble-outside-horizon phase. 
It consists of two parts, one in the AdS spacetime (region I), and the other in the black hole spacetime (region II). 
The AdS geodesic solution (green) is given in~\eqref{eq:general solution of final differential equation for geodesic for used metric sum}, while the black hole geodesic solution (orange) is the unusual one~\eqref{eq:second solution for general solution of final differential equation for geodesic for used metric black hole sum}.
(b) Zooming the portion of geodesics in (a). 
The geodesic solution is symmetric in the coordinate $x$.}
\end{figure}

We can focus on the entanglement entropy of a subregion in the AdS space, as shown in figure~\ref{fig:naive_geodesic}.
Let's try to solve the geodesic that crosses the brane with the connection conditions: the geodesic and its derivative are continuous on the crossing point.

We assume that the subregion in the boundary of time reversal invariant slice is located at $x \in (-\delta,\delta)$.
For the limit we considered in the following, the RT surface will cross the interface brane.
We assume the crossing point to be $(r,x) = (r_0, -x_0)$. 
Then for the geodesic in the AdS spacetime (region I in figure~\ref{fig:naive_geodesic}), the geodesic solution is given by \eqref{eq:general solution of final differential equation for geodesic for used metric thermal AdS sum}, where the parameters can be determined by the crossing point
\begin{subequations}
\label{eq:bubble-outside-horizon geodesic equation}
\begin{equation}
\label{eq:bubble-outside-horizon geodesic equation 1}
    r_0^{-2}+(L_1^{-2}+c_1)\sin^2{x_0}=c_1,
\end{equation}
where we have set parameter $c_2=0$ because the subregion is symmetric w.r.t $x=0$.

For the geodesic in the black hole spacetime (region II in figure~\ref{fig:naive_geodesic}), we can show that the geodesic equation must be the unusual form \eqref{eq:second solution for general solution of final differential equation for geodesic for used metric black hole sum}.
If it is a usual geodesic in region II, in appendix~\ref{details of calculation for geodesic in the bubble-outside-horizon phase}, we can show that the connection condition for the geodesic cannot be satisfied.
Intuitively, it is because the derivative $\frac{{\rm d}r^{-1}}{{\rm d}x}$ of the usual geodesic in region II is much smaller than that in region I.
However, for the unusual geodesic, we can have the derivative on the crossing point that is big enough to satisfy the connection condition.
Therefore, we use the unusual geodesic given in~\eqref{eq:second solution for general solution of final differential equation for geodesic for used metric black hole sum}.
We change the unknown parameter to $b_{1,2}$ to avoid confusion, and that the crossing point is located at the geodesic leads to the relation:
\begin{equation}
\label{eq:bubble-outside-horizon geodesic equation 2}
    r_0^{-2}+\left(\frac{1}{L_2^2(\mu-1)}-b_1\right)\sinh^2{\sqrt{\mu-1}(x_0-b_2)}=\frac{1}{L_2^2(\mu-1)}.
\end{equation}
Here we keep the parameter $b_2$ because, in general, the geodesic in region II is not symmetric to $x=0$ (see figure~\ref{fig:naive_geodesic} for an illustration). 
Besides, the geodesic in region II has the endpoint on the boundary $(r,x) = (\varepsilon^{-1},-\delta)$, which gives another relation
\begin{equation}
\label{eq:bubble-outside-horizon geodesic equation 3}
    \varepsilon^{2}+\left(\frac{1}{L_2^2(\mu-1)}-b_1\right)\sinh^2{\sqrt{\mu-1}(\delta-b_2)}=\frac{1}{L_2^2(\mu-1)}.
\end{equation}
Finally, as mentioned before, we require the geodesic to be smooth when crossing the interface brane, which leads to the connection condition
\begin{equation}
\label{eq:bubble-outside-horizon geodesic equation 4}
\begin{split}
    &\left.\frac{{\rm d}r^{-1}}{{\rm d}x}\right|_{r=r_0, {\rm I}}=\frac{(L_1^{-2}+c_1)\sin{x_0}\cos{x_0}}{\sqrt{c_1-(L_1^{-2}+c_1)\sin^2{x_0}}}\\
    =&\left.\frac{{\rm d}r^{-1}}{{\rm d}x}\right|_{r=r_0, {\rm II}}=\sqrt{\mu-1}\frac{(\frac{1}{L_2^{2}(\mu-1)}-b_1)\sinh{[\sqrt{\mu-1}(x_0-b_2)]}\cosh{[\sqrt{\mu-1}(x_0-b_2)]}}
    {\sqrt{b_1-(\frac{1}{L_2^{2}(\mu-1)}-b_1)\sinh^2{[\sqrt{\mu-1}(x_0-b_2)]}}}.
\end{split}
\end{equation}
\end{subequations}

These four equations are enough to determine the four unknown parameters $x_0, c_1, b_1$ and $b_2$.
However, it is difficult to obtain a general analytical expression for these parameters.
In the following, we assume that $L_i =1/\sqrt{\lambda_i}$ for $i=1,2$ are order-one numbers $\mathcal O(1)$, and we focus on the limit 
\begin{equation} \label{eq:limit approx}
\varepsilon\ll r_H^{-1}\ll x_0\ll 1,
\end{equation}  
to give analytical results. 
This limit is justified as follows:
\begin{itemize}
    \item $\varepsilon\ll r_H^{-1}$ is due to the requirement that the UV-cutoff should be the minimal length scale.
    
    \item Recall that $r_H = \sqrt{\frac{\mu-1}{\lambda_2}} $ . 
    Since we take the measurement regularization parameter to be small $\epsilon \ll 1$, it means $\mu \propto \epsilon^{-2} \gg 1$ according to~\eqref{eq:relation of mu and epsilon}.~\footnote{Note that we use $\epsilon$ to denote the regularization of the weak measurement, and $\varepsilon$ to denote the UV cutoff near the boundary.}
    See also the discussion in equation~\eqref{eq:relation of mu and epsilon}.
    Hence, $r_H^{-1} \propto 1/\sqrt{\mu} \propto \epsilon$.
    This justifies the limit $r_H^{-1} \ll x_0$. 
    
    \item $x_0\ll 1$ is because the original CFT is defined in a circle with radius $1$, and we require that the subsystem is smaller than the total system.~\footnote{While $x_0$ is not the length of the subsystem, we will see that at this limit, $\delta \sim x_0$. 
    Therefore, the limit~\eqref{eq:limit approx} is equivalent to $\varepsilon \ll r_H^{-1} \ll \delta \ll 1$. }
\end{itemize}

Intuitively, the limit is about the relation between the UV cutoff parameter $\varepsilon$, the measurement regularization parameter $r_H^{-1} \propto \epsilon$, and the subsystem size $\delta$.
It is also useful to remember that $\mu \propto \epsilon^{-2}$.
In such a limit, we first estimate the order of $r_0$, which is the location of the brane. 
From \eqref{eq:horizon_position} and \eqref{eq:zero point for Veff}, up to $\mathcal O(\mu^{-1})$ we have
\begin{equation}
\label{eq:relation between r0 and rH approx}
    \left(\frac{r_H}{r_0}\right)^2\approx \lambda_2^{-1}(- T+\sqrt{\lambda_1}+\sqrt{\lambda_2})( T-\sqrt{\lambda_1}+\sqrt{\lambda_2})=1-\left(\frac{ T-\sqrt{\lambda_1}}{\sqrt{\lambda_2}}\right)^2\equiv\eta^2,
\end{equation}
which leads to $r_H/r_0 \rightarrow \eta>0$.
In the bubble phase, $\eta<1$ and is an order-one number.
There is a subtlety: we take the limit $\mu\rightarrow\infty$ to determine the order of $r_0$, but when we solve equations \eqref{eq:bubble-outside-horizon geodesic equation}, we do not directly take this limit but leave it at the end of the calculation.
However, it is unimportant because  the results won't change in the order we consider in the discussion below.

In the limit~\eqref{eq:limit approx}, we expand variables in the order of $x_0$ and $r_H$.
From \eqref{eq:bubble-outside-horizon geodesic equation}, we can express $c_1$, $b_1$, and $b_2$ as a function of $x_0$, i.e.,
\begin{equation} \label{eq:c1_b1_b2 results}
    \begin{split}
    c_1 = & L_1^{-2}x_0^2, \\ 
    b_1=&-L_2^{2}(1-\eta^2)^{-1}L_1^{-4} x_0^2
    , \\
    b_2= & x_0 - L_1^{2}(1-\eta^2)\frac{r_H^{-2}}{{x_0}}.
    \end{split}
\end{equation}
We then plug them into \eqref{eq:bubble-outside-horizon geodesic equation} to get an equation that only involves $x_0$.
With $x_0\approx\delta$, the relevant solution reads 
\begin{equation}
\label{eq:solution of x0 with c1_b1_b2}
    x_0\approx\delta-L_1^2\sqrt{1-\eta^2}(1-\sqrt{1-\eta^2})r_H^{-2}\delta^{-1},
\end{equation}
where the second term has the order $\mathcal O(\frac{r_H^{-2}}{x_0^2}\cdot x_0)\ll \mathcal O(x_0)$.
This means to the leading order, $x_0 \approx \delta$.
Having obtained $x_0$, we then substitute it back into \eqref{eq:c1_b1_b2 results} to get the other parameters.
The detailed calculation is in appendix~\ref{details of calculation for geodesic in the bubble-outside-horizon phase}.

To verify our analytical geodesic solution, we also numerically solve~\eqref{eq:bubble-outside-horizon geodesic equation}.
With parameters $L_1=1, L_2=1.1, \mu=1000, \varepsilon=0.0001,  T=0.6, \delta=0.1$, the solutions are 
\begin{equation}
\label{eq:numerical solution for outside horizon}
    x_0=0.0979312, b_1=-0.0593183, b_2=0.0962979, c_1=0.0103255.
\end{equation}
From our analytical solution, we can get the following results 
\begin{equation}
\label{eq:analytical solution for outside horizon}
    x_0=0.0979612, b_1=-0.0598777, b_2=0.0963236, c_1=0.0095964.
\end{equation}
Thus, our analytical results are close to numerical results in the limit~\eqref{eq:limit approx}.

With the solution above, we can calculate the geodesic length.
The geodesic can be split into two halves, each of which ranges from the boundary point $x= \pm \delta$ to the symmetric point at $x = 0$. 
See figure~\ref{fig:naive_geodesic} for an illustration.
We focus on the length of one half of the geodesic, then the total length is simply $\Delta d = 2(\Delta d_1 + \Delta d_2)$, where $\Delta d_{1,2}$ denotes the length in region I and II, respectively.
For the geodesic in region I and region II, from \eqref{eq:solution of integral of geodesic length for used metric sum} and \eqref{eq:nonphysical solution of integral of geodesic length for used metric for black hole sum} we can derive the corresponding length $\Delta d_{1,2}$, and the total length of the geodesic is 
\begin{equation}
\label{eq:total length of geodesic for bubble-outside-horizon}
\begin{split}
    \Delta d=&2L_2\left\{\log{\left(\frac{2r_H^{-1}}{\varepsilon}\right)}-\tanh^{-1}{(\sqrt{1-\eta^2})}\right\}+2 L_1\log{\left(\frac{2L_1^{-1}\delta}{r_0^{-1}}\right)},
\end{split}
\end{equation}
where we keep the leading terms in the limit \eqref{eq:limit approx}.

It is easy to see that the last term in~\eqref{eq:total length of geodesic for bubble-outside-horizon} originates from the geodesic section in the AdS spacetime (region I). 
It can be understood as the length of the usual geodesic~\eqref{eq:total length of geodesic with unused metric for sued metric for thermal AdS sum}, with the cutoff shifted to $r_0^{-1}$.
On the other hand, the first two terms in~\eqref{eq:total length of geodesic for bubble-outside-horizon} originate from the geodesic section in the black hole spacetime (region II).~\footnote{We will discuss these two terms more in the following subsection. 
See the discussion of figure \ref{fig:different_part}.}
If the weak measurement parameter is fixed, namely, $\lambda_{1,2}$, and $T$ are fixed, then these terms are also fixed and do not change with $\delta$.
The only dependence w.r.t. the subsystem size $2\delta$ comes from the last term. Hence, the corresponding entanglement entropy satisfies a $\log$-law.
Namely, the entanglement entropy of the subsystem is given by
\begin{equation}
    S = \frac{c_1}{3} \log (2\delta),
\end{equation}
where we only keep the dependence of $\delta$.
The prefactor of the logarithm is the same as the central charge of the unmeasured CFT, which means the weak measurement is irrelevant.

\begin{figure}[tbp]
\centering 
{\includegraphics[width=.45\textwidth]{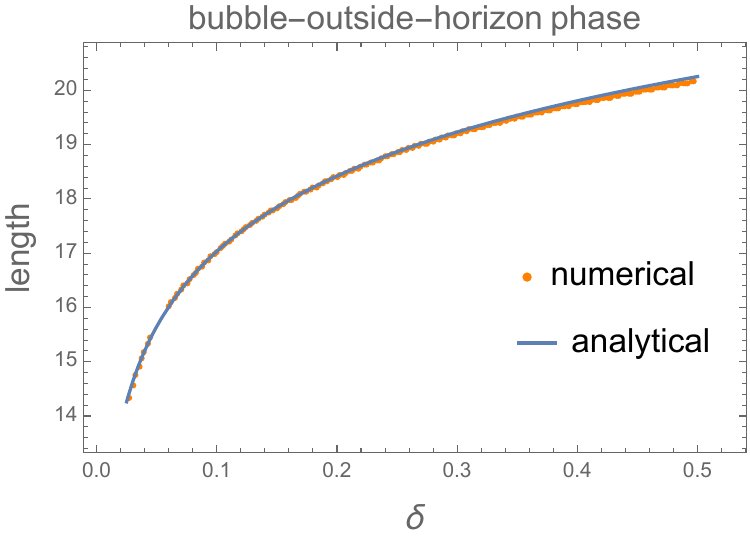}}
\caption{\label{fig:outside_horizon_geodesic_length} Geodesic length as a function of the subsystem size in the bubble-outside-horizon phase. 
The parameters are chosen to be $L_1=1, L_2=1.1, \mu=1000, \varepsilon=0.0001,  T=0.6$.}
\end{figure}

We numerically compute the geodesic length for different sizes of the subsystem, ranging from $0.025$ to $0.5$.
In figure~\ref{fig:outside_horizon_geodesic_length} we plot the numerical results and analytical results of \eqref{eq:total length of geodesic for bubble-outside-horizon}, it indicates that they are consistent with each other.

\subsection{Geodesic in the bubble-inside-horizon phase}
\label{sec:Geodesic in the bubble-inside-horizon phase}

\begin{figure}[tbp]
\centering 
{\includegraphics[width=.6\textwidth]{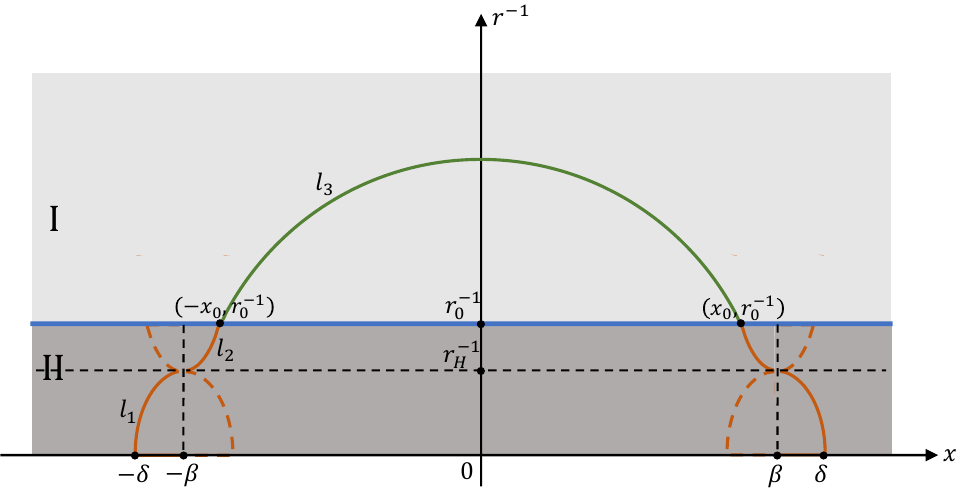}}
\caption{\label{fig:geodesic_for_inside_horizon} Illustration of the RT surface in the bubble-inside-horizon phase.
The blue line indicates the interface brane, and the dashed line indicates the horizon.
The RT surface has three parts, denoted by $l_{1,2,3}$.
The region II has two parts $l_{1,2}$ that connect at the horizon.
They are given by the unusual geodesic solution. 
$\pm\beta$ is the symmetric point of the solution of the parts $l_{1,2}$.
$(-x_0, r_0^{-1})$ is the connecting point between the geodesics of the part $l_3$ and $l_2$. 
}
\end{figure}

Now we consider the entanglement entropy of a subregion in the bubble-inside-horizon phase. 
In this phase, the bubble is inside the horizon, in other words, the black hole horizon exists in the bulk solution, as shown in the left panel of figure~\ref{fig:length_of_circle}.
A zooming-in figure is given by figure~\ref{fig:geodesic_for_inside_horizon}.
The blue (dashed) line represents the interface brane (horizon), and the light (dark) shaded region represents the AdS (black hole) spacetime. 
We still use the region I and region II to denote the AdS and the black hole spacetimes, respectively.

Again, we assume that the subregion in the boundary of time reversal invariant slice is located at $x \in (-\delta,\delta)$.
Naively, we would expect that the geodesic starting from the boundary would be located outside the horizon, and contained in the back hole geometry (region II).
If this was true, then the entanglement entropy would satisfy a volume law as discussed in equation~\eqref{eq:total length of geodesic with unused metric for sued metric for black hole sum}.
However, in the following, we will show that there exists a new kind of geodesic that can cross the horizon.
The new geodesic has a shorter length and leads to a $\log$-law with prefactor $\frac{c_1}{3}$, and, therefore, is preferred.

The new geodesic has two sections located in the AdS geometry (region I), and the black hole geometry (region II), respectively.
In region II, the usual geodesic in black hole geometry will never reach the horizon, so the unusual geodesic~\eqref{eq:second solution for general solution of final differential equation for geodesic for used metric black hole sum} that has a fixed point on the horizon $(x,r) = (-c_2, r_H)$ is needed to cross the horizon and connect to the section in the region I.
Besides, this unusual geodesic \eqref{eq:second solution for general solution of final differential equation for geodesic for used metric black hole sum} is symmetric w.r.t the fixed point on the horizon $x=-c_2$, so the derivative vanishes at the horizon $\left.\frac{{\rm d}r^{-1}}{{\rm d}x}\right|_{x=-c_2}=0$.
See figure~\ref{fig:geodesic_for_inside_horizon} for an illustration (the symmetric point is denoted by $\beta$ instead of $c_2$ in the figure).
We emphasize that the section of this new geodesic in the region I contributes to the reduction in length, compared to a geodesic that never crosses the horizon.

In figure~\ref{fig:geodesic_for_inside_horizon}, the geodesic has three parts, denoted as $l_{1,2,3}$.
For the parts $l_1$ and $l_2$ located within the black hole spacetime (region II), the corresponding geodesics are given by \eqref{eq:second solution for general solution of final differential equation for geodesic for used metric black hole sum}, with the same $c_2$ but different $c_1$.
The part $l_3$ is the usual geodesic \eqref{eq:general solution of final differential equation for geodesic for used metric thermal AdS sum} in the AdS metric.~\footnote{Actually, there is another possibility that between part $l_1$ and $l_2$ there can be an additional geodesic along the horizon, but in the appendix~\ref{details of calculation for geodesic in the bubble-inside-horizon phase} we show that this possibility always leads to a longer geodesic. 
So we only consider three parts $l_{1,2,3,}$.}
Assuming that two endpoints on the boundary are $(r^{-1},x)=(\varepsilon,\pm\delta)$, for any $\beta$ (which is defined as the symmetric point of the geodesic of part $l_{1,2}$ as shown in the figure) we have geodesic solutions as shown in figure~\ref{fig:geodesic_for_inside_horizon}.
In the following, we will first solve the general geodesic equations for any given $\beta$ and then determine the $\beta$ such that the total length of the geodesic is minimal.
To obtain an analytical result, we again consider the limit,
\begin{equation}  \label{eq:limit2}  \varepsilon\ll r_H^{-1}\ll \delta\ll1.
\end{equation}

The geodesics for the part $l_1$ and $l_2$ within the black hole spacetime (region II) have the form of equation~\eqref{eq:second solution for general solution of final differential equation for geodesic for used metric black hole sum}.
We denote the unknown parameters in the $l_1$ part by $b'_{1,2}$, and denote the unknown parameters in the $l_2$ part by $b_{1,2}$. 
For the geodesic of the part $l_3$ in the AdS spacetime (region I) given by~\eqref{eq:general solution of final differential equation for geodesic for used metric thermal AdS sum}, we denote the unknown parameters by $c_{1,2}$. 
As shown in figure~\ref{fig:geodesic_for_inside_horizon}, the symmetric point of the part $l_{1,2}$ is given by $b_2 = b'_2 = \beta$, while the symmetric point of the part $l_3$ is given by $c_2 = 0$. 

The other parameters $b_1$, $b'_1$, and $c_1$ can be determined in the following.
For the part $l_1$, that the endpoint $(r^{-1},x)=(\varepsilon,-\delta)$ is located on it gives the following equation for any given $\beta$
\begin{subequations}
\label{eq:geodesic in l123}
\begin{equation}
\label{eq:geodesic in l123 part 1}
    \varepsilon^{2}+\left(\frac{1}{L_2^2(\mu-1)}-b_1'\right)\sinh^2{[\sqrt{\mu-1}(\delta-\beta)]}=\frac{1}{L_2^2(\mu-1)}.
\end{equation}
For the parts $l_2$ and $l_3$, we have the following equations, where the first two come from the crossing point $(r^{-1},x)=(r_0^{-1},-x_0)$, and the last one is from the connection condition,
\begin{equation}
\label{eq:geodesic in l123 part 2}
    r_0^{-2}+(L_1^{-2}+c_1)\sin^2{x_0}=c_1,
\end{equation}
\begin{equation}
\label{eq:geodesic in l123 part 3}
    r_0^{-2}+\left(\frac{1}{L_2^2(\mu-1)}-b_1\right)\sinh^2{[\sqrt{\mu-1}(x_0-\beta)]}=\frac{1}{L_2^2(\mu-1)},
\end{equation}
\begin{equation}
\label{eq:geodesic in l123 part 4}
\begin{split}
    &\left.\frac{{\rm d}r^{-1}}{{\rm d}x}\right|_{r=r_0, {\rm I}}=\frac{(L_1^{-2}+c_1)\sin{x_0}\cos{x_0}}{\sqrt{c_1-(L_1^{-2}+c_1)\sin^2{x_0}}}\\
    =&-\left.\frac{{\rm d}r^{-1}}{{\rm d}x}\right|_{r=r_0, {\rm II}}=-\sqrt{\mu-1}\frac{(\frac{1}{L_2^{2}(\mu-1)}-b_1)\sinh{[\sqrt{\mu-1}(x_0-\beta)]}\cosh{[\sqrt{\mu-1}(x_0-\beta)]}}
    {\sqrt{\frac{1}{L_2^2(\mu-1)}-(\frac{1}{L_2^{2}(\mu-1)}-b_1)\sinh^2{[\sqrt{\mu-1}(x_0-b_2)]}}}.
\end{split}
\end{equation}
\end{subequations}
Notice that the minus sign in the last equation is because the definition of the geodesic in region II is opposite in the $r$ direction.
It is clear if we compare the part $l_2$ and the part $l_1$ in figure~\ref{fig:geodesic_for_inside_horizon}.

From the four equations~\eqref{eq:geodesic in l123}, we can determine four parameters, $b_1$, $b'_1$, $c_1$, and $x_0$. 
Recall that the parameter $\beta$ will be instead determined by the requirement of the minimal geodesic length.
From \eqref{eq:geodesic in l123 part 2} and \eqref{eq:geodesic in l123 part 3}  we can get $c_1$ and $b_1$ as a function of $x_0$
\begin{equation}
\label{eq:represent c1/b1 with x0}
    c_1=\frac{r_0^{-2}+L_1^{-2}\sin^2{x_0}}{\cos^2{x_0}},\qquad b_1=\frac{1}{L_2^2(\mu-1)}-\frac{\frac{1}{L_2^2(\mu-1)}-r_0^{-2}}{\sinh^2{[\sqrt{\mu-1}(x_0-\beta)]}}.
\end{equation}
Substituting these two expressions into \eqref{eq:geodesic in l123 part 4}, we will get an equation for $x_0$,
\begin{equation}
\label{eq:transformed eq for x0 exact}
    \tan{x_0}\tanh{[\sqrt{\mu-1}(\beta-x_0)]}=\frac{\sqrt{\mu-1} (\frac{1}{L_2^{2}(\mu-1)}-r_0^{-2})}{(L_1^{-2}+r_0^{-2})}.
\end{equation}
In appendix~\ref{details of calculation for geodesic in the bubble-inside-horizon phase}, we give a detailed analysis of $\beta$ and find that $\beta$ must have the same order of $\delta$ and $|\delta-\beta|\ll\delta$.
Then we can apply this approximation to solve the equation for $x_0$, which deduces that
\begin{equation}
\label{eq:solution of x0 for inside horizon approx}
    x_0=\beta-(1-\eta^2)L_1^2 r_H^{-2}\beta^{-1}.
\end{equation}
Here, the parameter $\eta$ is still given by the equation~\eqref{eq:relation between r0 and rH approx}:
\begin{equation}
    \eta = \sqrt{1-\left(\frac{ T-\sqrt{\lambda_1}}{\sqrt{\lambda_2}}\right)^2}.
\end{equation}
With this solution~\eqref{eq:solution of x0 for inside horizon approx}, the other parameters \eqref{eq:represent c1/b1 with x0}, and thus the geodesic length can be expressed as a function of $\beta$.

We can verify our analytical result by numerically calculation.
For instance, using parameters $L_1=1, L_2=1.1, \mu=10000, \varepsilon=0.00001,  T=1.6, \beta=0.1$, the numerical result of \eqref{eq:transformed eq for x0 exact} is 
$x_0=0.0996397$, and the analytical solution, \eqref{eq:solution of x0 for inside horizon approx} gives $x_0=0.09964$.
Hence, our analytical result is close to the numerical calculation in the limit~\eqref{eq:limit2}.

Now we can calculate the geodesic length with the solution above.
Using \eqref{eq:geodesic in l123 part 1} and \eqref{eq:total length of nonphysical geodesic for black hole sum} for the part $l_1$, \eqref{eq:geodesic in l123 part 3} and \eqref{eq:total length of nonphysical geodesic for black hole sum} for the part $l_2$ and \eqref{eq:geodesic in l123 part 2} and \eqref{eq:total length of geodesic with unused metric for sued metric for thermal AdS sum} for the part $l_3$, we can arrive at the total length of the geodesic
\begin{equation}
\label{eq:final total length of 123 with approx}
\begin{split}
    \Delta d_{l_1}+\Delta d_{l_2}+ \Delta d_{l_3} 
    \approx & L_2\log{\left(2\cosh{(\sqrt{\mu-1}(\delta-\beta))}\right)}
    +\frac{1}{2}L_2^{-1}(1-\eta^2)^{\frac{3}{2}} L_1^4 r_H^{-2}\beta^{-2}
    \\
    &+L_1\log{\left(\beta-(1-\eta^2)L_1^2 r_H^{-2}\beta^{-1}\right)}+{\rm const},
\end{split}
\end{equation}
where ${\rm const}$ denotes the terms independent of $\beta$. 
And we have used the approximation before and \eqref{eq:solution of x0 for inside horizon approx}.

Now the last thing we need to do is minimizing the geodesic length \eqref{eq:final total length of 123 with approx} w.r.t. $\beta$.
In appendix~\ref{details of calculation for geodesic in the bubble-inside-horizon phase}, we show that the minimal point of the geodesic length is given by
\begin{equation}
\label{eq:minimize length with beta}
    \beta \approx \delta-L_1 L_2 r_H^{-2} \delta^{-1}.
\end{equation}
Therefore, substituting \eqref{eq:minimize length with beta} into \eqref{eq:final total length of 123 with approx}, the geodesic length $\Delta d = 2 (\Delta d_{l_1} +\Delta d_{l_2} + \Delta d_{l_3})$ reads
\begin{equation}
\label{eq:final total length of 123 with beta}
    \Delta d \approx 2L_2\left\{\log{\left(\frac{2r_H^{-1}}{\varepsilon}\right)}+\tanh^{-1}{(\sqrt{1-\eta^2})}\right\}+2L_1\log{\left(\frac{2L_1^{-1}\delta}{r_0^{-1}}\right)},
\end{equation}
where in the derivation we have ignored higher orders in $\mathcal O(\frac{\delta^2}{r_H^{-2}})$.

It is easy to see that the last term in~\eqref{eq:final total length of 123 with beta} is from the geodesic section in the AdS geometry (region I). It can be understood as the length of the usual geodesic~\eqref{eq:total length of geodesic with unused metric for sued metric for thermal AdS sum} with the cutoff shifted to $r_0^{-1}$.
On the other hand, the first two terms in~\eqref{eq:final total length of 123 with beta} originate
from the geodesic section in the black hole spacetime (region II). 
It is clear that when the measurement parameters are fixed, these two terms do not depend on the size of the subsystem $2\delta$. 
The only dependence w.r.t. the subsystem size comes from the last term. 
The corresponding entanglement entropy satisfies a $\log$-law:
\begin{equation}
    S = \frac{c_1}{3} \log (2\delta).
\end{equation}
The prefactor of the logarithmic entanglement entropy will be the same as the central charge of the unmeasured CFT, which means the weak measurement is irrelevant.

\begin{figure}[tbp]
\centering 
\subfigure[]{\includegraphics[width=.3\textwidth]{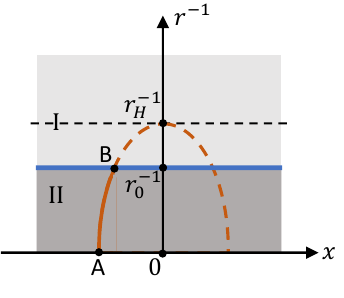}}\qquad
\subfigure[]{\includegraphics[width=.3\textwidth]{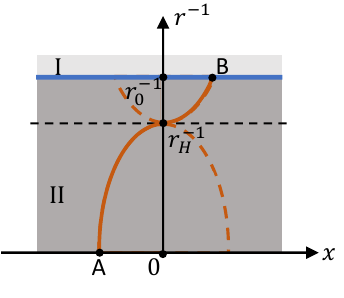}}
\caption{\label{fig:different_part} (a) The geodesic in the region II in the bubble-outside-horizon phase. 
(b) The geodesic in the region II in the bubble-inside-horizon phase.}
\end{figure}

Here is one additional remark for the first two terms in equation~\eqref{eq:total length of geodesic for bubble-outside-horizon} and~\eqref{eq:final total length of 123 with beta}. 
Apparently, they correspond to the length of the geodesic section in the black hole geometry. 
The first term $2L_2 \log{\left(\frac{2r_H^{-1}}{\varepsilon}\right)}$, present in both equations, is the leading contribution of the geodesic length from the horizon to the boundary. 
While the second term $2L_2 \tanh^{-1}{(\sqrt{1-\eta^2})}$ has opposite signs in the equation~\eqref{eq:total length of geodesic for bubble-outside-horizon} and~\eqref{eq:final total length of 123 with beta}.
It is because for \eqref{eq:total length of geodesic for bubble-outside-horizon} we have a part of the geodesic as shown in the orange solid curve in figure~\ref{fig:different_part} (a), while for \eqref{eq:final total length of 123 with beta} we have a part of the geodesic in figure~\ref{fig:different_part} (b).
It is clear from the figure that the difference comes from the geodesic length from the horizon (the dashed line) to the interface brane (the blue solid line).
This length gives exactly the second term.
Hence, for figure~\ref{fig:different_part} (a) we have $2L_2 \log{\left(\frac{2r_H^{-1}}{\varepsilon}\right)}-2L_2 \tanh^{-1}{(\sqrt{1-\eta^2})}$, while for figure~\ref{fig:different_part} (b) we have $2L_2 \log{\left(\frac{2r_H^{-1}}{\varepsilon}\right)}+2L_2 \tanh^{-1}{(\sqrt{1-\eta^2})}$.

\begin{figure}[tbp]
\centering 
{\includegraphics[width=.45\textwidth]{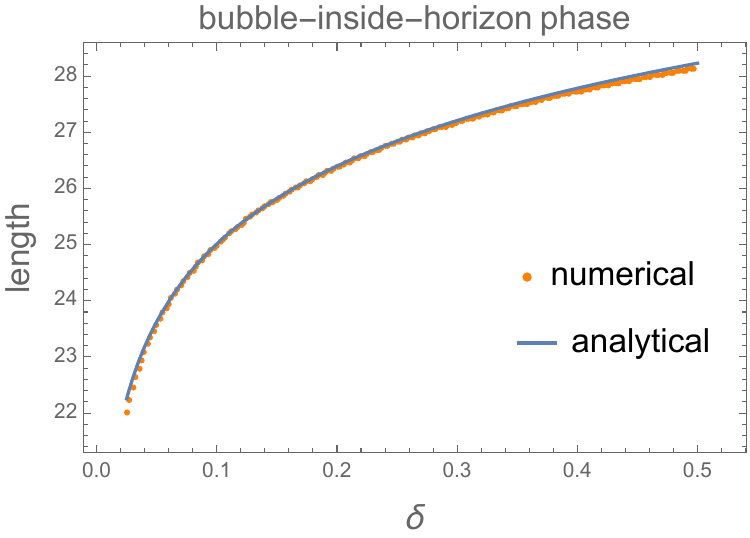}}
\caption{\label{fig:inside_horizon_geodesic_length} Geodesic length as a function of the subsystem size in the bubble-inside-horizon phase. 
The parameters are chosen to be $L_1=1, L_2=1.1, \mu=10000, \varepsilon=0.00001,  T=1.6$.
We do not use the variation to determine $\beta$ from $\delta$, instead, we use the approximation~\eqref{eq:minimize length with beta}.}
\end{figure}

Finally, we numerically compute the length of the geodesic for different sizes of the subsystem, ranging from $0.025$ to $0.5$.
In figure~\ref{fig:inside_horizon_geodesic_length}, we plot the numerical results and analytical results of \eqref{eq:final total length of 123 with beta}, and it strongly suggests that they are consistent with each other.

\section{Several extensions}
\label{sec:several extensions}

\subsection{Thick brane description}

In section~\ref{sec:measurement/measurement geodesic in infinite system}, we have considered weak measurements in an infinite system using a bottom-up thin-brane description. 
Here, we discuss the thick-brane description of the weak measurement and the relation to spatial interface CFT in an infinite system.
In the following, we set $L=1$ for simplicity.

\begin{figure}
    \centering
    \subfigure[]{\includegraphics[width=0.5\textwidth]{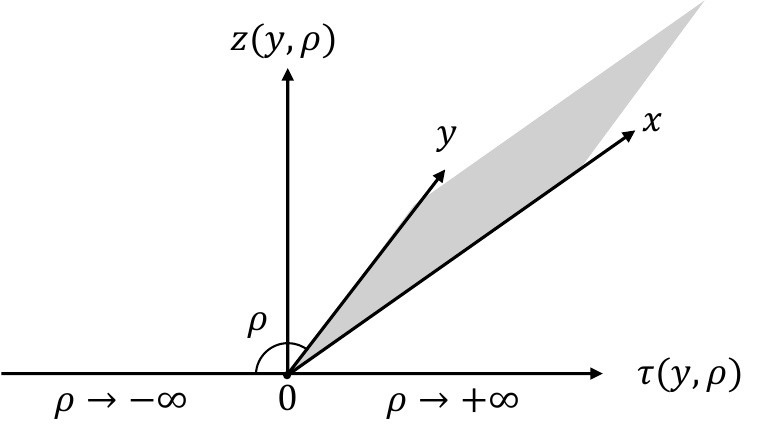}} \\
    \subfigure[]{\includegraphics[width=0.45\textwidth]{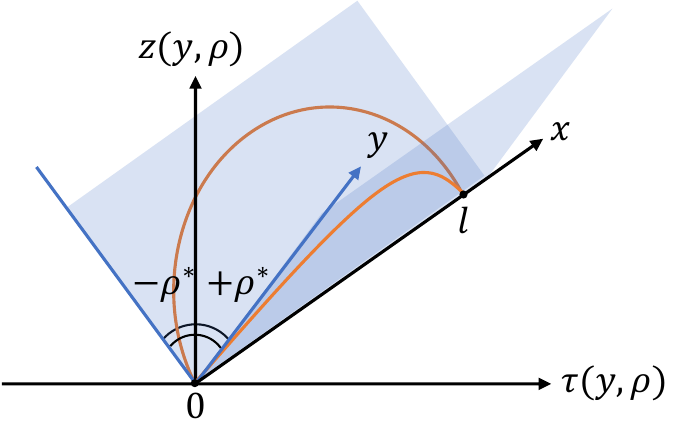}} \qquad 
    \subfigure[]{\includegraphics[width=0.45\textwidth]{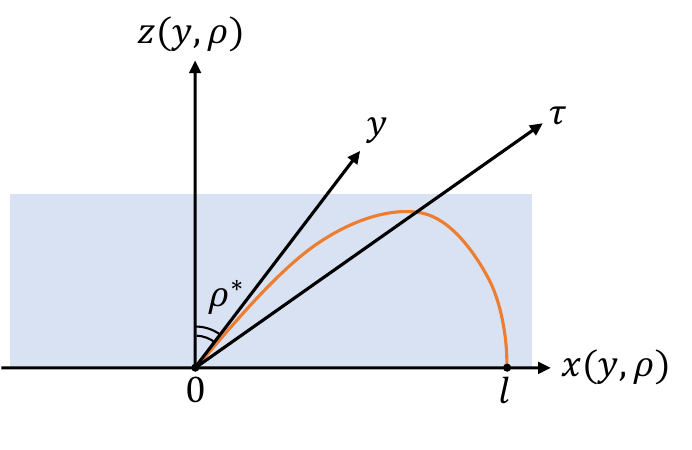}}
    \caption{(a) An illustration of the foliation coordinate~\eqref{eq:foliation_metric}. 
    Weak measurements occur at $\tau = 0$.
    For $\tau \ne 0$, the asymptotic boundary is given by the unmeasured CFT $\rho \rightarrow \pm \infty$.
    (b) The RT surface colored in orange connecting the two ends of the interval $x\in[0,l]$ is located in the $\rho = \pm \rho^\ast$ slice colored in blue for the weak measurement case.
    (c) The RT surface colored in orange connects two endpoints of $x \in [0, l]$ for the spatial interface CFT at the $\tau = 0$ slice colored in blue. 
    Note that the coordinate $x$ and $\tau$ are exchanged.}
    \label{fig:thick_brane}
\end{figure}

We will first discuss the subsystem entanglement entropy of the weak measurement, and then implement a spacetime rotation to investigate the spatial interface CFT. 
To this end, we consider a foliation of a general 3d Euclidean metric
\begin{equation} \label{eq:foliation_metric}
    {\rm d}s^2 = {\rm d}\rho^2 + e^{2A(\rho)} \frac{{\rm d}x^2 + {\rm d}y^2}{y^2},
\end{equation}
where $A(\rho)$ is a warpfactor that controls the size of each ${\rm AdS_2}$ slice.
Figure~\ref{fig:thick_brane} (a) illustrates the foliation. 
The Euclidean Poincare coordinate is related to the foliation by $z(y,\rho)$ and $\tau(y,\rho)$. 
As before, the weak measurement occurs at $\tau=0$. 
The post-measurement state has an effective central charge denoted by $c_{\rm eff}$.
For $\tau \ne 0$, the asymptotic region $z \rightarrow 0$ is given by the unmeasured CFT with central charge $c_1$.
In the foliation coordinate, the measurement region at the boundary is reached by $\lim_{y\rightarrow 0} \tau(y,\rho) = 0$, and the unmeasured CFT region at the boundary is reached by $ \rho \rightarrow \pm \infty$ as shown in figure~\ref{fig:thick_brane} (a).

It is helpful to look at the pure ${\rm AdS_3}$ metric: when
\begin{equation}
    e^{A(\rho)} = \cosh \rho,
\end{equation}
the metric is reduced to an ${\rm AdS_3}$ metric.
One can check this via a coordinate transformation similar to \eqref{eq:coordinate transformation},
\begin{equation}
    z=\frac{y}{\cosh\rho},\qquad \tau=y \tanh{\rho}.
\end{equation}
The only difference is that $x$ and $\tau$ are exchanged in~\eqref{eq:coordinate transformation}.
From this coordinate transformation, because $y \ge 0$, if the metric has time reversal symmetry, we can deduce that the metric is invariant under $\rho \rightarrow - \rho$.
Of course, this is a trivial example because there is no measurement in the pure ${\rm AdS_3}$ metric.
Nevertheless, we expect this property is general in the case of weak measurement. 
Namely, the imaginary time satisfies $\tau(y, -\rho) = - \tau(y, \rho)$. 
We make the following assumptions for the warpfactor: 
\begin{enumerate}
\item $A(\rho)$ is an even function of $\rho$, i.e., $A(\rho) = A(-\rho)$. This is because of the time reversal invariance. 
\item For $\rho \rightarrow \pm \infty$, we have $e^{A(\rho)} \sim \cosh \rho$, i.e., we recover ${\rm AdS_3}$ away from the measurement in the unmeasured CFT region.
\item The minimal value of the warpfactor is located at $\rho = \pm \rho^\ast$ with $A^* = A(\rho^\ast)$. 
We assume $\rho^\ast \ge 0$ without loss of generality. 
At this minimal point, the first-order derivative vanishes $A'(\rho^\ast)=\frac{{\rm d}A}{{\rm d}\rho} = 0$.
Notice that $\rho^\ast = 0 $ is a special case of this.
\end{enumerate}

We consider entanglement entropy of a subsystem $x \in [0,l]$. 
The corresponding RT surface is parametrized by $(y(x), \rho(x) )$.
Then the surface area of the RT surface can be rewritten as an action $\mathcal A = \int {\rm d}x \mathcal L $ with the Lagrangian,
\begin{equation}
    \mathcal L = \sqrt{e^{2A(\rho)} \frac{\dot y^{2} + 1}{y^2} + \dot \rho^{2}}, \quad \dot y = \frac{{\rm d}y}{{\rm d}x}, \quad \dot \rho = \frac{{\rm d}\rho}{{\rm d}x}.
\end{equation}

The RT surface is a saddle point of this action functional, so let's evaluate the equation of motion from the Lagrangian: 
\begin{equation}
    \begin{split} 
        &  e^{2A(\rho)}A'(\rho)(1+\dot y^2)^2 -y[(1+\dot y^2)(\dot\rho\dot y+y(-2A'(\rho)\dot \rho^2+\Ddot{\rho}))- y \dot \rho \dot y \Ddot{y}]= 0, \\
        & e^{2A(\rho)}[(1+\dot y^2)(1+yA'(\rho)\dot\rho\dot y)+y\Ddot{y}]+y^2\dot\rho[2yA'(\rho)\dot\rho^2\dot y-y\dot y\Ddot{\rho}+\dot\rho(1-\dot y^2+y\Ddot{y})]= 0, 
    \end{split}
\end{equation}
where the first and second equation comes from the variation of $\rho(x)$ and $y(x)$, respectively.
It is straightforward to check that a solution to the first equation is given by $\rho(x) = \rho^\ast$ because $A'(\rho^\ast) = 0$ and $\dot \rho = 0$.
It means that the RT surface is located at the slice with the minimal warpfunction $A^\ast \equiv A(\rho^\ast)$.  
Then, the equation for the geodesic is 
\begin{equation}
    1+\dot y^2+y\Ddot{y}=0,
\end{equation}
which gives a well-known solution 
\begin{equation}
\label{eq:geodesic for general case eA}
    y^2+(x-c_1)^2=c_2^2,
\end{equation}
where $c_{1,2}$ are integration constants determined by the endpoints $[0,l]$.
Therefore, the geodesic is an arc in $\rho=\rho^*$ slice.
In this slice, the metric is ${\rm AdS_2}$ with radius $e^{A^*}$.
With the results before, the area of the RT surface \eqref{eq:geodesic for general case eA} at the leading order of $l$ is 
\begin{equation}
\label{eq:geodesic length for general case eA}
    \mathcal A = 2 e^{A^\ast} \log l,
\end{equation}
which gives the effective central charge $c_{{\rm eff}}=\frac{3e^{A^\ast}}{2G_{(3)}}$, and $S_{l}=\frac{c_{\rm eff}}{3}\log{l}$. 
The factor of two in front is because there are two endpoints at the boundary of $y \rightarrow 0$. 
Notice that for $\rho^\ast \ne 0$, the RT surface is not located at the time reversal invariant slice. 
There is also a degenerate RT surface at $\rho = - \rho^\ast$. 
This is illustrated in figure~\ref{fig:thick_brane} (b).~\footnote{ 
There is also a solution at the time reversal invariant slice because $A'(0) = 0$ for the even function $A(\rho)$. 
This solution has the area given by $2 e^{A(0)} \log l > 2 e^{A^\ast} \log l$ because $A(0) > A^\ast$.}

Now, let's consider a spatial interface CFT via a spacetime rotation $\tau \leftrightarrow x$. 
The rotated metric from~\eqref{eq:foliation_metric} is 
\begin{equation}
\label{eq:thick brane defect metric for 3d}
    {\rm d}s^2 = {\rm d}\rho^2 + e^{2A(\rho)} \frac{{\rm d}\tau^2 + {\rm d}y^2}{y^2}. 
\end{equation}
An illustration of this coordinate is shown in figure~\ref{fig:thick_brane} (c). 
Note that the difference of coordinates between figure~\ref{fig:thick_brane} (b) and figure~\ref{fig:thick_brane} (c) is the exchange of $x$ and $\tau$ axis.
The spatial coordinate $x$ is given by $x = x( y, \rho)$.
Due to the time reflection symmetry in the case of weak measurements, the corresponding interface CFT after the spacetime rotation is symmetric under reflection along the interface. 
Hence, the interface is located at $x = 0$ and the spatial coordinate $x$ satisfies $x(y,-\rho) = - x (y,\rho)$. 
The asymptotic boundary $z \rightarrow 0$ for $x\ne 0$ is given by CFTs to the left and right of the interface.
These two boundaries are given by $\rho \rightarrow \pm \infty$, respectively.

We consider the entanglement entropy of an interval $x\in [0,l]$, where one endpoint is located on the interface and the other endpoint is located on the CFT, as shown in figure~\ref{fig:thick_brane}. 
In this case, the RT surface is parametrized by $y(\rho)$ at a fixed time slice $\tau = 0$.
In the following, we use a similar approach in reference~\cite{karch2021universal}.
The area functional is $\mathcal A = \int {\rm d}\rho \mathcal L$ with the Lagrangian
\begin{equation}
    \mathcal L = \sqrt{1 + e^{2A} \frac{\dot y^2}{y^2}}, \quad \dot y = \frac{{\rm d}x}{{\rm d}\rho}.
\end{equation}
The Lagrangian leads to the geodesic equation 
\begin{equation} \label{eq:geodesic_equation}
    \frac{\dot y}y = \frac{c_s e^{-A}}{\sqrt{e^{2A} - c_s^2}}, 
\end{equation}
where $0 \le c_s \le e^{A^\ast}$ is an integration constant.

In general, $\rho$ can run from $-\infty $ to $\infty$, and $c_s$ determines the endpoints of the interval $l_L$ and $l_R$. 
An exceptional case is that when $c_s = e^{A^\ast}$, the right-hand-side of~\eqref{eq:geodesic_equation} diverges at $\rho = \rho^\ast$.
The only possible solution is $y = 0 $ at $\rho^\ast$.
Therefore, the left endpoint of the interval is right on the interface $l_L = 0$, and for the interval we considered, $x\in [0,l]$, we also have $l_R = l$. 
In this spatial case, the parameter runs in the range $\rho \in [\rho_0, \rho_+]$ such that $\rho_0 \rightarrow \rho^*$, $\rho_+ \rightarrow \infty$.~\footnote{We do not directly take the limit because we need to regularize the area.} 
An illustration of the RT surface is shown in figure~\ref{fig:thick_brane} (c).

With a solution of $A(\rho)$, the RT surface satisfies ($c_s = e^{A^\ast}$)
\begin{equation} \label{eq:differential_equation_icft}
    \frac{\dot y}{y} = \frac{e^{A^\ast - A(\rho)}}{\sqrt{e^{2A(\rho)}- e^{2A^\ast}}},
\end{equation}
and its area reads
\begin{equation}
    \mathcal A = \int_{\rho_0}^{\rho_+} {\rm d}\rho \mathcal L, \quad \mathcal L = \frac1{\sqrt{1- e^{2(A^\ast - A(\rho))}}} .
\end{equation}

We are interested in the leading order of the area as a function of the interval length. 
To proceed, we consider the variation of the area w.r.t. $l$.
The dependence comes from the integration cutoffs: 
\begin{equation} 
\label{eq:area_variation}
    \frac{\delta \mathcal A}{\delta l} = \mathcal L|_{\rho = \rho_+} \frac{\delta \rho_+}{\delta l} - \mathcal L|_{\rho = \rho_0} \frac{\delta \rho_0}{\delta l}. 
\end{equation}

Let's first evaluate the variation of the integration cutoffs w.r.t. $l$.
Near the cutoff, $\rho_+ \rightarrow \infty$, which corresponds to the endpoint located on the CFT, the metric asymptotes to an ${\rm AdS_3}$: $e^{A(\rho_+)} \approx \cosh \rho_+ $.
We impose a conventional regularization $z=\varepsilon$ at $x=l$.
Then according to the coordinate transformation~\eqref{eq:coordinate transformation},  $\varepsilon \approx \frac{y}{\cosh \rho_+}$, $l = y \tanh \rho_+ $,
and this leads to
\begin{equation} \label{eq:variation_right}
    \rho_+ = \log \frac{2l}{\varepsilon}, \quad \frac{\delta \rho_+}{\delta l} = \frac1l.
\end{equation}

For the other endpoint on the interface, we have $\rho_0 \rightarrow \rho^\ast$. 
The calculation of its dependence on $l$ is more involved. 
Consider the AdS$_2$ slice at $\rho^\ast$, the time component of the metric is ${\rm d}s^2 = e^{2A^*} \frac{{\rm d}\tau^2}{y^2}$.
We should impose the conventional cutoff in this ${\rm AdS_2}$ Poincare coordinate, i.e., ${\rm d}s^2 = \frac{{\rm d}\tau^2}{ \varepsilon^2}$~\cite{karch2021universal}.
This will lead to $y(\rho_0) \approx e^{A^\ast} \varepsilon$.
Combined with the $y$ near the other endpoints at $\rho_+$, we have the relations $y(\rho_0) \approx e^{A^*} \varepsilon $ and $y(\rho_+) \approx l$ at the two ends. 
Integrating over the differential equation~\eqref{eq:differential_equation_icft} (note that this is not the area), we arrive at 
\begin{equation}
    \log  \frac{l}{e^{A^\ast} \varepsilon } = \int_{\rho_0}^{\rho_+} {\rm d} \rho \frac{e^{A^\ast - A(\rho)}}{\sqrt{e^{2A(\rho)}- e^{2A^\ast}}}.
\end{equation}
Taking the derivative with respect to $l$ for both the left-hand side and the right-hand side, we obtain
\begin{equation}
    \frac1l =  \frac{e^{A^\ast - A(\rho_+)}}{\sqrt{e^{2A(\rho_+)}- e^{2A^\ast}}} \frac{\delta \rho_+}{\delta l} - \frac{e^{A^\ast - A(\rho_0)}}{\sqrt{e^{2A(\rho_0)}- e^{2A^\ast}}} \frac{\delta \rho_0}{\delta l}.
\end{equation}
The first term on the right-hand side vanishes when $\rho_+ \rightarrow \infty$ as $\lim_{\rho_+ \rightarrow \infty} e^{-A(\rho_+)} = \lim_{\rho_+ \rightarrow \infty} \frac1{\cosh \rho_+} = 0 $.
Consequently, we have the variation of $\rho_0$ from the second term:
\begin{equation} \label{eq:variation_left}
    \frac{\delta \rho_0}{\delta l} = -\frac1l \frac{\sqrt{e^{2A(\rho_0)}- e^{2A^\ast}}}{e^{A^\ast - A(\rho_0)}} \approx - \frac1l \sqrt{e^{2A(\rho_0)}- e^{2A^\ast}} + \mathcal{O}(\rho_0 - \rho^\ast). 
\end{equation}

Now we have the variation of the integration cutoffs with respect to $l$ in equation~\eqref{eq:variation_right} and~\eqref{eq:variation_left}. 
Next, the Lagrangian at the cutoffs that is needed in equation~\eqref{eq:area_variation} is straightforward to get.
For the Lagrangian at $\rho_+ \rightarrow \infty$, because $e^{A(\rho_+)} = \cosh \rho_+$, $\mathcal L|_{\rho =\rho_+} = 1$.
For Lagrangian at $\rho_0$, we have
\begin{equation}
\mathcal L_{\rho= \rho_0} = \frac{e^{A^\ast}}{\sqrt{e^{2A(\rho_0)}- e^{2A^\ast}}} + \mathcal O(\rho_0 - \rho^\ast).
\end{equation}
Now put everything together, the variation of the area with respect to $l$, i.e., equation~\eqref{eq:area_variation}, reads
\begin{equation}
    \frac{\delta \mathcal A}{\delta l} = (1 + e^{A^\ast}) \frac1l. 
\end{equation}
Then by integration, we get the leading contribution of area
\begin{equation}
    \label{eq:area_thick_icft}
    \mathcal A = (1 + e^{A^\ast}) \log l,
\end{equation}
which gives the entanglement entropy $S_{l}=\frac{c_1+c_{\rm eff}}{6}\log{l}$ with $c_{{\rm eff}}=\frac{3e^{A^\ast}}{2G_{(3)}}$ and $c_1=\frac{3}{2G_{(3)}}$ denoting the effective central charge induced by the defect and the central charge of the original CFT (corresponding to the unmeasured CFT in our case), respectively.

A few remarks follow: 
\begin{itemize}
    \item Comparing~\eqref{eq:area_thick_icft} with the result of weak measurements in \eqref{eq:geodesic length for general case eA}, we can see the weak measurement induced effective central charge is the same as the effective central charge from the interface. 
    Clearly, the two endpoints of the interval in the measurement case are both located at $\tau = 0$, leading to a factor $2\times e^{A^\ast}$, and in the case of spatial interface CFT, one of the endpoints is located at $x=0$ while the other is located at $x>0$, leading to a factor $1 + e^{A^\ast}$. 
    \item While our calculation is valid for a general $\rho^\ast$, we expect that the weak measurement will lead to a post-measurement metric with $\rho^\ast = 0$ and $A'(\rho) > 0 $ for $\rho > 0$.
    The intuition is that weak measurements gradually decrease the entanglement of the state.
    Then, according to reference~\cite{karch2023universality}, there is a c-theorem stating that $c_{\rm eff} \le c_1$. 
    It means that the effective central charge induced by weak measurements is not greater than that of the unmeasured CFT.
    This is consistent with the calculation in the CFT side in reference~\cite{sun2023new}.  
\end{itemize}

\subsection{Higher-dimensions}

Let's discuss the higher-dimension case. 
Instead of entanglement entropy, which involves codimension-two geometric quantity, we can use correlation functions to reveal the relationship between weak measurement and spatial interface.
In particular, the heavy operator correlation function is dominated by geodesics in the bulk dual.

To this end, as in reference~\cite{Anous_2022}, we consider scalar fields $\phi_{1,2}$ as the AdS dual of the heavy operators $O_{1,2}$ in ${\rm CFT_{1,2}}$ (see figure~\ref{fig:ICFT_general} for the AdS bulk).
The scalar action is
\begin{equation}
\label{eq:scalar field action in AdS}
    S[\phi]=\int_{M_1} {\rm d}^3x\ \sqrt{g_1}\left(\frac{1}{2}g^{\mu\nu}\partial_\mu\phi_1\partial_\nu\phi_1+\frac{1}{2}m^2\phi_1^2\right)+\int_{M_2} {\rm d}^3x\ \sqrt{g_2}\left(\frac{1}{2}g^{\mu\nu}\partial_\mu\phi_2\partial_\nu\phi_2+\frac{1}{2}m^2\phi_2^2\right),
\end{equation}
where $g_{1,2}$ is the metric with radii $L_{1,2}$, and $\phi_{1,2}$ are free scalar fields in ${AdS_{1,2}}$ denoted by $M_{1,2}$.
Solving the equation of motion for $\phi_{1,2}$ and focusing on the interface brane, we can derive connection conditions for $\phi_{1,2}$ that
\begin{equation}
    \phi_1(x)=\phi_2(x), \qquad \partial_n\phi_1(x)=\partial_n\phi_2(x),
\end{equation}
with $x$ the coordinate on the interface brane.
Thus, we can define a global scalar field $\phi$ as $\phi(x)|_{x\in{AdS_{1,2}}}=\phi_{1,2}(x)$.
For the large mass limit, with a saddle point approximation, the correlation function of scalar fields reads~\cite{balasubramanian2000holographic}
\begin{equation}
\label{eq:scalar field correlation and geodesic approx}
    \left<\phi(x_1)\phi(x_2)\right>\sim \sum_{\rm P}e^{-m d_{\rm P}(x_1,x_2)},
\end{equation}
where we sum over the contribution of all geodesics and $d_{\rm P}(x_1,x_2)$ denotes the geodesic with two endpoints $x_{1,2}$.
The shortest geodesic will be the leading term in \eqref{eq:scalar field correlation and geodesic approx}.

Under ${\rm AdS/CFT}$ duality, for heavy operators $O_{1,2}$ on the CFT side, we have the scaling dimension relation $m^2L_{1,2}^2=\Delta_{1,2}(\Delta_{1,2}-2)\approx\Delta_{1,2}^2$, and the corresponding correlation function
\begin{equation}
\label{eq:relation of heavy operator and scalar field correlation function}
    \left<O_1(x_1)O_2(x_2)\right>= \lim_{z_1,z_2\rightarrow0}\frac{1}{z_1^{\Delta_1} z_2^{\Delta_2}}\left<\phi_1(x_1)\phi_1(x_2)\right>= \lim_{z_1,z_2\rightarrow0}\frac{1}{z_1^{\Delta_1} z_2^{\Delta_2}}\left<\phi(x_1)\phi(x_2)\right>,
\end{equation}
where $z$ is a UV cutoff in the Poincare coordinates.
Consequently, according to \eqref{eq:scalar field correlation and geodesic approx} and \eqref{eq:relation of heavy operator and scalar field correlation function}, the correlation function of heavy operators can be an indicator to show the relation of spatial defect and weak measurement.
In the following, we will work in this regime.

Generically, consider a CFT in $\mathbbm{R}^n = \{ (x^0,..., x^{n-1})|x^i \in (-\infty, \infty), i= 0, ..., n-1 \}$ with $n>2$, where $x^0$ denotes the imaginary time.
Analogous to \eqref{eq:metric of coordinate 2}, the corresponding metric in AdS space is 
\begin{equation}
\label{eq:metric of coordinate 2 for d dimension}
    {\rm d}s^2=L^2\frac{\sum_{i=0}^{n-1}{\rm d}(x^i)^2+{\rm d}z^2}{z^2},
\end{equation}
where $L$ is the AdS radius. 
In this geometry, the length of a geodesic connecting $x^i$ and $(x')^{i}$ is 
\begin{equation}
\label{eq:length of arc geodesic for d dimension}
    d=L\cosh^{-1}{\left(\frac{\sum_i(x^i-(x^i)')^2+z^2+z'^2}{2zz'}\right)}.
\end{equation}

To discuss an interface brane along $x^i$, $i\ge 1$ direction, the foliation coordinates like~\eqref{eq:metric of coordinate 1} are still valid here.
For instance, a foliation in $x^1$ is given by
\begin{equation}
\label{eq:metric of coordinate 1 for d dimension}
    {\rm d}s^2={\rm d}\rho^2+L^2\cosh^2{\left(\frac{\rho}{L}\right)}\left(\frac{{\rm d}(x^0)^2+\sum_{i=2}^{n-1}{\rm d}(x^i)^2+{\rm d}y^2}{y^2}\right),
\end{equation}
under the coordinate transformation, $z=\frac{y}{\cosh{\left(\frac{\rho}{L}\right)}}$, $x^1=y \tanh{\left(\frac{\rho}{L}\right)}$.

Without loss of generality, we consider a higher dimensional interface CFT separated by $x^1=0$: the gravity dual is given by two AdS spacetime with different radii $L_{1,2}$ with an interface brane terminated at $x^1= 0$ and $z = 0$. 
To this end, generalizing the action \eqref{eq:bottom-up model} to ${\rm AdS_{n+1}}$, we can solve the location of the interface brane.
Assuming the radii to be $L_{1,2}$ and the tensor to be $T$ for these two AdS spacetimes, and denoting their coordinates by $y_{1,2}$, $\rho_{1,2}$, $x_{1,2}^{i}$, $i\ne 1$, 
the junction condition leads to $y_1=y_2$, $x_1^i=x_2^i, i\neq1$, and $\rho_{1,2}$ which satisfy \eqref{eq:solution of matching condition initial}.
\footnote{Here, the connection condition and corresponding solution rely on the definition of tension in the action. 
For example, if we directly extend the action \eqref{eq:bottom-up model} to a higher-dimension ${\rm AdS_{n+1}}$, there will be a prefactor $\frac{1}{n-1}$ in front of the tension $T$ in the solution of brane.
Nevertheless, if we define the action by substituting $T$ with $(n-1)T$, then the connection condition and solution will be the same for all dimension cases.}

In Euclidean space, the time and space coordinates are related by a rotation, as is manifest in~\eqref{eq:metric of coordinate 2 for d dimension}. 
Hence, the holographic weak measurement can be similarly established by a spacetime rotation.
Therefore, our discussion of the weak measurements in infinite systems can be generalized into higher dimensions. 
Nevertheless, in higher dimensions, we can consider equal-time correlation functions for both the weak measurement and the spacial interface.

In the case of spatial defect, we consider the correlation functions of $O_1(x_1^i)$ and $O_2(x_2^i)$ located on the defect $x^1_{1,2} = 0$. 
Different from the 3d case, where operators located on the defect will be in different time slices, here we can consider a fixed time slice because of the additional spatial direction in higher dimensions.
It also means that we can compare the correlation function for both weak measurement and interface brane cases. 
Therefore, in the following, we will mainly focus on this case: $x^i_{1,2} = 0$, $i\neq n-1$ and  $x^{n-1}_1=0$, $x^{n-1}_2 = l$. 
Similar to figure~\ref{fig:infinite_EE} (a), but now the $\tau$ direction should be $x^1$ and $x$ direction should be $x^{n-1}$.
Then the corresponding shortest length of the geodesic which contributes to the leading term in \eqref{eq:relation of heavy operator and scalar field correlation function} is the same as before, $d_{\rm P}(x_1,x_2)=2L_2\log{l/\varepsilon}$.
And the correlation function is~\footnote{We omit subleading contributions originated from other geodesics.} 
\begin{equation}
\label{eq:correlation function}
    \left<O_1(x_1)O_2(x_2)\right>= \lim_{\varepsilon\rightarrow0}\frac{1}{\varepsilon^{2\Delta_2}}e^{-2m L_2\log{l/\varepsilon}}\approx\frac{1}{l^{2\Delta_2}},
\end{equation}
where $\Delta_2\approx m L_2$.

For the weak measurement case, analog to figure~\ref{fig:infinite_EE} (a), here we consider the correlation functions of $O_1(x_1^i)$ and $O_2(x_2^i)$ located in the same imaginary time slice $x^0_{1,2}=0$ and general space coordinates $x^i_{1,2}, i\geq1$. 
With space rotational symmetry, we can redefine the space coordinate such that $x^1_{1}=0$, $x^1_{2}=l$ and $x^i_{1}=x^i_2, i\geq2$.
Therefore, the geodesic will be the same as \eqref{eq:length of geodesic for measurement/CFT geodesic on infinite system}.
Plugging it in \eqref{eq:relation of heavy operator and scalar field correlation function} will give the same result \eqref{eq:correlation function}.
Therefore, in this special setup, we can directly establish the connection between the weak measurement and the spatial interface brane using the heavy operator correlation function.
Recently, on the CFT side, the relation between the temporal boundary and the spatial boundary has been explored in reference~\cite{lee2023quantum}.

Finally, let's notice that the thick brane description can result from higher-dimension theories.
In reference~\cite{Karch_2023}, authors consider a general ${\rm AdS_3}\times M^d$ geometry where $M^d$ is a compact internal space, e.g., $S^3$.
We did not address this construction in the main text, as our primary focus is on the measurement applied to a CFT with dimension $n$ instead of a top-down model.
Nevertheless, exploring the meaning of the measurement in that content remains an intriguing problem.

\subsection{Python's lunch}

In section~\ref{sec:Geodesic in the bubble-inside-horizon phase}, we consider the bubble-inside-horizon phase and its geodesics.
Although it has the same behavior of entanglement entropy as the bubble-outside-horizon phase, its geometry structure is different with the bubble-outside-horizon phase, and will lead to much larger complexity.

\begin{figure}[tbp]
\centering 
\subfigure[]
{\includegraphics[width=.35\textwidth]{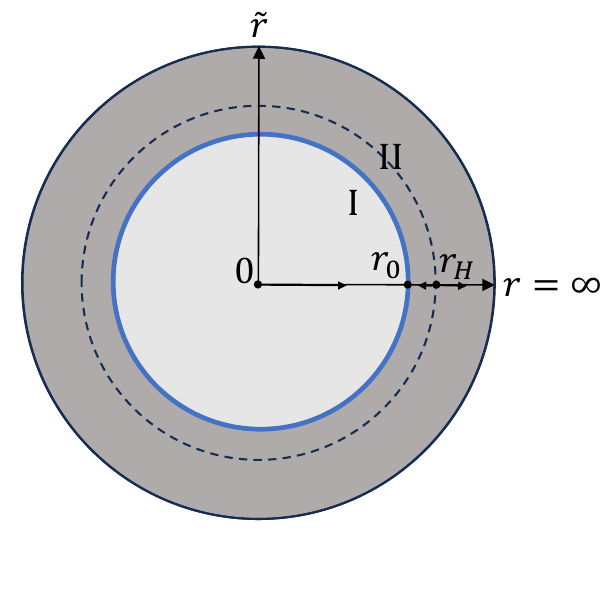}}\qquad
\subfigure[]
{\includegraphics[width=.25\textwidth]{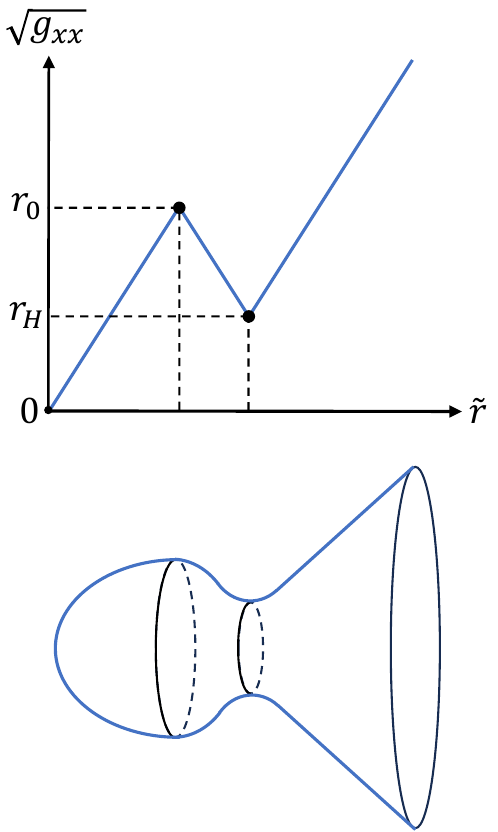}}
\caption{\label{fig:length_of_circle} (a) The time reversal invariance slice for the bubble-inside-horizon phase.
The blue circle indicates the interface brane and the dashed circle indicates the horizon.
(b) The metric component in the $x$ direction as a function of a global coordinate $\tilde r$.}
\end{figure}

The time reversal invariant slice of the bubble-inside-horizon phase is shown in figure~\ref{fig:length_of_circle} (a).
We define $\Tilde{r}$ as the global coordinate that corresponds to the distance to center and increases monotonically.
Then the relation between the local coordinate and the global coordinate is 
\begin{equation}
    r = \begin{cases}
        \tilde r  \quad & \tilde r \le r_0, \\
        2r_0 - \tilde r \quad & r_0 < \tilde r \le 2r_0- r_H, \\
        \tilde r-2r_0+2r_H \quad  & 2r_0- r_H < \tilde r.
    \end{cases}
\end{equation}
Here $r$ is the local AdS coordinate in region I, and the local black hole coordinate in region II. 
Notice that $r$ is also the metric component in the $x$ direction, i.e., $\sqrt{g_{xx}} = r$.
The function $r(\Tilde{r})$ is monotonic in the bubble-outside-horizon phase, but is not monotonic in the bubble-inside-horizon phase.
As shown in figure~\ref{fig:length_of_circle}, there are two minimal points at the center $r=0$ of the AdS spacetime and at the horizon $r=r_H$ of the black hole spacetime, respectively, and a maximal point at the interface brane $r=r_0$.
This realizes a Python's lunch geometry, and the details of the definition are given in appendix~\ref{sec:tensor network realization of Python's Lunch}.

A simple example of Python's lunch geometry is shown in figure~\ref{afig:diagram_of_Python's_lunch}, with a ``minmaxmin'' structure.
For the complexity of this tensor network, beside the contribution from the total number of tensors, the post-selection will play an important role and lead to exponentially large complexity.
An intuitive understanding is that, for the tensor network from maximum to minimum, we need to apply post-selection to decrease the number of legs and realize the final state.
In this process, we must select one state from the whole Hilbert space, with exponentially small probability.
Crudely, to have one successful outcome, we must repeat this experiment for exponential times, which leads to exponentially large complexity.

Similarly, for the bubble-inside-horizon phase, the region from maximum to minimum between $r=r_0$ and $r=r_H$ corresponds to the post-selection part and leads to exponentially big complexity.
While for the bubble-outside-horizon phase, $r(\Tilde{r})$ is a monotonic function so that no post-selection is needed.
Therefore, we conclude that, two irrelevant phases with and without a horizon will have completely different behaviors in terms of their complexity.
Especially, the bubble-inside-horizon phase has the geometry of Python's lunch.

\section{Discussion}
\label{sec:discussion}

To summarize, in this paper, we consider the holographic dual of weak measurements in conformal field theory. 
Because of the logarithmic scaling entanglement entropy (characterized by a distinct effective central charge) supported by post-measurement states from marginal measurements, the holographic dual involves interface branes separating spacetimes dual to the post-measurement state and the unmeasured CFT, respectively, generalizing the holographic dual of the conformal boundary state.
We also establish the correspondence between the weak measurement and the spatial interface.
In a finite system, while the irrelevant measurements will not change the entanglement scaling for the post-measurement state, it may create a Python's lunch. 
Besides, here we consider the phase with continuous central charge as a marginal phase, which corresponds to a marginal measurement term in XXZ model for CFT side~\cite{sun2023new}, but we expect that for more general cases, the marginal phase can also be considered as a stable fixed point.

We conclude this paper by mentioning a few open questions we would like to explore in the future:
(1) From an information perspective, the weak measurements result in several interesting scenarios for the post-measurement holographic dual. 
For the irrelevant case, the weak measurement can create a Python's lunch, greatly increasing the complexity for bulk reconstructions. 
How is the reconstruction map related to the measurement operators?
For the marginal case, while the AdS spacetime dual to the unmeasured CFT is replaced by a new different AdS metric, is the bulk information erased by the measurements? 
(2) In this paper, we briefly discuss the thick-brane description of the weak measurements, where the bulk solution is continuous, and find that it is consistent with the thin-brane description.
However, it is worthwhile to explore in more detail about the universal features of weak measurements using the general thick brane construction. 
Moreover, a top-down construction of weak measurements is also of great interest because we can get a handle from both sides of the theory. 
A classic example is the so-called Janus solution~\cite{bak2003a,bak2007three}. 
While such a solution is not directly related to weak measurements via a spacetime rotation, it would be interesting to explore other deformations or scenarios that can be related to measurements.

\acknowledgments

We thank Zhuo-Yu Xian and Cenke Xu for fruitful discussions. 
We are grateful to Sanjit Shashi for helpful communication regarding the thick brane construction.
SKJ would like to thank Stefano Antonini, Brianna Grado-White, and Brian Swingle for many useful discussions on related topics in previous collaborations.
XS acknowledges the support from Tsinghua Visiting Doctoral Students Foundation.
XS is also supported by the Lavin-Bernick Grant during the visit to Tulane university.
The work of SKJ is supported by a start-up fund at Tulane university.


\newpage 

\appendix

\section{Geodesic with spatial point defect}
\label{sec:Geodesic with spatial point defect}
In this section, we will use the method in reference~\cite{Anous_2022} to derive the bipartite entanglement entropy for finite or infinite systems. 
With AdS/CFT duality, we just need to calculate the length of geodesic in different AdS geometry.

We first introduce a few coordinate systems which will be useful later.
With \eqref{eq:embedded condition}, we define
\begin{equation}
\label{aeq:coordinates 3}
\begin{split}
    X^0=&L\frac{r_b}{r_H}\cosh{\frac{r_H}{L}\theta},\qquad X^1=L\sqrt{\frac{r_b^2}{r_H^2}-1}\cos{\frac{r_H\tau_b}{L^2}},\\
    X^2=&L\frac{r_b}{r_H}\sinh{\frac{r_H}{L}\theta},\qquad X^3=L\sqrt{\frac{r_b^2}{r_H^2}-1}\sin{\frac{r_H\tau_b}{L^2}},
\end{split}
\end{equation}
with corresponding metric
\begin{equation}
\label{aeq:metric of coordinate 3}
    {\rm d}s^2=\left(\frac{r_b^2-r_H^2}{L^2}\right){\rm d}\tau_b^2+\left(\frac{r_b^2-r_H^2}{L^2}\right)^{-1}{\rm d}r_b^2+r_b^2{\rm d}\theta^2.
\end{equation}
The parametrization of \eqref{aeq:coordinates 3} requires $\tau_b\sim \tau_b+\frac{2\pi L^2}{r_H}$, which means there is a periodic boundary condition on the $\tau_b$ direction.

Now we want to solve the junction condition \eqref{eq:matching condition 1} and \eqref{eq:matching condition 2} with coordinate \eqref{eq:coordinates 1}.
Consider a brane located on $\rho_i=\rho_i^*$ where $i=1,2$.
The first condition gives \eqref{eq:special form of matching condition 1}, which leads to $y_1=y_2,\ \tau_1=\tau_2$ and $L_1\cosh{\left(\frac{\rho_1^*}{L_1}\right)}=L_2\cosh{\left(\frac{\rho_2^*}{L_2}\right)}$.
The second condition can be derived below.
The extrinsic curvature can be expressed as $K_{ab}=\frac{1}{2}n_ig^{ij}\partial_j g_{ab}$ where $g_{ij}$ is metric in the original space and $g_{ab}$ is metric on the hypersurface.
We define $\Vec{n}=(n_{\rho}=1,0,0)$ with metric \eqref{eq:metric of coordinate 1}, then $K_{ab}=\frac{1}{2}g^{\rho\rho}\partial_{\rho}g_{ab}=\frac{1}{2}\partial_{\rho}g_{ab}=\frac{1}{2}\partial_{\rho}(\frac{L^2}{y^2} \cosh^2{\frac{\rho}{L}})=\frac{1}{L}\tanh{\frac{\rho}{L}}g_{ab}$.
Because for two regions 1 and 2, the directions of $\Vec{\rho}$ are opposite and $g_{ab}\equiv h_{ab}$, then \eqref{eq:matching condition 2} leads to \eqref{eq:special form of matching condition 2}.
Solving these two conditions, we have \eqref{eq:solution of matching condition}.
Finally, we can apply RT formula \eqref{eq:entanglement entropy with RT formula} to calculate the entanglement entropy with the length of different geodesics, and we will consider different cases in the following.

In reference~\cite{Anous_2022}, the authors give a general geodesic solution for any $\sigma_{1,2}$ with two regions ${1,2}$ in AdS space.
Assuming the interface brane is located at $\chi_{1,2}=\psi_{1,2}$ and two endpoints are located on $x_1=-\sigma_1$ and $x_2=\sigma_2$, the length of the geodesic is 
\begin{equation}
\label{aeq:general solution of geodesic length}
    d(\sigma_1,\sigma_2)=L_1\log{\left[\frac{2r}{\varepsilon_1}\tan{\left(\frac{\varphi}{2}\right)}\right]}+L_2\log{\left[\frac{2R}{\varepsilon_2}\tan{\left(\frac{\theta}{2}\right)}\right]},
\end{equation}
where 
\begin{equation}
\label{aeq:expression of R r}
\begin{split}
    r=\frac{1}{2}\csc{\left(\frac{\varphi}{2}\right)}\sec{\left(\frac{\psi_1+\psi_2}{2}\right)}\left[\sigma_2\cos{\left(\frac{\theta}{2}\right)}-\sigma_1\cos{\left(\frac{\theta}{2}+\varphi\right)}\right],\\
    R=\frac{1}{2}\csc{\left(\frac{\theta}{2}\right)}\sec{\left(\frac{\psi_1+\psi_2}{2}\right)}\left[\sigma_1\cos{\left(\frac{\varphi}{2}\right)}-\sigma_2\cos{\left(\frac{\varphi}{2}+\theta\right)}\right],
\end{split}
\end{equation}
$\varepsilon_{1,2}$ are UV-cutoff, $\varphi=\pi+\psi_1+\psi_2-\theta$ and $\theta$ is expressed as
\begin{equation}
\label{aeq:expression of theta}
\begin{split}
    &\cos{\theta}=\frac{\cos{\left(\frac{\psi_1-\psi_2}{2}\right)}}{\sigma_1^2+\sigma_2^2+2\sigma_1\sigma_2\cos{(\psi_1+\psi_2)}}\left\{-\sigma_2^2\cos{\left(\frac{\psi_1-\psi_2}{2}\right)}+\sigma_1^2\cos{\left(\frac{\psi_1+3\psi_2}{2}\right)}\right.\\
    &+2\sigma_1\sigma_2\sin{\psi_2}\sin{\left(\frac{\psi_1+\psi_2}{2}\right)}-\left[\sigma_1\sin{\left(\frac{\psi_1+3\psi_2}{2}\right)}-\sigma_2\sin{\left(\frac{\psi_1-\psi_2}{2}\right)}\right]\times\\
    &\left.\sqrt{\left[\frac{(\sigma_1+\sigma_2)^2-(\sigma_1-\sigma_2)^2\cos(\psi_1-\psi_2)+4\sigma_1\sigma_2\cos{(\psi_1+\psi_2)}}{2\cos^2{\frac{\psi_1-\psi_2}{2}}}\right]} \right\}.
\end{split}
\end{equation}
Latter, we can use these expressions to get the similar results in the main text.

Generically, the geodesic equation of equal time slice is 
\begin{equation}
\label{aeq:general geodesic equation of pure AdS in reference}
    \left(x-\frac{\sigma_1+\sigma_2}{2}\right)^2+z^2=\left(\frac{\sigma_1-\sigma_2}{2}\right)^2,
\end{equation}
where $\sigma_{1,2}$ are determined by two endpoints $(\tau,x,z)$ and $(\tau',x',z')$, and the length of the geodesic is \eqref{eq:length of arc geodesic}.

\subsection{Defect/CFT geodesic in an infinite system}
\subsubsection{General solution of defect/CFT geodesic in an infinite system}

The first case is a subsystem with two endpoints on the defect and the boundary, respectively, as shown in figure~\ref{fig:defect_CFT_geodesic_infinite_system}.
Here, we label the AdS dual of two half-plane CFTs as region 1, and the dual of the defect as region 2.
To show the geodesic explicitly, we extend the width of the defect, which is approximately zero in CFT.
Now we want to calculate the entanglement entropy of the subsystem using the RT formula.
The orange curve constituted by two arcs is a smooth geodesic which connects two endpoints.

With \eqref{aeq:general solution of geodesic length}, we can take the limit $\sigma_2\rightarrow0$ and $\sigma_1=l$.
Then \eqref{aeq:expression of theta} can be simplified as $\cos{\theta}=-\cos{(\psi_2-\psi_1)}$ for $\psi_2>\psi_1$, otherwise $\cos{\theta}=-1$.
Actually, $\cos{\theta}=-1$ with $\psi_2<\psi_1$ is not relevant because for $\theta=\pi$, we have $r=0$ in \eqref{aeq:expression of R r}, which means the whole geodesic is in region $1$.
With \eqref{eq:relation between psi and L}, $\psi_2<\psi_1$ corresponds to $L_1<L_2$, which means $c_1<c_2$.
Therefore, the induced effective central charge of entanglement entropy by defect cannot be larger than the original CFT without defect, which is consistent with the results we get in the main text.

Now let's focus on the nontrivial case with $\psi_2>\psi_1$ and $L_2<L_1$.
With $\theta=\pi-\psi_2+\psi_1,\ \varphi=2\psi_2$, (compared with reference we exchange label 1 and 2), we have
\begin{equation}
\label{aeq:solution of R r for defect/CFT geodesic on infinite system}
    R=\frac{\cos{\psi_1\sigma_1}}{\cos{\psi_1}+\cos{\psi_2}},\quad
    r=\frac{\sin{(\psi_2-\psi_1)}\sigma_1}{2\sin{\psi_2}(\cos{\psi_1}+\cos{\psi_2})}.
\end{equation}
So the geodesic length with the same UV-cutoff for two regions is 
\begin{equation}
\label{aeq:length of geodesic for defect/CFT geodesic on infinite system}
    d_{0,\sigma_1}=(L_1+L_2)\log{\frac{\sigma_1}{\varepsilon}}+d_{\rm sub},
\end{equation}
where $d_{\rm sub}$ corresponding to the sub-leading term of entanglement entropy induced by the brane in AdS space.
\begin{equation}
\label{aeq:boundary geodesic for defect/CFT geodesic on infinite system}
\begin{split}
    d_{\rm sub}=&L_2\log{\left(\frac{\sin{(\psi_2-\psi_1)}}{\cos{\psi_2}(\cos{\psi_1}+\cos{\psi_2})}\right)}+L_1\log{\left(\frac{2\cos{\psi_1}}{\tan{\frac{\psi_2-\psi_1}{2}}(\cos{\psi_1}+\cos{\psi_2})}\right)}\\
    =&L_2\log{\left(\frac{\sin{\frac{\psi_2-\psi_1}{2}}}{\cos{\psi_2}\cos{\frac{\psi_1+\psi_2}{2}}}\right)}+L_1\log{\left(\frac{\cos{\psi_1}}{\sin{\frac{\psi_2-\psi_1}{2}}\cos{\frac{\psi_1+\psi_2}{2}}}\right)}.
\end{split}
\end{equation}
We can find that the results of \eqref{aeq:length of geodesic for defect/CFT geodesic on infinite system} and \eqref{aeq:boundary geodesic for defect/CFT geodesic on infinite system} are consistent with the results of \eqref{eq:geodesic length of OA} and \eqref{eq:geodesic length of AB}.

Therefore, with \eqref{eq:entanglement entropy with RT formula} we have
\begin{equation}
\label{aeq:entanglement entropy for defect/CFT geodesic on infinite system}
    S_{0,\sigma_1}=\frac{c_1+c_2}{6}\log{\frac{l}{\varepsilon}}+S_{\rm sub},
\end{equation}
where $S_{\rm sub}=\frac{d_{\rm sub}}{4G_{(3)}}$ and $c_2=c_{\rm eff}$.
Here, the prefactor of the $\log$-term is $\frac{c_1+c_2}{6}$ because the locations of two endpoints are in region $1$ and $2$, respectively.

\subsubsection{Large size limit for the original region}
\begin{figure}[tbp]
\centering 
\subfigure[]{\includegraphics[width=.35\textwidth]{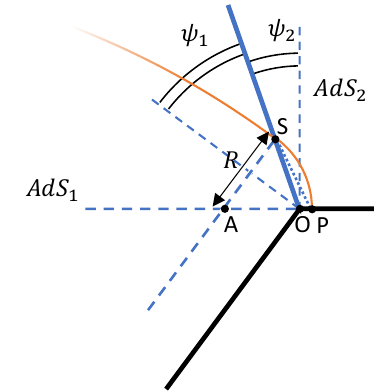}} \quad\quad\quad\quad
\subfigure[]{\includegraphics[width=.4\textwidth]{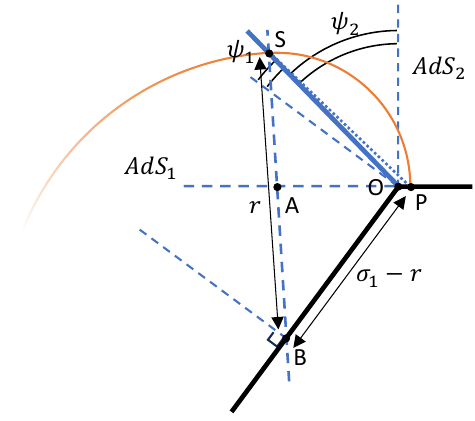}}
\caption{\label{afig:Figure/sailing_geodesic_limit} AdS dual of CFT with interface defect for infinite system and the defect/CFT geodesic.}
\end{figure}

In the following, we consider the limit case $\sigma_1\rightarrow\infty$ with fixed finite $\sigma_2$ in figure~\ref{afig:Figure/sailing_geodesic_limit}.
We denote the endpoints of the geodesic to be Q in region 1 and P in region 2, respectively.
In the limit $\sigma_1 \rightarrow \infty$, the figure does not show Q explicitly.
To have an intuition about the limit, we can calculate the limit of \eqref{aeq:expression of theta}.
To the first order of $\sigma_1^{-1}$,
\begin{equation}
\label{aeq:limit expression of theta}
\begin{split}
    &\cos{\theta}=\left\{
    \begin{aligned}
        \cos{(\psi_1+\psi_2)}+\frac{4\sigma_2\cos{\psi_1}(\cos{\psi_2}-\cos{\psi_1})\cos^2{\frac{\psi_1+\psi_2}{2}}}{-1 + \cos{(\psi_1-\psi_2)}}\sigma_1^{-1},\quad &\psi_1>\psi_2,\\
        \cos{(2\psi_2)}+4\sigma_2 \cos^2{\psi_2}\cos{\frac{\psi_1+\psi_2}{2}}\csc{\frac{\psi_1-\psi_2}{2}}\sin{\psi_2}\sigma_1^{-1},\quad &\psi_1<\psi_2.
    \end{aligned}
    \right.
\end{split}
\end{equation}
Therefore, the relation between $\psi_1$ and $\psi_2$ plays an important role for the limit case, and corresponds to relevant and irrelevant phases.
In the following, we will discuss this property in detail.

We first consider $\psi_2<\psi_1$ case shown in figure~\ref{afig:Figure/sailing_geodesic_limit} (a).
The key is that, for $\sigma_1\rightarrow\infty$ we have ${\rm SA}\parallel{\rm CFT_{left}}$, which can be proven that it only exists for $\psi_2<\psi_1$ as follows.
For $\triangle {\rm ASO}$, because $\sigma_2>0$, we have $\angle {\rm SOA}=\frac{\pi}{2}-\psi_2>\angle {\rm SPA}=\frac{\pi-(\psi_1+\psi_2)}{2}$, which requires $\psi_2<\psi_1$.
Then we can compute the geodesic.
With the law of sines, because $\angle {\rm ASO}=\pi-(\psi_1+\psi_2)-(\frac{\pi}{2}-\psi_2)=\frac{\pi}{2}-\psi_1$, we have
\begin{equation}
\label{aeq:law of sines for SOA}
    \frac{R-\sigma_2}{\sin{\left(\frac{\pi}{2}-\psi_1\right)}}=\frac{R}{\sin{\left(\frac{\pi}{2}-\psi_2\right)}}.
\end{equation}
The solution is $R=\frac{\cos{\psi_2}\sigma_2}{\cos{\psi_2}-\cos{\psi_1}}$.
It satisfies $R>\sigma_2$ because only one of $\psi_1$ and $\psi_2$ can be negative, which means $\cos{\psi_2}-\cos{\psi_1}>0$ with $\sin{\psi_1}+\sin{\psi_2}>0$.
Then we can get $|{\rm SO}|$ in $\triangle {\rm SAO}$.
With the law of sines, we have $|{\rm SO}|=\frac{R\sin{(\psi_1+\psi_2)}}{\cos{\psi_2}}$, which is a constant for the fixed $\sigma_2$.
Therefore, for the limit case $\sigma_1\rightarrow\infty$, the geodesic length in region 2 is a constant and contributes a sub-leading term.

Now we can calculate the entanglement entropy for $\psi_2<\psi_1$.
For the left part of the geodesic ${\rm QS}$ with endpoints ${\rm Q}$ and ${\rm S}$ we have ${\rm Q}=(-\sigma_1,\varepsilon)$ and ${\rm S}=(|{\rm OS}|\sin{\psi_1},|{\rm OS}|\cos{\psi_1})$.
Up to the leading term $\sigma_1^2$ in the numerator, the length of the geodesic QS is 
\begin{subequations}
\label{aeq:sailing case limit results L1smallerthanL2}
\begin{equation}
\label{aeq:sailing case limit results L1smallerthanL2 eq1}
    d_{\rm QS}=L_1\cosh^{-1}{\frac{(|{\rm OS}|\sin{\psi_1}+\sigma_1)^2+(|{\rm OS}|\cos{\psi_1})^2+\varepsilon^2}{2|{\rm OS}|\cos{\psi_1}\varepsilon}}\approx L_1\log{\frac{\sigma_1^2/\sigma_2}{\varepsilon}}+d_{\rm QS, sub}.
\end{equation}
where $d_{\rm QS, sub}=L_1\log{\frac{\cos{\psi_2}-\cos{\psi_2}}{ \sin{(\psi_1+\psi_2)}\cos{\psi_1}}}$.
For the right part of the geodesic ${\rm SP}$ we have ${\rm S}=(-|{\rm OS}|\sin{\psi_2},|{\rm OS}|\cos{\psi_2})$ and ${\rm P}=(\sigma_2,\varepsilon)$, which gives
\begin{equation}
\label{aeq:sailing case limit results L1smallerthanL2 eq2}
    d_{\rm SP}=L_2\cosh^{-1}{\frac{(|{\rm OS}|\sin{\psi_2}+\sigma_2)^2+(|{\rm OS}|\cos{\psi_2})^2+\varepsilon^2}{2|{\rm OS}|\cos{\psi_2}\varepsilon}}\approx L_2\log{\frac{\sigma_2}{\varepsilon}}+d_{\rm SP, sub},
\end{equation}
\end{subequations}
where $d_{\rm SP, sub}=L_2\log{\frac{\sin^2{(\psi_1+\psi_2)}+(\cos{\psi_2}-\cos{\psi_1})^2+2\sin{\psi_2}\sin{(\psi_1+\psi_2)}(\cos{\psi_2}-\cos{\psi_1})}{\cos{\psi_2}\sin{(\psi_1+\psi_2)}(\cos{\psi_2}-\cos{\psi_1})}}$.
Hence, the total length of the geodesic reads
\begin{equation}
\label{aeq:total length of geodesic sailing case limit results L1smallerthanL2}
    d=L_1\log{\frac{\sigma_1^2/\sigma_2}{\varepsilon}}+L_2\log{\frac{\sigma_2}{\varepsilon}}+d_{\rm sub},
\end{equation}
where $d_{\rm sub}=d_{\rm SP, sub}+d_{\rm QS, sub}$.
Therefore, for $\sigma_1 \gg 1 $ with a fixed $\sigma_2$, the prefactor of $\log{\sigma_1}$ is $2L_1$, which leads to the prefactor $\frac{c_1}{3}$ for the entanglement entropy.

Now we consider $\psi_2>\psi_1$ case shown in figure~\ref{afig:Figure/sailing_geodesic_limit} (b).
For $\sigma_1\rightarrow\infty$ we expect $|{\rm OS}|\rightarrow\infty$, which means $|{\rm AO}|\approx|{\rm AP}|$.
Then $\triangle {\rm SAO}$ is an isosceles triangle with $\angle {\rm ASP}=\angle {\rm APS}=\frac{\pi}{2}-\psi_2$.
To solve the geodesic, let's look at $\triangle {\rm SBO}$.
With the law of sines, we have
\begin{equation}
\label{aeq:law of sines for SBO}
    \frac{\sigma_1-r}{\sin{(\frac{\pi}{2}-\psi_2)}}=\frac{r}{\sin{(\frac{\pi}{2}+\psi_1})}=\frac{|{\rm SO}|}{\sin{(\psi_2-\psi_1)}}.
\end{equation}
The solutions are $r=\frac{\cos{\psi_1}\sigma_1}{\cos{\psi_2}+\cos{\psi_1}}$ and $|\rm SO|=\frac{\sin{(\psi_2-\psi_1)}\sigma_1}{\cos{\psi_2}+\cos{\psi_1}}$.

With the results above, we just need to plug new $|{\rm OS}|$ into \eqref{aeq:sailing case limit results L1smallerthanL2 eq1} and \eqref{aeq:sailing case limit results L1smallerthanL2 eq2}.
Finally, the total geodesic length is
\begin{equation}
\label{aeq:total length of geodesic sailing case limit results L1largerthanL2}
    d=L_1\log{\frac{\sigma_1}{\varepsilon}}+L_2\log{\frac{\sigma_1}{\varepsilon}}+d_{\rm sub},
\end{equation}
where $d_{\rm sub}=L_1\log{(\frac{A^2+1+2A\sin{\psi_1}}{A\cos{\psi_1}})}+L_2\log{(\frac{A}{\cos{\psi_2}})}$ and $A=\frac{\sin{(\psi_2-\psi_1)}}{\cos{\psi_2}+\cos{\psi_1}}$.
Therefore, in this case, the prefactor of $\log{\sigma_1}$ is $L_1+L_2$, which leads to the prefactor $\frac{c_1+c_2}{6}$ for the entanglement entropy.

Here is an additional remark about the results above.
For two cases $L_1<L_2$ and $L_1>L_2$, the prefactor of $\log{\sigma_1}$ are different at the large $\sigma_1$ limit.
The prefactor of $\log$-term in entanglement entropy is always the smaller one:
for $c_1<c_2$ we have $\frac{c_1}{6}+\frac{c_1}{6}$, but for $c_1>c_2$ we have $\frac{c_1}{6}+\frac{c_2}{6}$.
In summary, the leading term of the entanglement entropy at $\sigma_1 \rightarrow \infty$ reads 
\begin{equation}
    S_{-\sigma_1, \sigma_2} = \left( \frac{c_1}6 + \frac16 \min(c_1, c_2) \right) \log \sigma_1 .
\end{equation}
This property also exists in more complicated cases, which will be considered later.

\subsection{CFT/CFT geodesic in an infinite system}
\label{CFT/CFT geodesic in infinite system}

\subsubsection{General case for CFT/CFT geodesic in an infinite system}
\label{general case for CFT/CFT geodesic in infinite system}

\begin{figure}[tbp]
\centering 
\subfigure[]{\includegraphics[width=.3\textwidth]{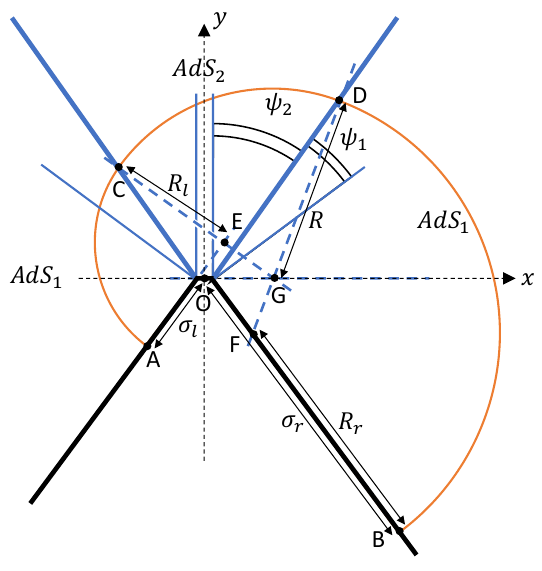}}\qquad
\subfigure[]{\includegraphics[width=.35\textwidth]{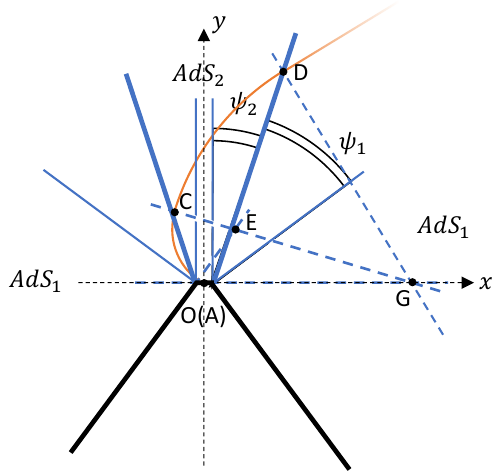}}
\caption{\label{afig:CFT_CFT_geodesic_infinite_system} AdS dual of CFT with interface defect for infinite system and the geodesic. (a) General CFT/CFT geodesic for $\psi_2>\psi_1$. (b) Geodesic for limit case $\sigma_l=0$ with $\psi_2<\psi_1$. }
\end{figure}

For a 1d quantum system with a defect located at $x=0$, we can choose a subsystem with two endpoints located on different sides.
In the following, we will construct the geodesic with several arcs and use \eqref{eq:length of arc geodesic} to calculate its length.
The diagram is shown in figure~\ref{afig:CFT_CFT_geodesic_infinite_system}.
Here we denote the location of two endpoints $(x=-\sigma_l,z=\varepsilon,\tau=0)$ and $(x'=\sigma_r,z'=\varepsilon,\tau'=0)$, and the radius of three arcs are $R_l,\ R,\ R_r$.
In the following, we will start with $R_r$ and $\sigma_r$, and finally express $\sigma_l$ with $R_r$ and $\sigma_r$.
Then we can express the geodesic length only with $\sigma_l$ and $\sigma_r$. 

We will calculate the geodesic in several steps.
(i) Assuming $|{\rm OD}|=\alpha$, and under the constraint $R_r>\frac{\sigma_r}{2}$, we have
\begin{subequations}
\label{aeq:CFT/CFT geodesic in infinite system general solution}
\begin{equation}
\label{aeq:CFT/CFT geodesic in infinite system general solution eq1}
\begin{split}
    |{\rm DF}|^2=[(\sigma_r-R_r)\cos{(\psi_1+\psi_2)}-\alpha\sin{\psi_2}]^2&+[(\sigma_r-R_r)\sin{(\psi_1+\psi_2)}+\alpha\cos{\psi_2}]^2=R_r^2,\\
    \alpha=(R_r-\sigma_r)\sin{\psi_1}+&\sqrt{R_r^2-(R_r-\sigma_r)^2\cos^2{\psi_1}}.
\end{split}
\end{equation}
(ii) Solving the equation of the line ${\rm FD}$, we have the point ${\rm G}=(x_0,0)$ with 
\begin{equation}
\label{aeq:CFT/CFT geodesic in infinite system general solution eq2}
    x_0=\alpha\sin{\psi_2}+\frac{(\sigma_r-R_r)\cos{(\psi_1+\psi_2)}-\alpha\sin{\psi_2}}{-(\sigma_r-R_r)\sin{(\psi_1+\psi_2)}-\alpha\cos{\psi_2}}(-\alpha\cos{\psi_2}).
\end{equation}
(iii) We define $|{\rm DG}|=\beta$ and $(\cdot)_{x,y}$ is $x,y$-coordinate of the point $(\cdot)$, then
\begin{equation}
\label{aeq:CFT/CFT geodesic in infinite system general solution eq3}
    \beta=R_r\frac{{\rm D}_y-{\rm G}_y}{{\rm D}_y-{\rm F}_y}=\frac{\alpha\cos{\psi_2}}{\alpha\cos{\psi_2-[-(\sigma_r-R_r)\sin{(\psi_1+\psi_2)}]}}R_r.
\end{equation}
(iv) With $|{\rm OC}|=\gamma$, then $|{\rm CG}|=|{\rm DG}|$ with the constraint $\beta>x_0$ gives
\begin{equation}
\label{aeq:CFT/CFT geodesic in infinite system general solution eq4}
    \gamma=\sqrt{\beta^2-x_0^2\cos^2{\psi_2}}-x_0\sin{\psi_2}.
\end{equation}
(v) With ${\rm E}$ located at $(x_1,\ y_1=\tan{(\psi_1+\psi_2)}x_1)$, it is also on the line ${\rm CG}$, which requires
\begin{equation}
\label{aeq:CFT/CFT geodesic in infinite system general solution eq5}
    x_1=\frac{\gamma \cos{\psi_2}}{\gamma \cos{\psi_2}+(\gamma\sin{\psi_2}+x_0)\tan{(\psi_1+\psi_2)}}x_0.
\end{equation}
(vi) Finally we have $|{\rm CE}|=|{\rm AE}|$, which gives
\begin{equation}
\label{aeq:CFT/CFT geodesic in infinite system general solution eq6}
    \sigma_l=-\frac{x_1}{\cos{(\psi_1+\psi_2)}}+\sqrt{\gamma^2+\frac{x_1}{\cos{(\psi_1+\psi_2)}}\left(\frac{x_1}{\cos{(\psi_1+\psi_2)}}-2\gamma\sin{\psi_1}\right)},
\end{equation}
where we use the constraint $\sigma_l>0$.
\end{subequations}

For a fixed $\sigma_r$, and any given $R_r$, we can express all variables above, including $\sigma_l$.
Then the length of the geodesic can be considered as below.
\begin{subequations}
\label{aeq:CFT/CFT geodesic in infinite system general solution of length}
For ${\rm AC}$, ${\rm A}=(-\sigma_l,\ \varepsilon)$ and ${\rm C}=(\gamma\sin{\psi_1},\ \gamma\cos{\psi_1})$, so the length is
\begin{equation}
\label{aeq:CFT/CFT geodesic in infinite system general solution of length eq1}
    d_{\rm AC}=L_1\cosh^{-1}{\left(\frac{\sigma_l^2+\gamma^2+2\sigma_l\gamma\sin{\psi_1}}{2\varepsilon\gamma\cos{\psi_1}}\right)} \approx L_1\log{\left(\frac{\sigma_l^2+\gamma^2+2\sigma_l\gamma\sin{\psi_1}}{\varepsilon\gamma\cos{\psi_1}}\right)}.
\end{equation}
Similarly, for ${\rm CD}$, ${\rm C}=(-\gamma\sin{\psi_2},\ \gamma\cos{\psi_2})$ and ${\rm D}=(\alpha\sin{\psi_2},\ \alpha\cos{\psi_2})$, so the length is
\begin{equation}
\label{aeq:CFT/CFT geodesic in infinite system general solution of length eq2}
    d_{\rm CD}=L_2\cosh^{-1}{\left(\frac{(\alpha+\gamma)^2-2\alpha\gamma\cos^2{\psi_2}}{2\alpha\gamma\cos^2{\psi_2}}\right)}.
\end{equation}
For ${\rm DB}$, ${\rm D}=(-\alpha\sin{\psi_1},\ \alpha\cos{\psi_1})$ and ${\rm B}=(\sigma_r,\ \varepsilon)$, so the length is
\begin{equation}
\label{aeq:CFT/CFT geodesic in infinite system general solution of length eq3}
    d_{\rm DB}=L_1\cosh^{-1}{\left(\frac{\sigma_r^2+\alpha^2+2\sigma_r\alpha\sin{\psi_1}}{2\varepsilon\alpha\cos{\psi_1}}\right)} \approx L_1\log{\left(\frac{\sigma_r^2+\alpha^2+2\sigma_r\alpha\sin{\psi_1}}{\varepsilon\alpha\cos{\psi_1}}\right)}.
\end{equation}
\end{subequations}
Therefore, the total length of the geodesic gives the entanglement entropy 
\begin{equation}
\label{aeq:entanglement entropy for CFT/CFT geodesic on infinite system}
\begin{split}
    &S_{R_r(\sigma_l),\sigma_r}=\frac{d_{\rm AC}+d_{\rm CD}+d_{\rm DB}}{4G_{(3)}}\\
    & =\frac{c_1}{6}\log{\left(\frac{\sigma_l^2+\gamma^2+2\sigma_l\gamma\sin{\psi_1}}{\varepsilon\gamma\cos{\psi_1}}\frac{\sigma_r^2+\alpha^2+2\sigma_r\alpha\sin{\psi_1}}{\varepsilon\alpha\cos{\psi_1}}\right)}+\frac{c_2}{6}\cosh^{-1}{\left(\frac{(\alpha+\gamma)^2-2\alpha\gamma\cos^2{\psi_2}}{2\alpha\gamma\cos^2{\psi_2}}\right)}.
\end{split}
\end{equation}
From the result above, it seems that the prefactor of $\log$-term is $\frac{c_1}{3}$, and the contribution of region 2 is only a constant.
Later we will show that it is not correct, and the behavior of the entanglement relies on the relative length of $\sigma_{l,r},\alpha$ and $\gamma$.
Besides, for different parameters, there are also two cases.

\subsubsection{Numerical results of the general case}
The equations above cannot be solved analytically with general variables $\sigma_l$ and $\sigma_r$, and only numerical calculation is available.
Here we show some numerical results.

We first consider the marginal case $L_1>L_2$ and $\psi_1<\psi_2$. 
With \eqref{aeq:CFT/CFT geodesic in infinite system general solution} and \eqref{aeq:CFT/CFT geodesic in infinite system general solution of length}, we consider $L_1=2,\ L_2=1,\ T=1,\ \varepsilon=1$ and $\sigma_r=10$.
Numerically, we can tune $R_r\in[5,20]$, and calculate the corresponding $\sigma_l$ and the total geodesic length.
In the calculation, we will find that there is a range  $R_r\in(R_r^{\rm min},R_r^{\rm max})$ for the solution to exist, and $R_r^{\rm min}>\frac{\sigma_r}{2}$.
\begin{figure}[tbp]
\centering    
    \subfigure[]
    {\includegraphics[width=.4\textwidth]{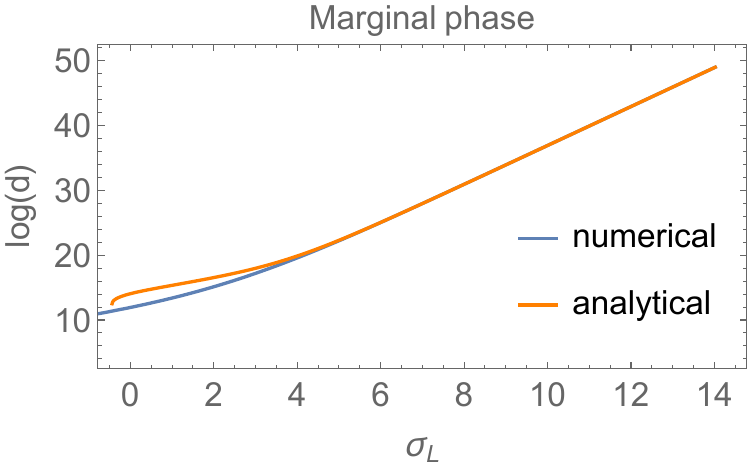}}
    \subfigure[]
    {\includegraphics[width=.4\textwidth]{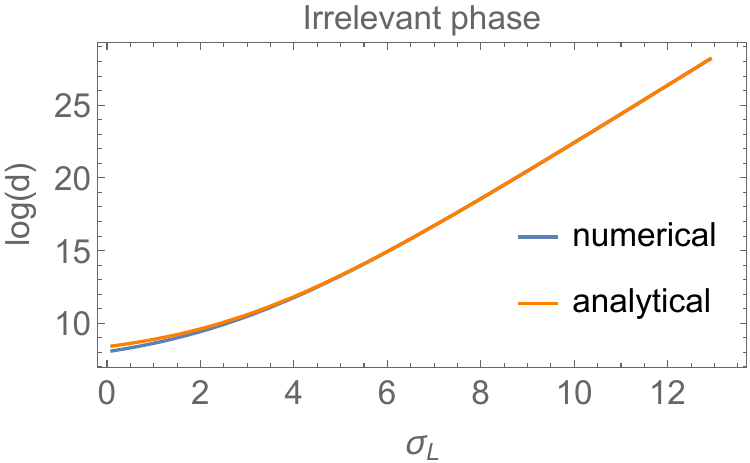}}
    \caption{\label{afig:general_solution_wrt_Rr_CFTCFT} (a) Numerical and analytical results of the length $d$ of the geodesic with respect to $\sigma_l$ for the marginal case. The parameters are $L_1=2,\ L_2=1,\ T=1,\ \varepsilon=1$, $\sigma_r=10$ and $R_r\in(R_r^{\rm min},R_r^{\rm max})$. (b) Numerical and analytical results of the length $d$ of the geodesic with respect to $\sigma_l$ for the irrelevant case. The parameters are $L_1=1,\ L_2=2,\ T=1,\ \varepsilon=1$, $\sigma_r=10$ and $R_r\in(R_r^{\rm min},\infty)$.}
\end{figure}
\begin{figure}[tbp]
\centering
    \subfigure[]{\includegraphics[width=.3\textwidth]{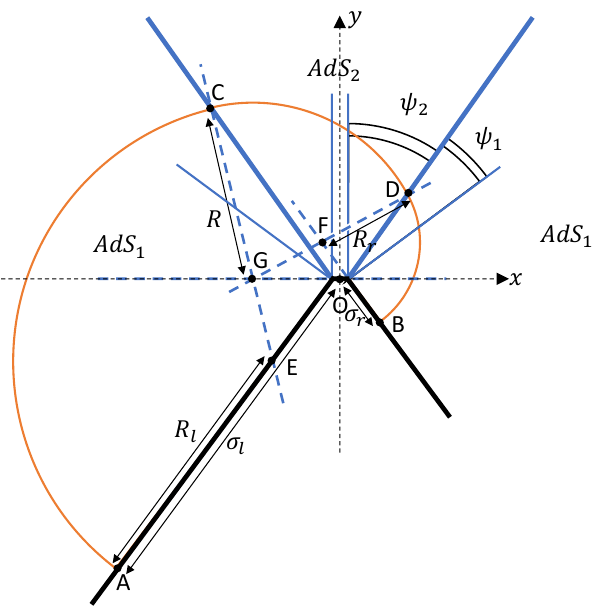}}\qquad
    \subfigure[]{\includegraphics[width=.3\textwidth]{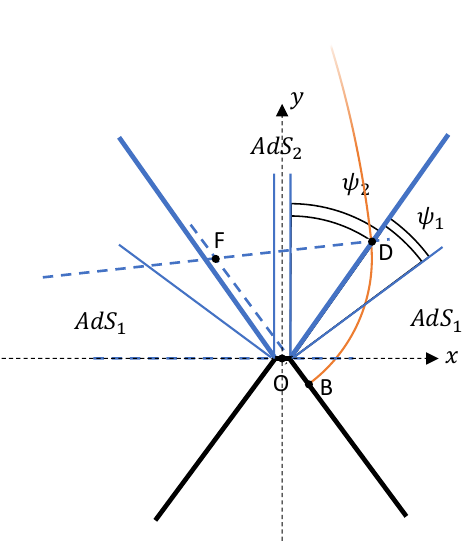}}
    \caption{\label{afig:geodesic_diagram_tmp} AdS dual of CFT with interface defect for infinite system and its geodesic. (a) General CFT/CFT geodesic for $\psi_2>\psi_1$. (b) Geodesic for large $\sigma_l$ limit case with $\psi_2>\psi_1$.}
\end{figure}
For $R_r=R_r^{\rm min}$, we have $\sigma_l=0$.
The geodesic length is shown in figure~\ref{afig:general_solution_wrt_Rr_CFTCFT} (a).
For the irrelevant case $L_1<L_2$ and $\psi_1>\psi_2$, the geodesic length is shown in figure~\ref{afig:general_solution_wrt_Rr_CFTCFT} (b).
With \eqref{aeq:CFT/CFT geodesic in infinite system general solution} and \eqref{aeq:CFT/CFT geodesic in infinite system general solution of length}, we consider $L_1=1,\ L_2=2,\ T=1,\ \varepsilon=1$ and $\sigma_r=10$.
In the numerical calculation, there is no $R_r^{\rm max}$ and $R_r^{\rm min}=\frac{\sigma_r}{2}$.

We first discuss the minimal value $R_r^{\rm min}$ in different cases.
It is unusual that $R_r^{\rm min}>\frac{\sigma_r}{2}$ for the marginal case, which means in figure~\ref{afig:CFT_CFT_geodesic_infinite_system} (a) although there is a finite length of $|{\rm DO}|$, the length of $|{\rm AO}|=0$.
In the following, we first prove the existence of $R_r^{\rm min}>\frac{\sigma_r}{2}$ for $L_1 > L_2$ case.
Actually, there are two possibilities.
One is $|{\rm CE}|=|{\rm OE}|\neq0$, and the other is $|{\rm OC}|=0$.
We can prove that only the second one is possible.
If $|{\rm CE}|=|{\rm OE}|\neq0$, then $\angle {\rm ECO }=\angle {\rm EOC}$, so $\angle {\rm CGO}=(\frac{\pi}{2}-\psi_2)-(\frac{\pi}{2}-\psi_1)=\psi_1-\psi_2<0$, which means there will be no crossing point ${\rm G}$ on the positive x-axis and no solution for ${\rm F}$.
For the second case we have $|{\rm DG}|=|{\rm OG}|$, $\angle {\rm DGO}=\pi-2(\frac{\pi}{2}-\psi_2)=2\psi_2$ and $\angle {\rm GOF}=\psi_1+\psi_2<\angle {\rm DGO}$, so there always exists the crossing point ${\rm F}$, which means the existence of $R_r^{\rm min}>\frac{\sigma_r}{2}$.

While, for $L_1<L_2$ and $\psi_1>\psi_2$, we have $\angle {\rm GOF}=\psi_1+\psi_2>\angle {\rm DGO}$, so there is no solution for crossing point and no $R_r^{\rm min}>\frac{\sigma_r}{2}$.
It can also be shown as follows on the other way with figure~\ref{afig:CFT_CFT_geodesic_infinite_system} (b).
With one fixed endpoint at the defect, we can prove that ${\rm DG}\parallel{\rm CFT_{right}}$ for another side, which means for finite $\sigma_r$ we cannot have $\sigma_l=0$.
The proof is in the following.
(i) Assuming $|{\rm CA}|=\alpha$, with $\angle {\rm CGA}=(\frac{\pi}{2}-\psi_2)-\angle {\rm GCA}=(\frac{\pi}{2}-\psi_2)-\angle {\rm EAC}=\psi_1-\psi_2$, we have the equation of line ${\rm CG}$ that $\frac{y-\alpha \cos{\psi_2}}{x+\alpha\sin{\psi_2}}=-\tan{(\psi_1-\psi_2)}$.
(ii) Setting $y=0$ we have $x_{\rm G}=x_0=\frac{\alpha\cos{\psi_2}}{\tan{\psi_1-\psi_2}}-\alpha\sin{\psi_2}$.
(iii) Defining $|{\rm AD}|=\beta$, with $|{\rm CG}|=|{\rm DG}|$ we have the equation $\beta^2-2\beta x_0 \sin{\psi_2}=\alpha^2+2\alpha x_0 \sin{\psi_2}$ and the solution $\alpha=\beta-2x_0\sin{\psi_2}$.
(iv) Then we can calculate the slope of the line ${\rm DG}$, which gives
\begin{equation}
\label{aeq:slope of DG}
    \tan{\angle {\rm DGA}}=\frac{\beta\cos{\psi_2}}{x_0-\beta\sin{\psi_2}}=\tan{(\psi_1+\psi_2)}.
\end{equation}
Because the slope of right CFT is $\tan{(\psi_1+\psi_2)}$, it means ${\rm CFT_{left}}\parallel{\rm DG}$.

To summarize, we have shown that for $L_1 > L_2$, $R_r^{\rm min}>\frac{\sigma_r}{2}$, while for $L_1 < L_2$, $R_r^{\rm min}=\frac{\sigma_r}{2}$.
Besides, we can calculate $R_r^{\rm min}$ analytically for $L_1 > L_2$.
In figure~\ref{afig:CFT_CFT_geodesic_infinite_system} (a), for $\triangle {\rm DFO}$, we have $|{\rm DF}|=R_r$ and $|{\rm OF}|=\sigma_r-R_r$.
For $R_r=R_r^{\rm min}$ with $|{\rm DG}|=|{\rm OG}|$, we have $\angle {\rm ODF}=\angle {\rm DOG}=\frac{\pi}{2}-\psi_2$, and $\angle {\rm DOF}=\frac{\pi}{2}+\psi_1$.
Then the law of sines gives
\begin{equation}
\label{aeq:the law of sines for DOF}
    \frac{R_r}{\sin{\left(\frac{\pi}{2}+\psi_1\right)}}=\frac{\sigma_r-R_r}{\sin{\left(\frac{\pi}{2}-\psi_2\right)}},
\end{equation}
which means $R_r^{\rm min}=\frac{\sigma_r\cos{\psi_1}}{\cos{\psi_1}+\cos{\psi_2}}>\frac{\sigma_r}{2}$.
If we plug the parameter above in it, we will get $R_r^{\rm min}\approx6.6667$, which is consistent with the numerical results.

Now we discuss the critical value $R_r^{\rm max}>\sigma_r$ for $L_1 > L_2$.
The key is that, if the crossing point ${\rm G}$ exists in figure~\ref{afig:geodesic_diagram_tmp} (a) for the limit case $R_r=R_r^{\rm max}$, which corresponds to ${\rm DF}\parallel{x-{\rm axis}}$, then $\angle {\rm DFB}=\psi_1+\psi_2$, and $\angle {\rm FDB}=\angle {\rm FBD}=\frac{\pi-(\psi_1+\psi_2)}{2}$.
Because we require $\angle {\rm FBD}<\angle {\rm FOD}=\psi_2+(\frac{\pi}{2}-(\psi_1+\psi_2))=\frac{\pi}{2}-\psi_1$, it means $\psi_1<\psi_2$.
This result is consistent with the conclusion that, for  $\psi_1<\psi_2$, there exists $R_r^{\rm max}$.
While for $\psi_1>\psi_2$, there does not exist $R_r^{\rm max}$ in the numerical results.
Similarly, we can also calculate $R_r^{\rm max}$ analytically.
With figure~\ref{afig:geodesic_diagram_tmp} (b), consider $\triangle {\rm FDO}$ and ${\rm DF}\parallel{x-{\rm axis}}$, where $|{\rm FD}|=|{\rm DB}|=R_r$ and $|{\rm FO}|=R_r-\sigma_r$.
Because $\angle {\rm FDO}=\frac{\pi}{2}-\psi_2$ and $\angle {\rm FOD}=\frac{\pi}{2}-\psi_1$, with the law of sines we have
\begin{equation}
\label{aeq:the law of sines for EFA}
    \frac{R_r}{\sin{\left(\frac{\pi}{2}-\psi_1\right)}}=\frac{R_r-\sigma_r}{\sin{\left(\frac{\pi}{2}-\psi_2\right)}},
\end{equation}
which gives $R_r^{\rm max}=\frac{\sigma_r\cos{\psi_1}}{\cos{\psi_1}-\cos{\psi_2}}>\sigma_r$. 
If we plug the parameters above in it, we will get $R_r^{\rm max}=20$ and it is consistent with numerical results.

There are several remarks for both cases.
(i) Although there is a finite range for $R_r$, for any $\sigma_l$ there always exists a solution for geodesics, which means $\sigma_l$ can range from zero to infinity.
(ii) We plot the geodesic length in a log-linear plot in figure~\ref{afig:general_solution_wrt_Rr_CFTCFT}.
For $L_1 > L_2$ the prefactor of $\log$-term for large $R_r$ is $3$, while for $L_1 < L_2$ the prefactor is $2$.
It can be understood as $3=L_1+L_2$ and $2=\frac{3}{2}L_1+\frac{1}{2}L_1$.

\subsubsection{Analytical approximation of the general case}
\begin{figure}[tbp]
\centering
    \subfigure[]{\includegraphics[width=.4\textwidth]{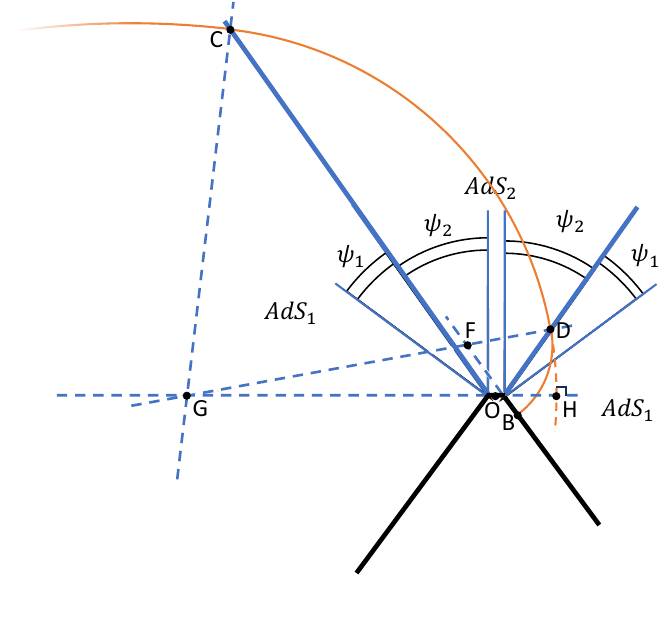}}\qquad
    \subfigure[]{\includegraphics[width=.4\textwidth]{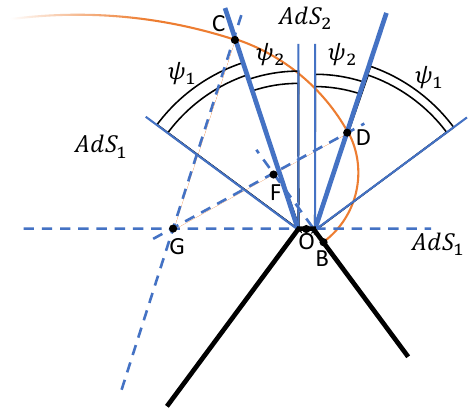}}
    \caption{\label{afig:general_limit} AdS dual of CFT with interface defect for infinite system and its geodesic. Here ${\rm A}$ is the crossing point of left CFT and geodesic, and ${\rm E}$ is the crossing point of left CFT and line ${\rm CG}$. For the discussion above, we consider the CFT boundary of ${AdS_2}$ is on a single line $x=0$.}
\end{figure}

In the following, we consider the limit case $\sigma_l\gg\sigma_r$ for different phases, which are shown in figure~\ref{afig:general_limit}.

For $\psi_1<\psi_2$ in figure~\ref{afig:general_limit} (a), the limit case is ${\rm DF}\parallel{\rm OG}$.
We first solve the geodesic.
Considering $\triangle {\rm FDB}$, (similar to figure~\ref{afig:Figure/sailing_geodesic_limit} (a)), we have $\angle {\rm DFB}=\psi_1+\psi_2, \angle {\rm FDO}=\frac{\pi}{2}-\psi_2$ and $\angle {\rm FOD}=\frac{\pi}{2}-\psi_1$.
With the law of sines and ${\rm DF}=R_r$, we have
\begin{subequations}
\label{aeq:the law of sines for L2smallerthanL1}
\begin{equation}
\label{aeq:the law of sines for FDO}
    \frac{R_r}{\sin{(\frac{\pi}{2}-\psi_1)}}=\frac{R_r-\sigma_r}{\sin{(\frac{\pi}{2}-\psi_2)}}=\frac{|\rm OD|}{\sin{(\psi_1+\psi_2)}}.
\end{equation}
The solutions are $R_r=\frac{\cos{\psi_1}\sigma_r}{(\cos{\psi_1}-\cos{\psi_2})}$ and $|{\rm OD}|=\frac{\sin{(\psi_1+\psi_2)}\sigma_r}{(\cos{\psi_1}-\cos{\psi_2})}=A_2\sigma_r$.
It means for $\sigma_l\rightarrow\infty$, $|{\rm OD}|\sim \mathcal O(\sigma_r)$ is a constant with a fixed $\sigma_r$, and $\angle {\rm GDH}=\angle {\rm GHD}=\frac{\pi}{2}$.
So $|{\rm OH}|\approx |{\rm OD}|\cos{(\frac{\pi}{2}-\psi_2)}\sim \mathcal O(\sigma_r)$ is also a constant.
Therefore, for $\triangle {\rm EBC}$, we have $\angle{\rm GCH}=\angle{\rm GHC}\approx\angle{\rm GOC}=\frac{\pi}{2}-\psi_2\approx\angle{\rm GCO}$ and $\angle{\rm COE}=\frac{\pi}{2}+\psi_1$.
Now with the law of sines (similar to figure~\ref{afig:Figure/sailing_geodesic_limit} (b)), we have
\begin{equation}
\label{aeq:the law of sines for EOC}
    \frac{R_l}{\sin{(\frac{\pi}{2}+\psi_1)}}=\frac{\sigma_l-R_l}{\sin{(\frac{\pi}{2}-\psi_2)}}=\frac{|\rm OC|}{\sin{(\psi_2-\psi_1)}}.
\end{equation}
\end{subequations}
The solutions are $R_l=\frac{\cos{\psi_1}\sigma_l}{(\cos{\psi_1}+\cos{\psi_2})}$ and $|{\rm OC}|=\frac{\sin{(\psi_2-\psi_1)}\sigma_l}{(\cos{\psi_1}+\cos{\psi_2})}=A_1 \sigma_l$.

Now we can consider the length of the geodesic.
For ${\rm AC}$ with ${\rm A}=(-\sigma_l,\varepsilon)$ and ${\rm C}=(|{\rm OC}|\sin{\psi_1},|{\rm OC}|\cos{\psi_1})$, we have 
\begin{subequations}
\label{aeq:limit case L1largerthanL2 CFT/CFT geodesic in infinite system general solution of length}
\begin{equation}
\label{aeq:limit case L1largerthanL2 CFT/CFT geodesic in infinite system general solution of length eq1}
\begin{split}
    d_{\rm AC}=&L_1 \cosh^{-1}{\frac{(|{\rm OC}|\sin{\psi_1}+\sigma_l)^2+\varepsilon^2+(|{\rm OC}|\cos{\psi_1})^2}{2\varepsilon |{\rm OC}|\cos{\psi_1}}}\approx L_1 \log{\frac{\sigma_l}{\varepsilon}\frac{A_1^2+1+2A_1^2\sin{\psi_1}}{A_1\cos{\psi_1}}}.
\end{split}
\end{equation}
For ${\rm CD}$ with ${\rm C}=(-|{\rm OC}|\sin{\psi_2},|{\rm OC}|\cos{\psi_2})$ and ${\rm D}=(|{\rm OD}|\sin{\psi_2},|{\rm OD}|\cos{\psi_2})$, we have 
\begin{equation}
\label{aeq:limit case L1largerthanL2 CFT/CFT geodesic in infinite system general solution of length eq2}
\begin{split}
    d_{\rm CD}=&L_2 \cosh^{-1}{\frac{(|{\rm OC}|+|{\rm OD}|)^2\sin^2{\psi_2}+(|{\rm OC}|\cos{\psi_2})^2+(|{\rm OD}|\cos{\psi_2})^2}{2|{\rm OC}||{\rm OD}|\cos^2{\psi_2}}}\approx L_2 \log{\frac{\sigma_l}{\sigma_r}\frac{A_1}{A_2\cos^2{\psi_2}}},
\end{split}
\end{equation}
where in the last equation we apply the limit $\sigma_l\gg\sigma_r$.
For ${\rm DB}$ with ${\rm B}=(\sigma_r,\varepsilon)$ and ${\rm D}=(-|{\rm OD}|\sin{\psi_1},|{\rm OD}|\cos{\psi_1})$, we have 
\begin{equation}
\label{aeq:limit case L1largerthanL2 CFT/CFT geodesic in infinite system general solution of length eq3}
\begin{split}
    d_{\rm DB}=&L_1 \cosh^{-1}{\frac{(|{\rm OD}|\sin{\psi_1}+\sigma_r)^2+\varepsilon^2+(|{\rm OD}|\cos{\psi_1})^2}{2\varepsilon |{\rm OD}|\cos{\psi_1}}}\approx L_1 \log{\frac{\sigma_r}{\varepsilon}\frac{A_2^2+1+2A_2^2\sin{\psi_1}}{A_2\cos{\psi_1}}}.
\end{split}
\end{equation}
\end{subequations}
Then the total length of the geodesic is 
\begin{equation}
\label{aeq:limit case L1largerthanL2 total CFT/CFT geodesic in infinite system general solution of length eq3}
    d\approx L_1 \log{\frac{\sigma_l}{\varepsilon}\frac{A_1^2+1+2A_1^2\sin{\psi_1}}{A_1\cos{\psi_1}}}+L_2 \log{\frac{\sigma_l}{\sigma_r}\frac{A_1}{A_2\cos^2{\psi_2}}}+L_1 \log{\frac{\sigma_r}{\varepsilon}\frac{A_2^2+1+2A_2^2\sin{\psi_1}}{A_2\cos{\psi_1}}},
\end{equation}
where the prefactor of $\log{\sigma_l}$ is $L_1+L_2$.
We can compare this analytical result \eqref{aeq:limit case L1largerthanL2 total CFT/CFT geodesic in infinite system general solution of length eq3} with the numerical results \eqref{aeq:entanglement entropy for CFT/CFT geodesic on infinite system}, and they are consistent in figure~\ref{afig:general_solution_wrt_Rr_CFTCFT} (a).

Now we consider another case $\psi_1>\psi_2$, which is more complicated and shown in figure~\ref{afig:general_limit} (b).
With the limit $\sigma_l\gg\sigma_r$, we expect that ${\rm CG}\parallel{\rm OE}$, which means $|{\rm OC}|\ll\sigma_l$.
While, ${\rm DG}$ will always cross ${\rm OG}$ at a finite angle $\angle {\rm DGO}$, which means $|{\rm OD}|\gg\sigma_r$.
Besides, with numerical results, we may expect that $\frac{|\rm OC|}{|\rm OD|}\sim \mathcal O(1)$ for $\sigma_l\rightarrow\infty$.
The key difference is that here we cannot directly get $|{\rm OC}|$ with $\sigma_l$ and $|{\rm OD}|$ with $\sigma_r$.
Instead, we expect $\sigma_r\ll R_r,\alpha,x_0,\beta,\gamma\ll x_1,\sigma_l$, which can be shown below.

Here we first define $|{\rm OD}|=\alpha$.
(i) With $|{\rm OD}|\gg\sigma_r$, we have $\angle {\rm FDB}=\angle {\rm DBF}\approx\angle {\rm DOF}=\frac{\pi}{2}-\psi_1$, $\angle {\rm DFO}=2\psi_1$ and $\angle {\rm DGO}=\psi_1-\psi_2$.
For $\triangle {\rm OGD}$ we apply the law of sines that
\begin{equation}
\label{aeq:the law of sines for OGD}
    \frac{|{\rm OG}|}{\sin{(\frac{\pi}{2}-\psi_1)}}=\frac{|{\rm DG}|}{\sin{(\frac{\pi}{2}+\psi_2)}}=\frac{|{\rm DO}|}{\sin{(\psi_1-\psi_2)}}.
\end{equation}
The solutions are $|{\rm OG}|=\frac{\cos{\psi_1}\alpha}{\sin{(\psi_1-\psi_2)}}$ and $|{\rm DG}|=\frac{\cos{\psi_2}\alpha}{\sin{(\psi_1-\psi_2)}}$.
(ii) With ${\rm CG}\parallel{\rm OE}$, we have $\angle {\rm CGO}=\angle {\rm GOA}=\psi_1+\psi_2$, and $\angle {\rm GCO}=\frac{\pi}{2}-\psi_1$.
Using the law of sines for $\triangle {\rm OCG}$ we have
\begin{equation}
\label{aeq:the law of sines for OCG}
    \frac{|{\rm OG}|}{\sin{(\frac{\pi}{2}-\psi_1)}}=\frac{|{\rm OC}|}{\sin{(\psi_1+\psi_2)}}=\frac{|{\rm CG}|}{\sin{(\frac{\pi}{2}-\psi_2)}}.
\end{equation}
The solutions are $|{\rm OC}|=\frac{\sin{(\psi_1+\psi_2)}\alpha}{\sin{(\psi_1-\psi_2)}}$ and $|{\rm CG}|=\frac{\cos{\psi_2}\alpha}{\sin{(\psi_1-\psi_2)}}$.
(iii) Finally, with $\angle {\rm DFO}=2\psi_1$ and $\angle {\rm DOF}=\frac{\pi}{2}-\psi_1$, we have $\frac{\alpha}{\sin{2\psi_1}}=\frac{|{\rm DF}|}{\sin{(\frac{\pi}{2}-\psi_1)}}$.
So $|{\rm DF}|=\frac{\cos{\psi_1}\alpha}{\sin{2\psi_1}}$.
To be consistent with the notations above, we have $|{\rm OG}|=x_0, |{\rm OC}|=\gamma, |{\rm CG}|=\beta, |{\rm FD}|=R_r$.
Therefore, $|{\rm OC}|=\frac{\sin{(\psi_1+\psi_2)}\sin{2\psi_1}}{\sin{(\psi_1-\psi_2)}\cos{\psi_1}}R_r$ and $|{\rm OD}|=\frac{\sin{2\psi_1}}{\cos{\psi_1}}R_r$.

However, the results above are not enough to solve the values of all unknown variables using $\sigma_l$ and $\sigma_r$.
We need \eqref{aeq:CFT/CFT geodesic in infinite system general solution}.
With the discussion above, $\sigma_l\gg R_r\gg\sigma_r$.
By symmetry, we assume that $R_r\sim \mathcal O(\sqrt{\sigma_l\sigma_r})\sim \mathcal O(\sqrt{\sigma_l})$ for $\sigma_l\rightarrow\infty$ and a fixed $\sigma_r$.
This assumption can also be checked numerically.
Then to simplify equations \eqref{aeq:CFT/CFT geodesic in infinite system general solution}, we can first rescale all variables with $\frac{1}{\sigma_l}$, which means $\sigma_l'=1$, $\sigma_r'=\frac{\sigma_r}{\sigma_l}\ll1$ and $R_r'=\frac{R_r}{\sigma_l}=c\cdot \sqrt{\frac{\sigma_r}{\sigma_l}}$, where $c$ is an unknown variable to be determined.
Now if we can get $c$, then all variables can be solved.
The constraint for $c$ comes from $\sigma_l'=1$.
For the limit case, with \eqref{aeq:CFT/CFT geodesic in infinite system general solution eq6} we have $\sigma_l'\approx-\frac{2x_1'}{\cos{(\psi_1-\psi_2)}}$ at the zero order. 
Then with the Taylor expansion of \eqref{aeq:CFT/CFT geodesic in infinite system general solution}, we can finally express $x_1'$ to the zero order of $\sigma_r'$, which gives $x_1'=-c^2\cdot\frac{\sin^2{\psi_1}\sin{[2(\psi_1+\psi_2)]}}{\sin{(\psi_1-\psi_2)}}+ \mathcal O(\sigma_r')$.
So we have $1=\sigma_l'\approx c^2\cdot\frac{2\sin^2{\psi_1}\sin{[2(\psi_1+\psi_2)]}}{\sin{(\psi_1-\psi_2)}\cos{(\psi_1+\psi_2)}}$, which means
\begin{equation}
\label{aeq:solution of c for limit case}
    c=\frac{1}{2\sin{\psi_1}}\sqrt{\frac{\sin{(\psi_1-\psi_2)}}{\sin{(\psi_1+\psi_2)}}}.
\end{equation}
Now we have $|{\rm OC}|=\frac{\sin{(\psi_1+\psi_2)}|{\rm OD}|}{\sin{(\psi_1-\psi_2)}}=\frac{\sin{(\psi_1+\psi_2)}}{\sin{(\psi_1-\psi_2)}}\cdot2\sin{\psi_2}R_r=\sqrt{\frac{\sin{(\psi_1+\psi_2)}}{\sin{(\psi_1-\psi_2)}}}\sqrt{\sigma_l\sigma_r}=b\sqrt{\sigma_l\sigma_r}$ and $|{\rm OD}|=b^{-1}\sqrt{\sigma_l\sigma_r}$.

Finally, we can calculate the length of the geodesic, which is similar to \eqref{aeq:limit case L1largerthanL2 CFT/CFT geodesic in infinite system general solution of length}.
For ${\rm AC}$ we have 
\begin{subequations}
\label{aeq:limit case L1smallerthanL2 CFT/CFT geodesic in infinite system general solution of length}
\begin{equation}
\label{aeq:limit case L1smallerthanL2 CFT/CFT geodesic in infinite system general solution of length eq1}
\begin{split}
    d_{\rm AC}\approx& L_1\log{\frac{b^2\sigma_l\sigma_r+\sigma_l^2+2b\sigma_l^{3/2}\sigma_r^{1/2}\sin{\psi_1}}{b(\sigma_l\sigma_r)^{1/2}\cos{\psi_1}\varepsilon}}\approx L_1\left(\log{\left(\frac{b^{-1}\sigma_r^{-1/2}\sigma_l^{3/2}}{\cos{\psi_1}\varepsilon}\right)}+2b\sin{\psi_1}\sqrt{\frac{\sigma_r}{\sigma_l}}\right).
\end{split}
\end{equation}
For ${\rm CD}$ we have 
\begin{equation}
\label{aeq:limit case L1smallerthanL2 CFT/CFT geodesic in infinite system general solution of length eq2}
\begin{split}
    d_{\rm CD}=& L_2 \cosh^{-1}{\frac{(b^{-1}+b)^2\sin^2{\psi_2}+(b^{-2}+b^2)\cos^2{\psi_2}}{2\cos^2{\psi_2}}}.
\end{split}
\end{equation}
For ${\rm DB}$ we have 
\begin{equation}
\label{aeq:limit case L1smallerthanL2 CFT/CFT geodesic in infinite system general solution of length eq3}
\begin{split}
    d_{\rm DB}\approx& L_1\log{\frac{b^{-2}\sigma_l\sigma_r+\sigma_r^2+2b^{-1}\sigma_l^{1/2}\sigma_r^{3/2}\sin{\psi_1}}{b^{-1}(\sigma_l\sigma_r)^{1/2}\cos{\psi_1}\varepsilon}}\approx L_1\left(\log{\left(\frac{b^{-1}\sigma_r^{1/2}\sigma_l^{1/2}}{\cos{\psi_1}\varepsilon}\right)}+2b\sin{\psi_1}\sqrt{\frac{\sigma_r}{\sigma_l}}\right).
\end{split}
\end{equation}
\end{subequations}
Then the total length of the geodesic is $d\approx d_{\rm AC}+d_{\rm CD}+d_{\rm DB}$, where the prefactor of $\log{\sigma_l}$ is $\frac{3}{2}L_1+\frac{1}{2}L_1$.
Actually, we can compare this analytical result with numerical results \eqref{aeq:entanglement entropy for CFT/CFT geodesic on infinite system}, and they are consistent in figure~\ref{afig:general_solution_wrt_Rr_CFTCFT} (b).

\subsection{ETW/defect geodesic with an ETW brane}

\subsubsection{Special case with \texorpdfstring{$\sigma_l\rightarrow0$}{}}

\begin{figure}[tbp]
\centering 
\subfigure[]{\includegraphics[width=.35\textwidth]{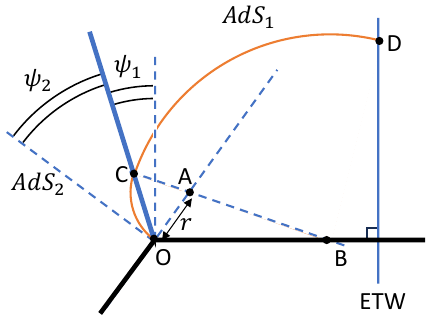}}\quad
\subfigure[]{\includegraphics[width=.35\textwidth]{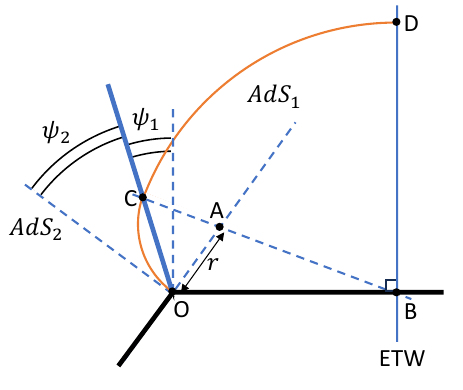}}
\caption{\label{afig:geodesic_diagram_ETW_defet_0} AdS dual of CFT with interface defect for half infinite system and ETW brane. Two endpoints of the geodesic are located on the defect and the ETW brane.}
\end{figure}

In the following, we consider the effect of the ETW brane.
More specifically, the two endpoints of the geodesic are located at $\sigma_l\rightarrow0$ and the ETW brane, as shown in figure~\ref{afig:geodesic_diagram_ETW_defet_0}.
It is obvious that there are more than one solution for the geodesic equation with an endpoint on the ETW, so we need to find a geodesic with the minimal length.
Similar to appendix~\ref{general case for CFT/CFT geodesic in infinite system}, we start by assuming $|{\rm OA}|=r$.
With $\angle {\rm ACO}=\frac{\pi}{2}-\psi_2$ and $\angle {\rm COB}=\frac{\pi}{2}+\psi_1$, we have $\angle {\rm CBO}=\psi_2-\psi_1$.
It is obvious that for $\psi_2<\psi_1$ the crossing point of lines ${\rm CA}$ and ${\rm OB}$ will be on the left of the point ${\rm O}$, which means there is no solution for the geodesic we want.

Therefore, in the following, we just consider $\psi_2 > \psi_1$.
(i) Starting from figure~\ref{afig:geodesic_diagram_ETW_defet_0} (a), we assume $|{\rm OA}|=r$, and the equation of the line ${\rm AB}$ is $y-r\sin{(\psi_1+\psi_2)}=\tan{(\pi+\psi_1-\psi_2)}(x-r\cos{(\psi_1+\psi_2)})$.
Setting $y=0$ we have 
\begin{subequations}
\label{aeq:defect/ETW geodesic in half infinite system special equation}
\begin{equation}
\label{aeq:defect/ETW geodesic in half infinite system special equation eq1}
    x_{\rm B}=\beta=\frac{-r\sin{(\psi_1+\psi_2)}}{\tan{(\psi_1-\psi_2)}}+r\cos{(\psi_1+\psi_2)}.
\end{equation}
(ii) Assuming $|{\rm OC}|=\alpha$, because the point ${\rm C}$ is located at the line ${\rm AB}$, we can solve and get 
\begin{equation}
\label{aeq:defect/ETW geodesic in half infinite system special equation eq2}
    \alpha=\frac{r\sin{(\psi_1+\psi_2)}-r\tan{(\psi_1-\psi_2)}\cos{(\psi_1+\psi_2)}}{\cos{\psi_1}+\sin{\psi_1}\tan{(\psi_1-\psi_2)}}.
\end{equation}
(iii) In $\triangle {\rm BCO}$, using the law of sines we have $|{\rm BC}|=\frac{\beta \cos{\psi_1}}{\cos{\psi_2}}$.
Assuming ${\rm D}=(\sigma,\delta)$, because $|{\rm BC}|=|{\rm BD}|$, we have
\begin{equation}
\label{aeq:defect/ETW geodesic in half infinite system special equation eq3}
    \delta=\sqrt{\beta^2\left(\frac{\cos{\psi_1}}{\cos{\psi_2}}\right)^2-(\sigma-\beta)^2}.
\end{equation}
\end{subequations}

Now we can express the geodesic length with the variables above.
For ${\rm OC}$, with ${\rm C}=(\alpha\sin{\psi_2},\alpha\cos{\psi_2})$ and ${\rm O}=(0,\varepsilon)$ we have 
\begin{subequations}
\label{aeq:defect/ETW geodesic in half infinite system special solution}
\begin{equation}
\label{aeq:defect/ETW geodesic in half infinite system special solution eq1}
    d_{\rm OC}=L_2\cosh^{-1}{\frac{(\alpha\sin{\psi_1})^2+\varepsilon^2+(\alpha\cos{\psi_1})^2}{2\varepsilon\alpha\cos{\psi_2}}}\approx L_2\log{\frac{\alpha}{\varepsilon\cos{\psi_2}}}.
\end{equation}
For ${\rm CD}$, with ${\rm C}=(-\alpha\sin{\psi_1},\alpha\cos{\psi_1})$ and ${\rm D}=(\sigma,\delta)$ we have 
\begin{equation}
\label{aeq:defect/ETW geodesic in half infinite system special solution eq2}
    d_{\rm CD}=L_1\cosh^{-1}{\frac{(-\alpha\sin{\psi_1}-\sigma)^2+\delta^2+(\alpha\cos{\psi_1})^2}{2\delta\alpha\cos{\psi_1}}}.
\end{equation}
\end{subequations}
Therefore, the total geodesic length is 
\begin{equation}
\label{aeq:total defect/ETW geodesic in half infinite system special solution}
    d=L_2\log{\frac{\alpha}{\varepsilon\cos{\psi_2}}}+L_1\cosh^{-1}{\frac{(-\alpha\sin{\psi_1}-\sigma)^2+\delta^2+(\alpha\cos{\psi_1})^2}{2\delta\alpha\cos{\psi_1}}}.
\end{equation}
Besides, we also mention that without a defect (region 2), the geodesic will cross the ETW brane perpendicularly.
So, the length is 
\begin{equation}
\label{aeq:total defect/ETW geodesic in half infinite system special solution without defect}
    d_0=L_1\cosh^{-1}{\frac{\sigma^2+\sigma^2+\varepsilon^2}{2\sigma\varepsilon}}\approx L_1\log{\frac{2\sigma}{\varepsilon}}.
\end{equation}

Now what we need to do is to minimize \eqref{aeq:total defect/ETW geodesic in half infinite system special solution} with respect to $r$.
Actually, with numerical check we can find that, after minimizing \eqref{aeq:total defect/ETW geodesic in half infinite system special solution}, the corresponding $r=r_0$ will lead $\beta=\sigma$, which means the geodesic will cross the ETW brane perpendicularly and the point ${\rm B}$ is also located at the ETW.
In the following, we will prove that the geodesic will always cross the ETW brane perpendicularly, and express the total geodesic length analytically.

The calculation is tedious.
With \eqref{aeq:defect/ETW geodesic in half infinite system special equation}, we can have $\beta=\beta(r),\ \alpha=\alpha(r),\ \delta=\delta(\beta)$.
Then we plug these equations into \eqref{aeq:total defect/ETW geodesic in half infinite system special solution} and calculate $\frac{\partial d}{\partial r}$.
For the perpendicular case we have $\beta=\sigma$, which means $r=r_0=\frac{\sigma}{\cos{(\psi_1 + \psi_2)} - \sin{(\psi_1 + \psi_2)}/\tan{(\psi_1 -\psi_2)}}$.
With $r=r_0$ in $\frac{\partial d}{\partial r}$ and simplifying it with $\psi_1<\psi_2 $, we have
\begin{equation}
\label{aeq:simplified d(d)/d(r) for r0}
    \left.\frac{\partial d}{\partial r}\right|_{r=r_0}=-\frac{2 \cos{\psi_2}\sin{\psi_2} (L_2 \csc{(\psi_1 - \psi_2)} - L_1 \cos{\psi_2} \csc{(\psi_1 - \psi_2)}\sec{\psi_1}) }{\sigma}.
\end{equation}
With the junction condition \eqref{eq:special form of matching condition 1} we have $\frac{L_1}{\cos{\psi_1}}=\frac{L_2}{\cos{\psi_2}}$, so $\left.\frac{\partial d}{\partial r}\right|_{r=r_0}=0$, which means that $r=r_0$ corresponds to the minimal point of $d(r)$.
Therefore, the minimal geodesic will cross the ETW brane perpendicularly.
Actually, we will show later that this property of minimal geodesics will also be true for more complicated cases, and give an intuitive proof in appendix~\ref{Proof of perpendicular crossing of geodesic}.

For the minimal geodesic above with $r=r_0$, we can plug it into \eqref{aeq:total defect/ETW geodesic in half infinite system special solution} and calculate the length of the minimal geodesic in figure~\ref{afig:geodesic_diagram_ETW_defet_0} (b).
Firstly, we consider $\triangle {\rm COB}$ with the law of sines, then we have
\begin{equation}
\label{aeq:COB with the law of sines}
    \frac{\sigma}{\sin{(\frac{\pi}{2}-\psi_2})}=\frac{|\rm CB|}{\sin{(\frac{\pi}{2}+\psi_1})}=\frac{|\rm OC|}{\sin{(\psi_2-\psi_1)}},
\end{equation}
which leads $|{\rm CB}|=|{\rm BD}|=\frac{\sigma\cos{\psi_1}}{\cos{\psi_2}}$ and $|{\rm OC}|=\frac{\sigma\sin{(\psi_2-\psi_1)}}{\cos{\psi_2}}$.
Plugging $|{\rm OC}|$ into \eqref{aeq:defect/ETW geodesic in half infinite system special solution}, we have the total geodesic length
\begin{equation}
\label{aeq:total defect/EOW geodesic in half infinite system with r0}
    d=L_2\log{\left(\frac{\sigma}{\varepsilon}\right)}+d_{\rm sub},
\end{equation}
with the sub-leading term
\begin{equation}
\begin{split}
\label{aeq:boundary contribution of defect/EOW geodesic in half infinite system with r0}
    d_{\rm sub}=&L_2\log{\left(\frac{\sin{(\psi_2-\psi_1)}}{\cos^2{\psi_2}}\right)}\\
    &+L_1 \cosh^{-1}{\left(\frac{\sin^2{(\psi_2-\psi_1)}+\cos^2{\psi_2}+\cos^2{\psi_1}+2\sin{(\psi_2-\psi_1)}\sin{\psi_1}\cos{\psi_2}}{2\sin{(\psi_2-\psi_1)}\cos^2{\psi_1}}\right)}.
\end{split}
\end{equation}
Therefore, the corresponding entanglement entropy is $S_{\sigma}=\frac{c_2}{6}\log{\frac{\sigma}{\varepsilon}}+S_{\rm sub}$ with $S_{\rm sub}=\frac{d_{\rm sub}}{4G_{(3)}}$.

\subsubsection{General case with \texorpdfstring{$\sigma_l\neq0$}{}}
\begin{figure}[tbp]
\centering 
\subfigure[]{\includegraphics[width=.35\textwidth]{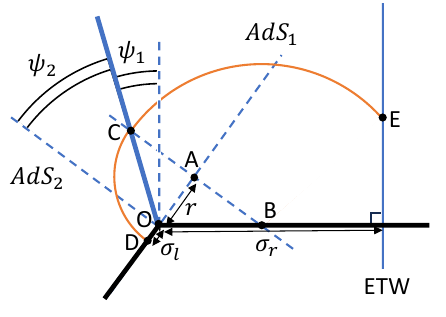}}\quad
\subfigure[]{\includegraphics[width=.35\textwidth]{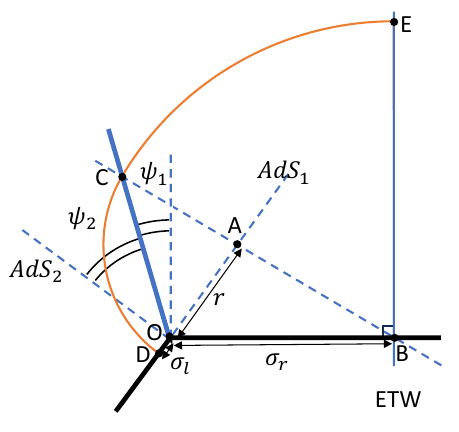}}
\caption{\label{afig:geodesic_diagram_EOW_defet_neq0} AdS dual of CFT with interface defect for half infinite system and ETW brane. The two endpoints are located on the left CFT and the ETW brane.}
\end{figure}

Before, we only considered the case with $\sigma_l\rightarrow0$, now we consider a more general case where $\sigma_l\neq0$.
Similar to the method above, in figure~\ref{afig:geodesic_diagram_EOW_defet_neq0} (a) we assume $|{\rm OA}|=r,\ |{\rm OC}|=\alpha$.
(i) Because $|{\rm OA}|=|{\rm AC}|$, we can solve $(\sigma_l+r)^2=\left[r\cos{(\psi_1+\psi_2)}+\alpha\sin{\psi_1}\right]^2+\left[r\sin{(\psi_1+\psi_2)}-\alpha\cos{\psi_1}\right]^2$ and get 
\begin{subequations}
\label{aeq:defect/EOW geodesic in half infinite system general equation}
\begin{equation}
\label{aeq:defect/EOW geodesic in half infinite system general equation eq1}
    \alpha=r\sin{\psi_2}+\sqrt{(r+\sigma_l)^2-r^2\cos^2{\psi_2}}.
\end{equation}
(ii) Then we have the equation of the line ${\rm AC}$ and $B=(x_0,0)$ on it, so
\begin{equation}
\label{aeq:defect/EOW geodesic in half infinite system general equation eq2}
    x_0=[-r\sin{(\psi_1+\psi_2)}]\frac{-\alpha\sin{\psi_1}-r\cos{(\psi_1+\psi_2)}}{\alpha\cos{\psi_1}-r\sin{(\psi_1+\psi_2)}}+r\cos{(\psi_1+\psi_2)}.
\end{equation}
(iii) Assuming ${\rm E}=(\sigma_r, \beta)$, with $|{\rm BC}|=|{\rm BE}|$, we have $(-\alpha\sin{\psi_1}-x_0)^2+(\alpha\cos{\psi_1})^2=(\sigma_r-x_0)^2+\beta^2$, which leads to
\begin{equation}
\label{aeq:defect/EOW geodesic in half infinite system general equation eq3}
    \beta=\sqrt{2\alpha\sin{\psi_1}x_0+\alpha^2-\sigma_r^2+2\sigma_r x_0}.
\end{equation}
\end{subequations}

Now we can calculate the length of the geodesic.
For ${\rm DC}$, with ${\rm D}=(-\sigma_l,\varepsilon)$ and ${\rm C}=(\alpha\sin{\psi_2},\alpha\cos{\psi_2})$, we have 
\begin{subequations}
\label{aeq:defect/EOW geodesic in half infinite system general solution}
\begin{equation}
\label{aeq:defect/EOW geodesic in half infinite system general solution eq1}
    d_{\rm DC}=L_2\cosh^{-1}{\frac{(\alpha\sin{\psi_2}+\sigma_l)^2+\varepsilon^2+(\alpha\cos{\psi_2})^2}{2\varepsilon\alpha\cos{\psi_2}}}\approx L_2\log{\frac{\alpha^2+2\alpha\sigma_l\sin{\psi_2}+\sigma_l^2}{\varepsilon\alpha\cos{\psi_2}}}.
\end{equation}
For ${\rm CE}$, with ${\rm C}=(-\alpha\sin{\psi_1},\alpha\cos{\psi_1})$ and ${\rm E}=(\sigma_r,\beta)$, we have 
\begin{equation}
\label{aeq:defect/EOW geodesic in half infinite system general solution eq2}
    d_{\rm CE}=L_1\cosh^{-1}{\frac{(\alpha\sin{\psi_1}+\sigma_r)^2+(\alpha\cos{\psi_1})^2+\beta^2}{2\alpha\beta\cos{\psi_1}}}= L_1\cosh^{-1}{\frac{\alpha^2+\beta^2+2\alpha\sigma_r\sin{\psi_1}+\sigma_r^2}{2\alpha\beta\cos{\psi_1}}}.
\end{equation}
\end{subequations}
Therefore, the total length is 
\begin{equation}
\label{aeq:total defect/EOW geodesic in half infinite system general solution}
    d=L_2\log{\frac{\alpha^2+2\alpha\sigma_l\sin{\psi_2}+\sigma_l^2}{\varepsilon\alpha\cos{\psi_2}}}+L_1\cosh^{-1}{\frac{\alpha^2+\beta^2+2\alpha\sigma_r\sin{\psi_1}+\sigma_r^2}{2\alpha\beta\cos{\psi_1}}}.
\end{equation}
Similar to the method above, we should minimize \eqref{aeq:total defect/EOW geodesic in half infinite system general solution} with respect to $r$.
Numerically, it is easy to check that the minimal geodesic crosses the ETW brane perpendicularly.
But analytical proof is difficult.

In the following, we will take the assumption above and calculate the length analytically.
With $x_0=\sigma_r$, expressing $\alpha$ with $r$ by \eqref{aeq:defect/EOW geodesic in half infinite system general equation eq2} and plugging it into \eqref{aeq:defect/EOW geodesic in half infinite system general equation eq1},  we will get an effective cubic equation for $r$.
What we need to do is solving this equation which has an analytical solution.
The effective cubic equation is $ar^3+br^2+cr+d=0$ with factors
\begin{equation}
\label{aeq:effective cubic equation for r}
\begin{split}
    a=&-2\sigma_r \cos{\psi_1}\cos{\psi_2}+2\sigma_l\cos^2{\psi_2} + 2\sigma_r\cos^2{\psi_2}\cos{(\psi_1 + \psi_2)},\\
    b =&\ \sigma_r^2\cos^2{\psi_1}-4\sigma_l\sigma_r\cos{\psi_1}\cos{\psi_2} +\sigma_l^2\cos^2{\psi_2}-\sigma_r^2\cos^2{\psi_2},\\
    c =&\ 2\sigma_l\sigma_r^2\cos^2{\psi_1}-2\sigma_l^2\sigma_r\cos{\psi_1}\cos{\psi_2},\\
    d =&\ \sigma_l^2\sigma_r^2\cos^2{\psi_1}.
\end{split}
\end{equation}
This equation has analytical solutions, and we will not show them here.
After solving this equation, we also need to choose one true solution from three solutions.
One criterion is that the corresponding $\alpha,\ \beta$ and $x_0$ are positive.
Finally, plugging the solution into \eqref{aeq:defect/EOW geodesic in half infinite system general equation} and \eqref{aeq:total defect/EOW geodesic in half infinite system general solution}, we will get the final minimal length.

Here we give some remarks about the results above.
(i) In the discussion we don't consider the relation between $L_1$ and $L_2$, which plays an important role before when we take $\sigma_l\rightarrow0$.
With reference~\cite{Anous_2022}, for any given $\sigma_l$ and $\sigma_r$, we can have a nontrivial solution.
Then for our case, we can consider there is an ETW brane across the center of the arc in the right region 1.
Therefore, the solution always exists in this case.
(ii) However, the prefactor of $\log$-term may be different for different relation between $L_1$ and $L_2$, and it may also be different at different limits.
For example, $\sigma_r\rightarrow\infty$ and $\sigma_l\rightarrow\infty$ may have different prefactors of $\log$-term.

\subsubsection{Proof of perpendicular crossing of a geodesic}
\label{Proof of perpendicular crossing of geodesic}
Before, we show that for some simple cases we can prove analytically and numerically that, with ETW brane, the minimal geodesic will end on the ETW brane perpendicularly.
And for more complicated cases, we may only check this property numerically.
Here we give an intuitive proof without any calculation for this property.

For example, considering the most complicated case shown in figure~\ref{afig:geodesic_diagram_EOW_defet_neq0}, the geodesic starts from the point ${\rm D}$ with fixed $\sigma_l$.
Now we want to find the minimal geodesic from D to the ETW brane.
We can construct a different geometry that includes the original part, plus an additional part symmetric with respect to the ETW brane.
Then we can image that, when we minimize the geodesic with endpoints D and its symmetric point, the corresponding geodesic length will always twice of the original geodesic.
If the original geodesic is not perpendicular to ETW brane, then the new geodesic in doubled geometry will not be smooth, which means it is not a ``geodesic''.
Then we can have a shorter one.
Therefore, with ETW brane, the geodesic will always be perpendicular to the ETW brane.

\section{Boundary entropy with path integral}
\label{sec:boundary entropy with path integral}

In this section, we focus on the boundary entropy induced by the interface brane.
By evaluating the partition function, we can obtain the boundary entropy without UV information, and the results are universal.

\subsection{A review of a single region boundary entropy and the conformal transformation}

Before discussing our cases, we first review the results in reference~\cite{Fujita_2011}.
They consider a single CFT region with a boundary.
The corresponding action is 
\begin{equation}
\label{aeq:AdS/BCFT action in ref}
    I=\frac{1}{16\pi G_N}\int_N \sqrt{-g}(R-2\Lambda)+\frac{1}{8\pi G_N}\int_Q \sqrt{-h}(K-T),
\end{equation}
where $R$ denotes Ricci scalar, $\Lambda$ is cosmological constant, $K$ is extrinsic curvature, and $T$ is the tension of the interface brane.
Generally, we have $\Lambda=-\frac{d(d-1)}{2L^2}$ where $L$ is the ${\rm AdS}_{d+1}$ radius (here we take $d=2$ and $\Lambda=-\frac{1}{L^2}$).
Besides, we have $R=\frac{2d+2}{d-1}\Lambda$, so for $d=2$ we have $R=6\Lambda=-\frac{6}{L^2}$.
While, for the integral on $Q$, we need to solve the equation of motion and fix the location of brane.

In the action we only have one region, so we only consider the second junction condition, which requires
\begin{equation}
\label{aeq:e.o.m. for K}
    K_{ab}=(K-T)h_{ab}.
\end{equation}
Taking trace gives $K=\frac{d}{d-1}T$, it means $K=2T$ for $d=2$.
Therefore, we still have 
\begin{equation}
\label{aeq:expression of extrinsic curvature}
    K_{ab}=\frac{1}{L}\tanh{\left(\frac{\rho}{L}\right)}g_{ab},
\end{equation}
which gives $K=g^{ba}K_{ab}=\frac{d}{L}\tanh{\left(\frac{\rho}{L}\right)}$.
For $d=2$, we have $LT=\tanh{\left(\frac{\rho}{L}\right)}$.
Now we can simplify the action to (with Euclidean time $\tau$)
\begin{equation}
\label{aeq:simplified AdS/BCFT action in ref}
    I=\frac{-1}{16\pi G_N}\left(-\frac{4}{L^2}\int_N \sqrt{g}+K\int_Q \sqrt{h}\right).
\end{equation}
We keep $K$ in action instead of $T$ as because we can consider $K$ as a variable that belongs to one region.
Then, we calculate $\int_N \sqrt{g}$ and $\int_Q \sqrt{h}$, where $h$ is the induced metric on the brane.

Before calculating the volume, we introduce a UV cutoff to make it finite.
To realize it, we apply a special conformal transformation (SCT) for the CFT.
The general form of transformation is \cite{Berenstein_1999}
\begin{equation}
\label{aeq:SCT for CFT in AdS}
    x_\mu'=\frac{x_\mu+c_\mu x^2}{1+2c\cdot x+c^2 x^2},
\end{equation}
where we define $c_\mu=(c_x,c_\tau,c_z=0)$, $c\cdot x=\sum_\mu c_\mu x_\mu$, $c^2=\sum_\mu c_\mu^2$ and $x^2=\sum_\mu x_\mu^2$.
In the equation (3.58) of reference~\cite{Anous_2022}, the corresponding transformation is $c_\mu=(0,c_\tau=c,0)$.
While, in the equation (3.5) of reference~\cite{Fujita_2011}, they consider a general transformation to $(x,\tau,z)$
\begin{equation}
\label{aeq:SCT for CFT in AdS for boundary entropy}
\begin{split}
    x'=\frac{x+c_x (x^2+\tau^2+z^2)}{1+2(c_x x+c_\tau \tau)+(c_x^2+c_\tau^2)(x^2+\tau^2+z^2)},\\
    \tau'=\frac{\tau+c_\tau (x^2+\tau^2+z^2)}{1+2(c_x x+c_\tau \tau)+(c_x^2+c_\tau^2)(x^2+\tau^2+z^2)},\\
    z'=\frac{z}{1+2(c_x x+c_\tau \tau)+(c_x^2+c_\tau^2)(x^2+\tau^2+z^2)}.
\end{split}
\end{equation}

Here we briefly discuss the SCT in $d=2$ CFT.
Defining $x_\mu=(x,\tau)$, $b=b_x+{\rm i}b_\tau$ and $z=x+{\rm i}\tau$, the transformation $x_\mu=\frac{x_\mu+b_\mu x^2}{1+2b\cdot x+b^2 x^2}$ can be expressed as $z'=\frac{z+b z \Bar{z}}{1+(b\Bar{z}+\Bar{b}z)+b\bar{b}z\bar{z}}$.
On the other hand, the transformation above can be written as $z'=\frac{z}{1+\bar{b}z}=\frac{1}{z^{-1}+\bar{b}}$. 
It is because
\begin{equation}
\label{aeq:property of SCT}
    z'=\frac{z\Bar{z}}{\Bar{z}+\bar{b}z\bar{z}}=\frac{z\Bar{z}(z+b z\bar{z})}{(\Bar{z}+\bar{b}z\bar{z})(z+b z\bar{z})}=\frac{z\Bar{z}(z+b z\bar{z})}{z\Bar{z}+\bar{b}z^2\bar{z}+bz\bar{z}^2+b\bar{b}z\bar{z} z\bar{z}}=\frac{z+b z \Bar{z}}{1+(b\Bar{z}+\Bar{b}z)+b\bar{b}z\bar{z}}.
\end{equation}
Then it is straightforward to check that the lines $x=x_0$ and $\tau=\tau_0$ will be mapped to circles except for some special lines that depend on $b$.

Now we consider the SCT in the AdS space.
To simplify the problem, we consider that $c_\mu=(c_0,0,0)$, and the location of brane is $x=z\sinh{\delta}+x_0, \delta=\frac{\rho}{L}$.
Then we can directly check that under \eqref{aeq:SCT for CFT in AdS for boundary entropy}, the brane is mapped to
\begin{equation}
\label{aeq:brane location after SCT}
    (x'-d_1)^2+\tau'^2+(z'-d_2)^2-d_3=0,
\end{equation}
where $d_1=\frac{1+2c_0 x_0}{2c_0(1+c_0 x_0)}$, $d_2=-\frac{\sinh{\delta}}{2c_0(1+c_0 x_0)}$ and $d_3=\frac{\cosh^2{\delta}}{4c_0^2 (1 + c_0 x_0)^2}$.
It is a sphere with radius $R=\sqrt{d_3}=\frac{\cosh{\delta}}{|2c_0 (1 + c_0 x_0)|}$ and center $(d_1,0,d_2)$.
For $z'=0$ and $\tau'=0$, we have $x'=\frac{1}{c_0}, \frac{x_0}{1+c_0 x_0}$, which means $(\frac{1}{c_0},0,0)$ is a fixed point for any $x_0$.
Besides, for $x_0=0$ we have $r_D=\frac{1}{2|c_0|}$ in \cite{Fujita_2011}.
Therefore, with a proper $c_\mu$, we can transform the region $x<x_0$ to a disk.

\begin{figure}[tbp]
    \centering 
    {\includegraphics[width=.35\textwidth]{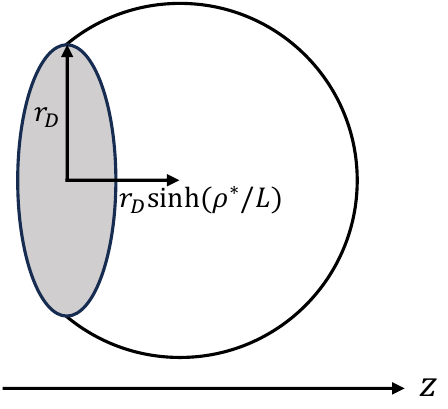}}
    \caption{\label{afig:integral_region} Location of the brane after conformal transformation.}
\end{figure}

After the SCT, the CFT region is on a disk with radius $r_D$ in figure~\ref{afig:integral_region}.
However, the disk radius is not essential for boundary entropy.
And the location of the center is $z=r_D \sinh{\frac{\rho^*}{L}}$.
Because conformal transformation won't change the angle, the angle between the brane and the asymptotic CFT boundary is $\frac{\pi}{2}-\psi$, and $\tan{\psi}=\sinh{\frac{\rho^*}{L}}$ \cite{Fujita_2011}.
This is also consistent with the result above, where the location of center is $z=\sqrt{d_3}$ and corresponding $r_D=\frac{1}{|2c_0(1+c_0 x_0)|}$.
\footnote{In the following subsection, to simplify the problem, we may add an ETW brane artificially, which only acts as a UV-cutoff.
However, from later results we will find that the value of radius $r_D$ will not change the boundary entropy, which means we can consider the region bounded by an ETW brane at $x_0\rightarrow+\infty$.
Then we can apply the corresponding conformal transformation with $c_0=-\frac{\delta}{x_0(x_0+\delta)}$ and $\delta\gtrsim0$.
It means the region on the left of $x=x_0$ will be mapped to a disk, and the corresponding radius is $r_D=\frac{(x_0+\delta)^2}{2\delta}$.
Assuming $\delta=1$ and $x_0\gg1$, we have $r_D\approx\frac{x_0^2}{2}$.
The corresponding brane after SCT will cross the x-axis on $x=\frac{1}{c_0}=-\frac{x_0(x_0+\delta)}{\delta}\approx -x_0^2$ and $x=\frac{x_0}{1+c_0 x_0}\approx x_0$, which means the disk region will cover the whole space for $x_0\rightarrow+\infty$.}

Now we discuss the integral $\int_N \sqrt{g}$ and $\int_Q \sqrt{h}$. 
With metric \eqref{eq:metric of coordinate 2}, we have $\sqrt{g}=\left(\frac{L}{z}\right)^3$.
Defining $\Bar{z}=r_D (\sinh{\frac{\rho^*}{L}}+\cosh{\frac{\rho^*}{L}})=r_D e^{\rho^*/L}$, we have
\begin{equation}
\label{aeq:integral of volume for boundary entropy}
\begin{split}
    \int_N \sqrt{g}=&L^3\int_\varepsilon^{\Bar{z}} {\rm d}z\ 
    \frac{1}{z^3}\int {\rm d}^2 S=L^3\int_\varepsilon^{\Bar{z}} {\rm d}z\ 
    \frac{1}{z^3}\pi\left((r_D \cosh{\frac{\rho^*}{L}})^2-(z-r_D \sinh{\frac{\rho^*}{L}})^2\right)\\
    =&\pi L^3\left(\frac{r_D^2}{2\varepsilon^2}-\frac{1}{2}e^{-2\rho^*/L}+2r_D\sinh{\frac{\rho^*}{L}}(\frac{1}{\varepsilon}-\frac{1}{r_D}e^{-\rho^*/L})+\log{\frac{\varepsilon}{r_D}e^{-\rho^*/L}}\right).
\end{split}
\end{equation}
As for $\int_Q \sqrt{h}$, because Q satisfies the equation $x^2+\tau^2+(z-r_D\sinh{\frac{\rho^*}{L}})^2=r_D^2\cosh^2{\frac{\rho^*}{L}}$, the induced metric is
\begin{equation}
\label{aeq:induced metric on Q}
    {\rm d} s^2=\frac{L^2}{z^2}\left(\left(1+\frac{(z-r_D\sinh{\frac{\rho^*}{L}})^2}{x^2}\right){\rm d} z^2+\left(1+\frac{\tau^2}{x^2}\right){\rm d}\tau^2+\frac{2\tau}{x^2}\left(z-r_D\sinh{\frac{\rho^*}{L}}\right){\rm d}z{\rm d}\tau\right).
\end{equation}
Therefore, we have volume form
\begin{equation}
\label{aeq:volume form of Q}
\begin{split}
    \sqrt{h}=&\frac{L^2}{z^2}\sqrt{\frac{r_D^2+z_0^2}{r_D^2-\tau^2-z^2+2z z_0}}.
\end{split}
\end{equation}
To simplify the notation, in the following we will denote $z_0=r_D\sinh{\frac{\rho^*}{L}}$, then the integral region is $Q:\ \tau\in[0,\sqrt{z_0^2+r_D^2-(z-z_0)^2}],\ z\in[\varepsilon,\sqrt{z_0^2+r_D^2}+z_0]$.
Thus, the integral is 
\begin{equation}
\label{aeq:integral of surface for boundary entropy}
\begin{split}
    \int_Q \sqrt{h}=&4\int_\varepsilon^{\sqrt{z_0^2+r_D^2}+z_0}{\rm d}z\int_0^{\sqrt{z_0^2+r_D^2-(z-z_0)^2}}{\rm d}\tau\frac{L^2}{z^2}\sqrt{\frac{r_D^2+z_0^2}{r_D^2-\tau^2-z^2+2z z_0}}\\
    =&2\pi L^2 r_D\cosh{\frac{\rho^*}{L}}\left(\frac{1}{\varepsilon}+\frac{-e^{-\rho^*/L}}{r_D}\right).
\end{split}
\end{equation}
Finally we get the integral of action \eqref{aeq:simplified AdS/BCFT action in ref}
\begin{equation}
\label{aeq:integral result of AdS/BCFT action in ref}
\begin{split}
    I=&\frac{-1}{16\pi G_N}\left[-\frac{4}{L^2}\left(\pi L^3\left(\frac{r_D^2}{2\varepsilon^2}-\frac{1}{2}e^{-2\rho^*/L}+2r_D\sinh{\frac{\rho^*}{L}}(\frac{1}{\varepsilon}-\frac{1}{r_D}e^{-\rho^*/L})+\log{(\frac{\varepsilon}{r_D}e^{-\rho^*/L})}\right)\right)\right.\\
    &\left.+K\left(2\pi L^2 r_D\cosh{\frac{\rho^*}{L}}\left(\frac{1}{\varepsilon}+\frac{-e^{-\rho^*/L}}{r_D}\right)\right)\right]\\
    =&\frac{L}{4 G_N}\left(\frac{r_D^2}{2\varepsilon^2}+\log{\frac{\varepsilon}{r_D}}-\frac{\rho^*}{L}-\frac{1}{2}+\frac{r_D}{\varepsilon}\sinh{\frac{\rho^*}{L}}\right),
\end{split}
\end{equation}
where we use $LT=\tanh{\left(\frac{\rho}{L}\right)}$.
The final result is the equation (3.7) in reference~\cite{Fujita_2011}.
The corresponding boundary entropy is 
\begin{equation}
\label{aeq:boundary entropy from integral of action}
    S_{\rm bdy}=-\left[I(\rho^*)-I(0)\right]=\frac{\rho^*}{4G_N}.
\end{equation}
With $LT=\tanh{\left(\frac{\rho}{L}\right)}$, we have $S_{\rm bdy}=\frac{L\tanh^{-1}{LT}}{4G_N}=\frac{c\tanh^{-1}{(LT)}}{6}$.
Later we will find it is universal and valid for other cases.
Besides, we can also express the boundary entropy as $S_{\rm bdy}=\frac{c\rho^*}{6L}=\frac{c}{6}\tanh^{-1}{(\sin{\psi})}$, where $\psi$ is defined below \eqref{eq:metric of coordinate 2}.
With \eqref{eq:helpful equation 2 for proof}, we find that $S_{\rm bdy}=\frac{c}{6}\log{\left(\tan{(\frac{\psi}{2}+\frac{\pi}{4})}\right)}$, which is consistent with \eqref{eq:simplified boundary entropy for symmetric CFT/CFT geodesic on infinite system}, where the prefactor $1/3$ is because there are two branes and three regions for that case.

\begin{figure}[tbp]
\centering 
\subfigure[]{\includegraphics[width=.8\textwidth]{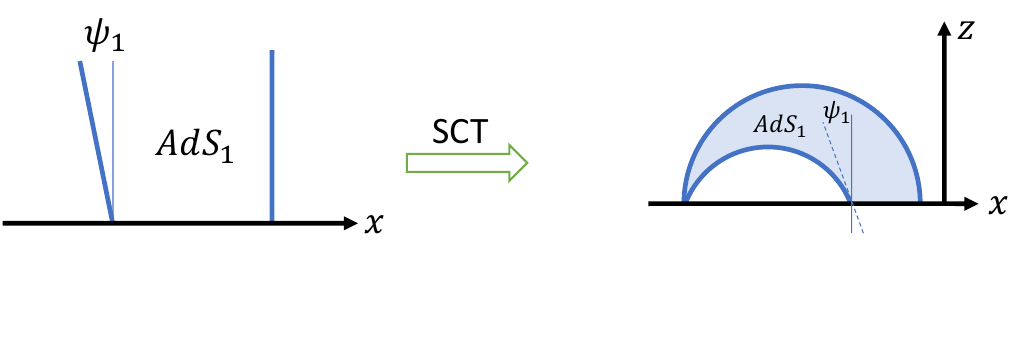}}\\
\subfigure[]{\includegraphics[width=.8\textwidth]{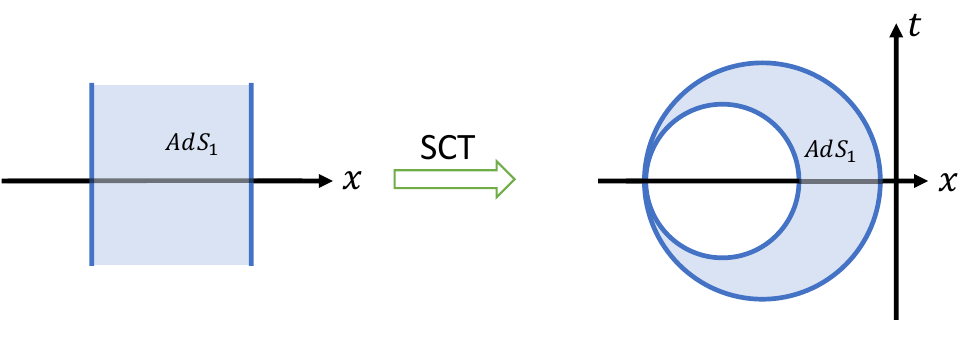}}
\caption{\label{afig:reference_right_figure} (a) AdS bulk dual of single CFT region before and after conformal transformation. (b) Single CFT region on the boundary before and after conformal transformation.}
\end{figure}

Let's consider two branes. 
Under the same SCT, on the boundary CFT, two branes will be mapped to two circles which are tangent at the fixed point, which is shown in figure~\ref{afig:reference_right_figure}.
From the figure~\ref{afig:reference_right_figure} (a), we can define the angle $\rho$ which corresponds to the inner angle in the blue region.
It means we define the direction of the ETW brane to be outwards to the bulk region.
Then the corresponding boundary entropy is $S_{\rm bdy}=\frac{\rho^*_1+\rho^*_2}{4G_N}$.
We can consider the integral as two parts.
One part is the larger disk and the other is the smaller disk, and the corresponding actions have opposite signs.
For each action, we can calculate the boundary entropy with the method above.
It is worth mentioning that if we set the larger ETW brane perpendicularly with angle $\rho^*=0$, then it will not contribute to the boundary entropy.

\subsection{Boundary entropy for more than one region}

\begin{figure}[tbp]
\centering 
\subfigure[]{\includegraphics[width=.9\textwidth]{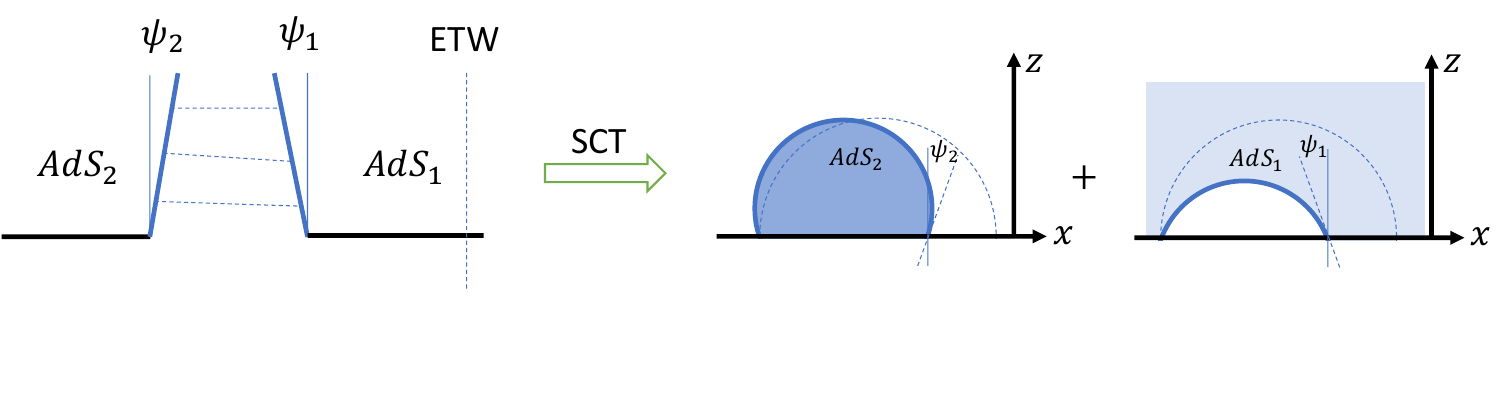}}\\
\subfigure[]{\includegraphics[width=.7\textwidth]{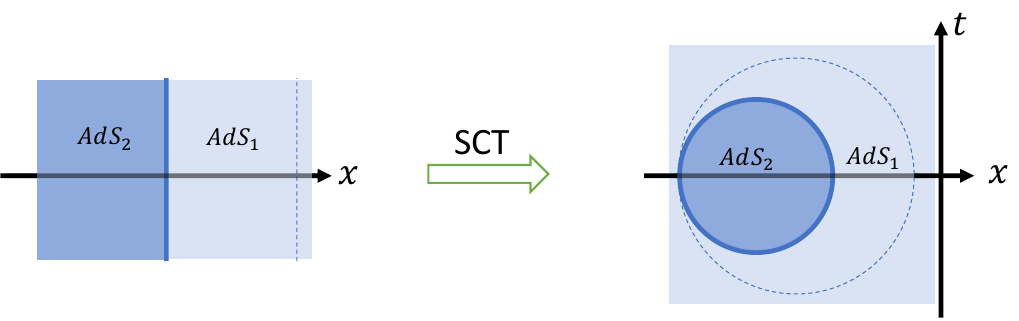}}
\caption{\label{afig:two_region_one_brane_reference} (a) AdS bulk dual of two CFT regions before and after conformal transformation. (b) Two CFT regions on the boundary before and after conformal transformation. Here we show ETW with blue dotted line, and for limit case we can take ETW to infinity.}
\end{figure}

In the following, we consider two regions with different AdS radii that are separated by an interface brane, as shown in figure~\ref{afig:two_region_one_brane_reference}. 
We expect that the boundary entropy of this case is also $S_{\rm bdy}=\frac{\rho^*_1+\rho^*_2}{4G_N}$, where $\rho^*_1,\ \rho^*_2$ are the corresponding angles of the brane in region 1 and 2.
Actually, we can obtain a result consistent with \eqref{eq:simplified boundary entropy for symmetric CFT/CFT geodesic on infinite system}.

We first consider figure~\ref{afig:two_region_one_brane_reference} (a).
If we connect two half planes of CFT, then obviously two bulk AdS regions will have an overlap.
The proper way to understand it is in figure~\ref{afig:two_region_one_brane_reference} (a), where we use some magic glue to connect two parts.
Then we must discuss two bulk regions individually with a proper connection condition and a local coordinate transformation.
After special conformal transformation, we have the geometry shown in figure~\ref{afig:two_region_one_brane_reference}.
To be concrete, we discuss the action \eqref{eq:bottom-up model}.
With connection condition \eqref{eq:matching condition 2}, we have $\Delta K=2T$, and the action can be written as 
\begin{equation}
\begin{split}
\label{aeq:transformed bottom-up model}
    S_{{\rm EH}}=&-\frac{1}{16\pi G_{(3)}} \left[\int_{\mathcal{M}_1}{\rm d}^3 x\ \sqrt{g_1}\left(R_1+\frac{2}{L^2_1}\right)+\int_{\mathcal{M}_2}{\rm d}^3 x\ \sqrt{g_2}\left(R_2+\frac{2}{L^2_2}\right)+\int_{\mathcal{S}} {\rm d}^2 y\ \sqrt{h}(K_1-K_2)\right]\\
    =&-\frac{1}{16\pi G_{(3)}} \left[\int_{\mathcal{M}_1}{\rm d}^3 x\ \sqrt{g_1}\left(R_1+\frac{2}{L^2_1}\right)+\int_{\mathcal{S}} {\rm d}^2 y\ \sqrt{h}K_1+\int_{\mathcal{M}_2}{\rm d}^3 x\ \sqrt{g_2}\left(R_2+\frac{2}{L^2_2}\right)\right.\\
    &\left.-\int_{\mathcal{S}} {\rm d}^2 y\ \sqrt{h}K_2\right]=-\frac{1}{16\pi G_{(3)}} \sum_{\alpha={1,2}}\left[\int_{\mathcal{M}_\alpha} {\rm d}^3 x\ \sqrt{g_\alpha}\left(R_\alpha+\frac{2}{L^2_\alpha}\right)+\int_{\mathcal{S}_\alpha} {\rm d}^2 y\ \sqrt{h}K_\alpha\right]=\sum_{\alpha={1,2}} S_\alpha.
\end{split}
\end{equation}
In reference~\cite{Anous_2022} the left region is $1$, the right region is $2$ and the direction of brane and the extrinsic curvature $K_{1,2}$ are from $1$ to $2$.
But here we introduce the surface $\mathcal{S}_{1,2}$, where $\mathcal{S}_1=\mathcal{S}=-\mathcal{S}_2$.
Therefore, directions of $\mathcal{S}_{1,2}$ are defined from inside to outside.
With this, we have $S_{{\rm EH}}=\sum_{\alpha={1,2}} S_\alpha$, and $S_\alpha$ is the same as \eqref{aeq:simplified AdS/BCFT action in ref}.
So from the form of the action we can also show that the boundary entropy can be expressed as the sum of contributions from the two regions.
Besides, here we only consider the brane between two regions, but not the larger ETW brane, because we set the corresponding angle $\rho=0$.

\section{Different phases and phase transition for measurements}
\label{sec:different phases and phase transition for measurement (quantum quench)}

In this appendix, we give (i) a detailed derivation of the geodesic solutions in AdS and black hole spacetime, and (ii) more details on the geodesic length calculation in the bubble phase.

\subsection{Convention of metric and corresponding geodesics}
\label{convention of metric and corresponding geodesic}

The geodesic equation is
\begin{equation}
\label{aeq:differential equation of geodesic}
    \frac{{\rm d}^2 x^\mu}{{\rm d}\lambda^2}+\Gamma^\mu_{\alpha\beta}\frac{{\rm d}x^\alpha}{{\rm d}\lambda}\frac{{\rm d}x^\beta}{{\rm d}\lambda}=f(\lambda)\frac{{\rm d}x^\mu}{{\rm d}\lambda},
\end{equation}
where $f(\lambda)$ is a function to be determined.
\footnote{Do not confuse with the function $f(r)$ in the metric \eqref{eq:general metric for AdS 3D}.}
With the metric \eqref{eq:general metric for AdS 3D} and the coordinate $(\tau=\tau_0,r=r(x),x=x)$, plugging the Christoffel coefficient into \eqref{aeq:differential equation of geodesic}, we have
\begin{subequations}
\begin{equation}
\label{aeq:specific equation of differential equation of geodesic 1 for used metric}
    \Ddot{x}+\frac{2}{r}\Dot{r}\Dot{x}=f(x)\Dot{x},\\
\end{equation}
\begin{equation}
\label{aeq:specific equation of differential equation of geodesic 2 for used metric}
    \Ddot{r}+\frac{r}{L^2(\mu-1)-r^2}\Dot{r}^2+r(\mu-1-\frac{r^2}{L^2})\Dot{x}^2=f(x)\Dot{r}.
\end{equation}
\end{subequations}
Therefore, with a fixed $f(x)$, we only need to solve the differential equation 
\begin{equation}
\label{aeq:final differential equation for geodesic for used metric}
    \Ddot{r}+\frac{r^2-2[L^2(\mu-1)-r^2]}{r[L^2(\mu-1)-r^2]}\Dot{r}^2+r(\mu-1-\frac{r^2}{L^2})=0.
\end{equation}
The general solution for this differential equation has two parameters, which are denoted as $c_1,c_2$.
For two situations $\mu>1$ and $\mu<1$, the general solutions are 
\begin{subequations}
\label{aeq:general solution of final differential equation for geodesic for used metric}
\begin{equation}
\label{aeq:general solution of final differential equation for geodesic for used metric black hole}
    r=\frac{1}{\sqrt{c_1-(\frac{1}{L^2(\mu-1)}-c_1)\sinh^2{[\sqrt{\mu-1}(x+c_2)]}}}, \qquad \mu>1,
\end{equation}
\begin{equation}
\label{aeq:general solution of final differential equation for geodesic for used metric thermal AdS}
    r=\frac{1}{\sqrt{c_1-(\frac{1}{L^2(1-\mu)}+c_1)\sin^2{[\sqrt{1-\mu}(x+c_2)]}}}, \qquad \mu<1.
\end{equation}
\end{subequations}
Here, we only take positive branch because $r>0$.
For $\mu>1$, we have the constraint $0<c_1<\frac{1}{L^2(\mu-1)}$, while for $\mu<1$, we have the constraint $c_1>0$.
It is obvious that for the limit $\mu\rightarrow1$, the geodesic equation will become a circle with redefinition $\Tilde{z}=1/r$.
However, there is no unique solution for differential equation \eqref{aeq:final differential equation for geodesic for used metric}.
Moreover, there is another solution for $\mu>1$
\begin{equation}
\label{aeq:second solution for general solution of final differential equation for geodesic for used metric black hole}
    r^{-2}+\left(\frac{1}{L^2(\mu-1)}-c_1 \right)\sinh^2{[\sqrt{\mu-1}(x+c_2)]}=\frac{1}{L^2(\mu-1)}.
\end{equation}

We calculate the geodesic length,
\begin{equation}
\label{aeq:integral of geodesic length for used metric}
    d=\int {\rm d}l\ \sqrt{g}=\int {\rm d}s=\int \sqrt{\frac{(r'(x))^2}{f(r)}+r^2}\ {\rm d}x,
\end{equation}
where $f(r)$ is the function defined in metric \eqref{eq:general metric for AdS 3D}, and $r'(x)$ is the derivative of the geodesic.
The integral can be solved and the final results for the geodesics in \eqref{aeq:general solution of final differential equation for geodesic for used metric} are
\begin{subequations}
\label{aeq:solution of integral of geodesic length for used metric}
\begin{equation}
\label{aeq:solution of integral of geodesic length for used metric for black hole}
    d=L\tanh^{-1}{\left[\frac{\tanh{(\sqrt{\mu-1}(x+c_2))}}{L\sqrt{c_1(\mu-1)}}\right]},\qquad \mu>1,
\end{equation}
\begin{equation}
\label{aeq:solution of integral of geodesic length for used metric for thermal AdS}
    d=L\tanh^{-1}{\left[\frac{\tan{(\sqrt{1-\mu}(x+c_2))}}{L\sqrt{c_1(1-\mu)}}\right]},\qquad \mu<1.
\end{equation}
\end{subequations}
Actually, in \eqref{aeq:general solution of final differential equation for geodesic for used metric} and \eqref{aeq:solution of integral of geodesic length for used metric} we can notice that the solutions of two cases are related.
For example, starting from \eqref{aeq:solution of integral of geodesic length for used metric for black hole}, if we now consider $\mu<1$, then $\sqrt{\mu-1}={\rm i}\sqrt{1-\mu}$ and ${\rm i}$ in $\tanh$ will change it to $-{\rm i}\tan$.

Assuming two endpoints located at $(r=\frac{1}{\varepsilon},x=\pm\frac{1}{2}x_0)$, for $\mu<1$, we have parameter $c_1=[\varepsilon^2+\frac{1}{L^2(1-\mu)}\sin^2{(\sqrt{1-\mu}\frac{x_0}{2})}]/\cos^2{(\sqrt{1-\mu}\frac{x_0}{2})}$ and $c_2=0$.
Then the total geodesic length is 
\begin{equation}
\label{aeq:total length of geodesic with unused metric for sued metric for thermal AdS}
\begin{split}
    \Delta d=2L\tanh^{-1}{\left[\frac{\tan{(\sqrt{1-\mu}\frac{x_0}{2})}}{L\sqrt{c_1(1-\mu)}}\right]}\approx2L\log{\left(\frac{2\sin{(\sqrt{1-\mu}\frac{x_0}{2})}}{L\sqrt{1-\mu}\varepsilon}\right)}.
\end{split}
\end{equation}
Therefore, for $x_0\ll \frac{1}{\sqrt{1-\mu}}$, we retain $\Delta d=2L\log{\frac{x_0/L}{\varepsilon}}$.
Similarly, for $\mu>1$, we have parameter $c_1=[\varepsilon^2+\frac{1}{L^2(\mu-1)}\sinh^2{(\sqrt{\mu-1}\frac{x_0}{2})}]/\cosh^2{(\sqrt{\mu-1}\frac{x_0}{2})}$ and $c_2=0$.
Then the total geodesic length is
\begin{equation}
\label{aeq:total length of geodesic with unused metric for sued metric for black hole}
\begin{split}
    \Delta d=2L\tanh^{-1}{\left[\frac{\tanh{(\sqrt{\mu-1}\frac{x_0}{2})}}{L\sqrt{c_1(\mu-1)}}\right]}
    \approx2L\log{\left(\frac{2\sinh{(\sqrt{\mu-1}\frac{x_0}{2})}}{L\sqrt{\mu-1}\varepsilon}\right)}.
\end{split}
\end{equation}
Therefore, for $x_0\ll \frac{1}{\sqrt{\mu-1}}$, we retain $\Delta d=2L\log{\frac{x_0/L}{\varepsilon}}$. While, for $x_0\gg \frac{1}{\sqrt{\mu-1}}$, we retain $\Delta d=2L\log{(\frac{1}{\varepsilon L\sqrt{\mu-1}})}+L\sqrt{\mu-1}x_0$, which corresponds to volume-law entanglement.

We focus on the unusual geodesic in \eqref{aeq:second solution for general solution of final differential equation for geodesic for used metric black hole}.
Before, we have argued that for \eqref{aeq:general solution of final differential equation for geodesic for used metric black hole}, $0<c_1<\frac{1}{L^2(\mu-1)}$, which leads to a circle in some limit.
For \eqref{aeq:second solution for general solution of final differential equation for geodesic for used metric black hole}, if $0<c_1<\frac{1}{L^2(\mu-1)}$, it is similar to \eqref{aeq:general solution of final differential equation for geodesic for used metric black hole}.
But we can also have $c_1<0$, which is special because the corresponding derivative $\frac{{\rm d}r^{-1}}{{\rm d}x}$ will be larger when $c_1$ is smaller.
Besides, there is a fixed point on \eqref{aeq:second solution for general solution of final differential equation for geodesic for used metric black hole} for $(x=-c_2,r^{-1}=\sqrt{\frac{1}{L^2(\mu-1)}})$, which is on the horizon of black hole.
It is one reason why we consider it unusual because usual geodesic will not touch the horizon.
Analogous to \eqref{aeq:integral of geodesic length for used metric} and \eqref{aeq:solution of integral of geodesic length for used metric for black hole}, we have the length of geodesic for \eqref{aeq:second solution for general solution of final differential equation for geodesic for used metric black hole} that
\begin{equation}
\label{aeq:nonphysical solution of integral of geodesic length for used metric for black hole}
    d=L\tanh^{-1}{\left[\sqrt{2-c_1 L^2(\mu-1)}{\tanh{(\sqrt{\mu-1}(x+c_2))}}\right]}.
\end{equation}

In the following, we compare the length of geodesic for two cases with black hole metric.
If we assume points $(r_0,\pm\frac{x_0}{2})$ on two geodesics, then $c_2=0$, and corresponding parameters $c_1^{\rm u}$ for usual geodesic \eqref{aeq:general solution of final differential equation for geodesic for used metric black hole} and $c_1^{\rm n}$ for unusual geodesic \eqref{aeq:second solution for general solution of final differential equation for geodesic for used metric black hole} are
\begin{equation}
\label{aeq:parameter c1 for two geodesic}
    c_1^{\rm u}=\frac{r^{-2}+\frac{1}{L^2(\mu-1)}\sinh^2{(\sqrt{\mu-1}\frac{x_0}{2})}}{\cosh^2{(\sqrt{\mu-1}\frac{x_0}{2})}},\qquad c_1^{\rm n}=\frac{r^{-2}+\frac{1}{L^2(\mu-1)}\left(\sinh^2{(\sqrt{\mu-1}\frac{x_0}{2})}-1\right)}{\sinh^2{(\sqrt{\mu-1}\frac{x_0}{2})}}.
\end{equation}
Plugging it in \eqref{aeq:solution of integral of geodesic length for used metric for black hole} and \eqref{aeq:nonphysical solution of integral of geodesic length for used metric for black hole}, we have 
\begin{subequations}
\label{aeq:comapre length od two geodesic}
\begin{equation}
\label{aeq:length of geodesic 1}
    \Delta d^{\rm u}=2L\tanh^{-1}{\left[{\left(\frac{L^2(\mu-1)r_0^{-2}+\sinh^2{[\sqrt{\mu-1}\frac{x_0}{2}]}}{\cosh^2{[\sqrt{\mu-1}\frac{x_0}{2}]}}\right)}^{-\frac{1}{2}} {\tanh{(\sqrt{\mu-1}(x+c_2))}}\right]},
\end{equation}
\begin{equation}
\label{aeq:length of geodesic 2}
    \Delta d^{\rm n}=2L\tanh^{-1}{\left[{\left(\frac{-L^2(\mu-1)r_0^{-2}+\cosh^2{[\sqrt{\mu-1}\frac{x_0}{2}]}}{\sinh^2{[\sqrt{\mu-1}\frac{x_0}{2}]}}\right)}^{\frac{1}{2}} {\tanh{(\sqrt{\mu-1}(x+c_2))}}\right]}.
\end{equation}
\end{subequations}
For the usual geodesic we have $c_1<\frac{1}{L^2(\mu-1)}$, which means $r_0^{-2}<c_1<\frac{1}{L^2(\mu-1)}$.
For the unusual geodesic we also have $c_1<\frac{1}{L^2(\mu-1)}$, which means $r_0^{-2}<\frac{1}{L^2(\mu-1)}$.
Therefore, we can directly check that, with $r_0^{-2}{L^2(\mu-1)}<1$, we have $\Delta d^{\rm u}<\Delta d^{\rm n}$, which means the usual geodesic is a global minimum with the shortest length.
Besides, taking $r_0^{-1}=\varepsilon$, we can simplify \eqref{aeq:length of geodesic 2} to get
\begin{equation}
\label{aeq:total length of nonphysical geodesic for black hole}
\begin{split}
    \Delta d=2L\tanh^{-1}{\left[\left(1-\frac{L^2(\mu-1)\varepsilon^{2}}{\cosh^2{[\sqrt{\mu-1}\frac{x_0}{2}]}}\right)^{\frac{1}{2}}\right]}
    \approx2L\log{\left(\frac{2\cosh{(\sqrt{\mu-1}\frac{x_0}{2})}}{L\sqrt{\mu-1}\varepsilon}\right)}.
\end{split}
\end{equation}
However, different from \eqref{aeq:total length of geodesic with unused metric for sued metric for black hole}, for $x_0\ll \frac{1}{\sqrt{\mu-1}}$, we cannot retain $\log$-law entanglement.
For $x_0\gg \frac{1}{\sqrt{\mu-1}}$, we retain $\Delta d=2L\log{(\frac{1}{\varepsilon L\sqrt{\mu-1}})}+L\sqrt{\mu-1}x_0$, which corresponds to volume law entanglement.

\subsection{Details of calculation for the geodesic in the bubble-outside-horizon phase}
\label{details of calculation for geodesic in the bubble-outside-horizon phase}

As shown in the main text, with the subregion in the boundary of time reversal invariant slice located at $(-\delta,\delta)$, if the geodesic in region II has the usual form, the equations of undetermined parameters are
\begin{subequations}
\label{eq:naive try for geodesic equation}
\begin{equation}
\label{eq:naive try for geodesic equation 1}
    r_0^{-2}+(L_1^{-2}+c_1)\sin^2{x_0}=c_1,
\end{equation}
\begin{equation}
\label{eq:naive try for geodesic equation 2}
    r_0^{-2}+\left(\frac{1}{L_2^2(\mu-1)}-b_1\right)\sinh^2{\sqrt{\mu-1}(x_0-b_2)}=b_1.
\end{equation}
\begin{equation}
\label{eq:naive try for geodesic equation 3}
    \varepsilon^{2}+\left(\frac{1}{L_2^2(\mu-1)}-b_1\right)\sinh^2{\sqrt{\mu-1}(\delta-b_2)}=b_1.
\end{equation}
\begin{equation}
\label{eq:naive try for geodesic equation 4}
\begin{split}
    &\left.\frac{{\rm d}r^{-1}}{{\rm d}x}\right|_{r=r_0, {\rm I}}=\frac{(L_1^{-2}+c_1)\sin{x_0}\cos{x_0}}{\sqrt{c_1-(L_1^{-2}+c_1)\sin^2{x_0}}}\\
    =&\left.\frac{{\rm d}r^{-1}}{{\rm d}x}\right|_{r=r_0, {\rm II}}=\sqrt{\mu-1}\frac{(\frac{1}{L_2^{2}(\mu-1)}-b_1)\sinh{[\sqrt{\mu-1}(x_0-b_2)]}\cosh{[\sqrt{\mu-1}(x_0-b_2)]}}
    {\sqrt{b_1-(\frac{1}{L_2^{2}(\mu-1)}-b_1)\sinh^2{[\sqrt{\mu-1}(x_0-b_2)]}}}.
\end{split}
\end{equation}
\end{subequations}

With the limit \eqref{eq:limit approx}, let's try to solve the equations \eqref{eq:naive try for geodesic equation}.
From \eqref{eq:naive try for geodesic equation 2} and \eqref{eq:naive try for geodesic equation 3}, the denominators of the two sides in \eqref{eq:naive try for geodesic equation 4} are both $r_0^{-1}$.
So we only need to look at the numerator.
From \eqref{eq:naive try for geodesic equation 1} we know $c_1=[r_0^{-2}+L_1^{-2} \sin^2{x_0}]/\cos^2{x_0}\sim \mathcal O(x_0^2)$, then the numerator on the left-hand side of \eqref{eq:naive try for geodesic equation 4} is in the order of $\mathcal O(x_0)$.
However, with the help of \eqref{eq:naive try for geodesic equation 2}, the numerator on the right-hand side of \eqref{eq:naive try for geodesic equation 4} can be simplified as
\begin{equation}
\label{eq:numerator on the rhs of 4}
   \sqrt{(\mu-1)(b_1-r_0^{-2})\left[\frac{1}{L_2^{2}(\mu-1)}-r_0^{-2}\right]} <\sqrt{(\mu-1)(r_H^{-2}-r_0^{-2})^2}=L_2^{-1}(1-\eta^2)r_H^{-1}\sim \mathcal{O}(r_H^{-1}) 
\end{equation}
where we have used $b_1<\frac{1}{L_2^{2}(\mu-1)}=r_H^{-2}$.
Therefore, in \eqref{eq:naive try for geodesic equation 4} we have $\mathcal O(x_0)$ for the numerator of the left-hand side and $\mathcal O(r_H^{-1})$ for the numerator of the right-hand side.
With the assumption $r_H^{-1}\ll x_0$ in \eqref{eq:limit approx}, there is no solution for \eqref{eq:naive try for geodesic equation}.
The naive geodesic we use does not have a solution at the limit.
The key is the derivative of geodesic in region II is too small to satisfy the connection condition.
This problem can be solved by considering an unusual geodesic in the region II.

With the unusual geodesic, we first check if \eqref{eq:bubble-outside-horizon geodesic equation} has a solution that satisfies the connection condition \eqref{eq:bubble-outside-horizon geodesic equation 4}.
Again, the denominators of the two sides in \eqref{eq:bubble-outside-horizon geodesic equation 4} are both $r_0$ owing to \eqref{eq:bubble-outside-horizon geodesic equation 1} and \eqref{eq:bubble-outside-horizon geodesic equation 2}.
The numerator for the left-hand side is the same as before, and is again in the same order of $\mathcal O(x_0)$.
But for the right-hand side, with \eqref{eq:bubble-outside-horizon geodesic equation 2} we have 
\begin{equation}
\label{eq:numerator on the rhs of 4 right}
\begin{split}
    &\sqrt{(\mu-1)\left[\frac{1}{L_2^{2}(\mu-1)}-r_0^{-2}\right]\left[\frac{2}{L_2^{2}(\mu-1)}-r_0^{-2}-b_1\right]}\\
    & = L_2^{-1}\sqrt{(1-\eta^2)(2 r_H^{-2}-r_0^{-2}-b_1)}.
\end{split}
\end{equation}
So we require $b_1\sim-x_0^2$ to satisfy the connection condition.

In the limit~\eqref{eq:limit approx}, we expand variables in the order of $x_0$ and $r_H$.
From \eqref{eq:bubble-outside-horizon geodesic equation 1}, \eqref{eq:bubble-outside-horizon geodesic equation 2}, and \eqref{eq:bubble-outside-horizon geodesic equation 4}, we can express $c_1$, $b_1$, and $b_2$ as a function of $x_0$, i.e., \eqref{eq:c1_b1_b2 results}.
We then substitute \eqref{eq:c1_b1_b2 results} into \eqref{eq:bubble-outside-horizon geodesic equation 3} to get an equation that only involves $x_0$.
\eqref{eq:bubble-outside-horizon geodesic equation 3} can be brought to $\sinh^2{[\sqrt{\mu-1}(\delta-b_2)]}=\frac{\frac{1}{L_2^2(\mu-1)}-\varepsilon^{2}}{\frac{1}{L_2^2(\mu-1)}-b_1}$.
We can use \eqref{eq:c1_b1_b2 results} to simplify both the left-hand side and right-hand side to get
\begin{equation}
\label{eq:simplify solve x0 eq 2}
\begin{split}
    &L_1^{2}\sqrt{1-\eta^2}r_H^{-2}x_0^{-1}\left[1+\frac{1}{2}L_2^{-2} L_1^{4}(1-\eta^2)^2\frac{r_H^{-2}}{x_0^2}\right]
    \\
    &=\delta-x_0+L_1^{2}(1-\eta^2)\frac{r_H^{-2}}{{x_0}}\left(1+\frac{1}{2}L_2^{-2} L_1^{4}(1-\eta^2)^2\frac{r_H^{-2}}{x_0^2}\right).
\end{split}
\end{equation}
Therefore, at the leading order, we obtain a quadratic equation for $x_0$ with two solutions. 
Because $x_0\approx\delta$, the relevant solution is \eqref{eq:solution of x0 with c1_b1_b2}, where the second term has the order $\mathcal O(\frac{r_H^{-2}}{x_0^2}\cdot x_0)\ll \mathcal O(x_0)$.

With the solution above, we can calculate the geodesic length $\Delta d = 2(\Delta d_1 + \Delta d_2)$, where $\Delta d_{1,2}$ denotes the length in region I and II, respectively.
For the geodesic in region II, from \eqref{eq:nonphysical solution of integral of geodesic length for used metric for black hole sum} we have $\Delta d_2 = (\Delta d_2)_1-(\Delta d_2)_2$ with
\begin{equation}
\label{eq:geodesic in eqgion II for outside black hole}
\begin{split}
    (\Delta d_2)_1 & = L_2\tanh^{-1}{\left[\sqrt{2-b_1 L_2^2(\mu-1)}{\tanh{(\sqrt{\mu-1}(\delta-b_2))}}\right]}, \\
    (\Delta d_2)_2 & = L_2\tanh^{-1}{\left[\sqrt{2-b_1 L_2^2(\mu-1)}{\tanh{(\sqrt{\mu-1}(x_0-b_2))}}\right]}.
\end{split}
\end{equation}
Given the solution \eqref{eq:solution of x0 with c1_b1_b2} and \eqref{eq:c1_b1_b2 results}, the approximation above leads to 
\begin{equation}
\label{eq:final result for geodesic in eqgion II for outside black hole}
\begin{split}
    \Delta d_2
    =&L_2\left\{\log{\left(\frac{2r_H^{-1}}{\varepsilon}\right)}+\frac{1}{2}L_2^{-2}L_1^4(1-\eta^2)(1-\sqrt{1-\eta^2})r_H^{-2}\delta^{-2}-\tanh^{-1}{(\sqrt{1-\eta^2})}\right\}. 
\end{split}
\end{equation}
Similarly, for the geodesic in region I, it is
\begin{equation}
\label{eq:geodesic in eqgion I for outside black hole}
\begin{split}
    \Delta d_1=&L_1\tanh^{-1}{\left[\frac{\tan{x_0}}{L_1\sqrt{c_1}}\right]}=L_1\tanh^{-1}{\left[\left(\frac{\sin^2{x_0}}{L_1^2 r_0^{-2}+\sin^2{x_0}}\right)^{\frac{1}{2}}\right]}\\
    \approx& L_1\tanh^{-1}{\left(1-\frac{1}{2}L_1^2\eta^2 r_H^{-2}\delta^{-2}\right)}\approx L_1\log{\left(\frac{2L_1^{-1}\eta^{-1}\delta}{r_H^{-1}}\right)},
\end{split}
\end{equation}
Finally, putting everything together, we have the total length of geodesic \eqref{eq:total length of geodesic for bubble-outside-horizon} in the main text.

\subsection{Details of calculation for the geodesic in the bubble-inside-horizon phase}
\label{details of calculation for geodesic in the bubble-inside-horizon phase}

\begin{figure}[tbp]
\centering {\includegraphics[width=.45\textwidth]{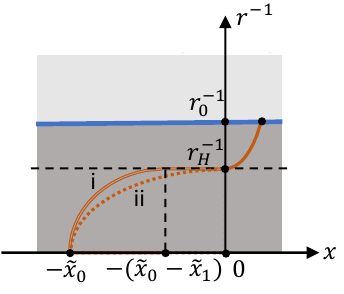}}
\caption{\label{afig:two_geodesic_cases} Two possible geodesics
in the region II in the bubble-inside-horizon phase.}
\end{figure}

With \eqref{eq:geodesic in l123}, we mention that the existence of geodesic along horizon will always lead to a longer geodesic length.
Now we compare the two cases shown in figure~\ref{afig:two_geodesic_cases}.
According to \eqref{aeq:second solution for general solution of final differential equation for geodesic for used metric black hole}, assuming $(r, x)=(\varepsilon^{-1},-\Tilde{x}_0)$ located on the geodesic, then from \eqref{aeq:total length of nonphysical geodesic for black hole} we have 
\begin{equation}
\label{aeq:length of 2 geodesic}
    \Delta d=L_2\tanh^{-1}{\left[\left(1-\frac{L_2^2(\mu-1)\varepsilon^{2}}{\cosh^2{[\sqrt{\mu-1}\Tilde{x}_0]}}\right)^{\frac{1}{2}}\right]}
    \approx L_2\log{\left(\frac{2\cosh{(\sqrt{\mu-1}\Tilde{x}_0)}}{L_2\sqrt{\mu-1}\varepsilon}\right)}.
\end{equation}
Now we show that $L_2\log{\left(\frac{2\cosh{(\sqrt{\mu-1}\Tilde{x}_0)}}{L_2\sqrt{\mu-1}\varepsilon}\right)}<L_2\log{\left(\frac{2\cosh{(\sqrt{\mu-1}\Tilde{x}_1)}}{L_2\sqrt{\mu-1}\varepsilon}\right)}+r_H(\Tilde{x}_0-\Tilde{x}_1)$ for $0<\Tilde{x}_1<\Tilde{x}_0$.
Define function $f(x)=\sqrt{\mu-1}x-\log{[\cosh{(\sqrt{\mu-1}x)}]}$, then the problem above is equivalent to $f(\Tilde{x}_1)<f(\Tilde{x}_0)$ for $\Tilde{x}_1<\Tilde{x}_0$.
Up to rescaling of $x$ we can define $\Tilde{f}(x)=x-\log{[\cosh{(x)}]}$, and $\Tilde{f}'(x)=1-\tanh{x}>0$.
Therefore, we only consider the geodesic with three parts as in the main text.

To solve \eqref{eq:transformed eq for x0 exact} in the main text, we mentioned that $\beta$ must have the same order of $\delta$ and $|\delta-\beta|\ll\delta$.
Here we give a proof.
An intuition argument is that for $|\delta-\beta|\sim \mathcal O(\delta)$, the geodesic corresponds to volume law, while we can find a geodesic with $\log$-law later.
We first construct a geodesic with $\log$-law for $\varepsilon\ll r_H^{-1}\ll \delta\ll1$.
Assume $\delta\approx\beta$ and $b_1'=b_1$ in \eqref{eq:geodesic in l123}, then the total length of part $l_1$ and $l_2$ is smaller than the length of total unusual geodesic in \eqref{aeq:total length of nonphysical geodesic for black hole} that
\begin{equation}
\label{aeq:construct log-law geodesic 1}
    {\rm Length}[l_1+l_2]<2L_2\log{\left(\frac{2\cosh{(\sqrt{\mu-1}(\delta-\beta))}}{L_2\sqrt{\mu-1}\varepsilon}\right)}.
\end{equation}
Assume $(\delta-\beta)\sim \mathcal O(r_H^{-1})$, then the right-hand side of \eqref{aeq:construct log-law geodesic 1} is $L_2\log{\left(2\cosh{(\sqrt{\mu-1}(\delta-\beta))}\right)}\sim \mathcal O(1)$ except UV-cutoff and the divergence of $\log{\sqrt{\mu-1}}$.
For part $l_3$, the point $(r_0,-x_0)$ is on it with $|\delta-x_0|<2|\delta-\beta|\sim \mathcal O(r_H^{-1})$, which means $x_0\sim \mathcal O(\delta)\gg r_0^{-1}$.
Therefore, with \eqref{aeq:total length of geodesic with unused metric for sued metric for thermal AdS}, we have 
\begin{equation}
\label{aeq:construct log-law geodesic 2}
\begin{split}
    {\rm Length}[l_3]\approx L_1\log{\left(\frac{2\sin{x_0}}{L_1 r_0^{-1}}\right)},
\end{split}
\end{equation}
which corresponds to $\log$-law entanglement for $x_0\ll1$.
Therefore, we have constructed a geodesic with $\log$-law entanglement.
Now we prove that for $|\delta-\beta|\sim \mathcal O(\delta)$ the geodesic corresponds to volume law.
We consider the length of part $l_1$, which corresponds to equation \eqref{eq:geodesic in l123 part 1}.
Assuming $\beta=\eta' \delta$ and $\eta'\sim \mathcal O(1)<1, (1-\eta')\sim \mathcal O(1)<1$, with \eqref{aeq:total length of nonphysical geodesic for black hole}, we have
\begin{equation}
\label{aeq:length of part a for volume law case}
\begin{split}
    {\rm Length}[l_1]\approx L_2\log{\left(\frac{2\cosh{(\sqrt{\mu-1}(1-\eta')\delta)}}{L_2\sqrt{\mu-1}\varepsilon}\right)},
\end{split}
\end{equation}
With limit $\sqrt{\mu-1}\delta\sim \mathcal O(\frac{\delta}{r_H^{-1}})\gg1$, the right-hand side of \eqref{aeq:length of part a for volume law case} can be simplified as ${\rm r.h.s.}\approx L_2\log{\left(\frac{\exp{(L_2^{-1} r_H(1-\eta')\delta)}}{L_2\sqrt{\mu-1}\varepsilon}\right)}=L_2\log{\left(\frac{r_H^{-1}}{\varepsilon}\right)}+r_H(1-\eta')\delta$, which corresponds to volume law entanglement.
Therefore, we finish the proof that $|\delta-\beta|\ll\delta$.

In the following, we show the details of solving the geodesic.
For \eqref{eq:transformed eq for x0 exact} in the main text, the right-hand side is of the order $\mathcal O(r_H^{-1})$.
Then the left-hand side implies that the solution is $x_0\sim r_H^{-1}$ or $x_0 \sim \beta $. 
If $x_0\sim  r_H^{-1}$, with \eqref{eq:geodesic in l123 part 3} and \eqref{eq:total length of nonphysical geodesic for black hole sum}, the length of the part $l_2$ is
\begin{equation}
\label{eq:wrong part b length}
\begin{split}
    {\rm Length}[l_2]=L_2\tanh^{-1}{\left[\left(1-\frac{L_2^2(\mu-1)r_0^{-2}}{\cosh^2{[\sqrt{\mu-1}(\beta-x_0)]}}\right)^{\frac{1}{2}}\right]}
    \approx r_H(\beta-x_0)-L_2\log{\eta}.
\end{split}
\end{equation}
Because $\beta-x_0\approx\delta$, the geodesic length leads to volume-law entanglement.
Therefore, we should consider the solution $x_0 \sim \beta $, which will later be shown to have $\log$-law entanglement.
In this case, \eqref{eq:transformed eq for x0 exact} can be expanded as
\begin{equation}
\label{eq:simplify x0 eq for x0 approx beta}
    L_2^{-1}r_H(\beta-x_0)x_0\approx L_2^{-1}L_1^2(1-\eta^2)r_H^{-1}.
\end{equation}
which leads to the solution \eqref{eq:solution of x0 for inside horizon approx} in the main text.

We can calculate the geodesic length with the solution above.
For the part $l_1$, with \eqref{eq:geodesic in l123 part 1} and \eqref{eq:total length of nonphysical geodesic for black hole sum}, we have 
\begin{equation}
\label{eq:length of part a for right x0}
\begin{split}
    \Delta d_{l_1}\approx
    &L_2\log{\left(\frac{r_H^{-1}}{\varepsilon}\right)}+L_2\log{\left(2\cosh{(\sqrt{\mu-1}(\delta-\beta))}\right)}.
\end{split}
\end{equation}
If $|\delta-\beta|\lesssim O(r_H^{-1})$, the second term in \eqref{eq:length of part a for right x0} will at most contribute to $\mathcal O(1)$.
For the part $l_2$ with \eqref{eq:geodesic in l123 part 3} and \eqref{eq:total length of nonphysical geodesic for black hole sum}, we have 
\begin{equation}
\label{eq:simplified length of part b for right x0}
\begin{split}
    \Delta d_{l_2}\approx& L_2\tanh^{-1}{\left[\left(1-\frac{\eta^2}{\cosh^2{[\sqrt{\mu-1}(\beta-x_0)]}}\right)^{\frac{1}{2}}\right]} \\
    \approx& L_2\left\{\tanh^{-1}{(\sqrt{1-\eta^2})}+\frac{(\mu-1)(\beta-x_0)^2}{2\sqrt{1-\eta^2}}\right\},
\end{split}
\end{equation}
where we have used $\sqrt{\mu-1}(\beta-x_0)\approx L_1^2 L_2^{-1}(1-\eta^2)r_H^{-1}\beta^{-1}\sim \mathcal O(\frac{r_H^{-1}}{\delta})\ll1$.
For the part $l_3$ with \eqref{eq:geodesic in l123 part 2} and \eqref{eq:total length of geodesic with unused metric for sued metric for thermal AdS sum}, we have 
\begin{equation}
\label{eq:length of part c for right x0}
\begin{split}
    \Delta d_{l_3}=& 
    L_1\tanh^{-1}{\sqrt{\frac{\sin^2{x_0}}{L_1^2 r_0^{-2}+\sin^2{x_0}}}}
    \approx L_1\log{\left(\frac{2L_1^{-1}\sin{x_0}}{r_0^{-1}}\right)},
\end{split}
\end{equation}
where we have used $x_0\approx\beta\sim \mathcal{O}(\delta)$ and $r_0^{-1}\sim \mathcal{O}(r_H^{-1})\ll \mathcal{O}(x_0)$. 
By summing them, we get \eqref{eq:final total length of 123 with approx} in the main text.

Now we need to minimize the geodesic length \eqref{eq:final total length of 123 with approx} w.r.t. $\beta$.
Denoting its right-hand side as $f(\beta)$,
its derivative is 
\begin{equation}
\label{eq:deriovative of fbeta}
\begin{split}
    f'(\beta)=&-r_H\tanh{(L_2^{-1} r_H(\delta-\beta))}
    -L_2^{-1}(1-\eta^2)^{\frac{3}{2}} L_1^4 r_H^{-2}\beta^{-3}
    +L_1\frac{1+(1-\eta^2)L_1^2 r_H^{-2}\beta^{-2}}{\beta-(1-\eta^2)L_1^2 r_H^{-2}\beta^{-1}}\\
    \approx & \frac{L_2}{\beta}\left\{\frac{L_1}{L_2}-L_2^{-1}\beta r_H \tanh{(L_2^{-1} r_H(\delta-\beta))}\right\},
\end{split}
\end{equation}
where in the last equation we only keep the order $\mathcal O(\beta^{-1})$ and ignore higher orders, like $O(\beta^{-1}\frac{r_H^{-2}}{\beta^2})\ll O(\beta^{-1})$.
It turns out that the only consistent solution of $f'(\beta) = 0$ is given by 
\begin{equation}
\label{eq:minimize length for beta}
    \beta \approx \delta-L_1 L_2 r_H^{-2} \delta^{-1},
\end{equation}
which is shown in the main text.

Finally, substituting \eqref{eq:minimize length for beta} into \eqref{eq:final total length of 123 with approx}, the geodesic length is \eqref{eq:final total length of 123 with beta}.

\section{Tensor network realization of Python's Lunch}
\label{sec:tensor network realization of Python's Lunch}

\subsection{Brief review of Python's Lunch property}
\label{sec:brief review of Python's Lunch property}

Here we briefly review some basics about complexity and Python's Lunch.
The first definition is the complexity of a unitary operator $U$, which is defined as the minimal number of 2-qubit gates $g$ to prepare $U$, e.g., $U=g_n g_{n-1}...g_1$.
The corresponding number is $\mathcal{C}(U)=\mathcal{C}(U^\dagger)$.
Similarly, we can define the relative complexity between two states $\left|\psi\right>$ and $\left|\phi\right>$, which is defined as the complexity of the unitary transformation satisfying $\left|\psi\right>=U\left|\phi\right>$.
Besides, with gauge redundancy there is more than one $U$, and we must minimize all possible $\mathcal{C}(U)$.
Therefore, we define $\mathcal{C}(\psi,\phi)=\min{\mathcal{C}(U)}=\mathcal{C}(\phi,\psi)$.

In this article, we always consider the black holes, which can be modeled as a quantum computer with $N$ qubits and evolve under some all-to-all Hamiltonian or discrete gates.
Here the original theory always has a Hamiltonian, while to realize some properties, we may construct it with some discrete gates.
Under this equivalence, we can consider the complexity of a black hole, or more generally, a configuration of spacetime geometry.
In reference~\cite{brown2019pythons}, they consider restricted and unrestricted complexity, and focus on the former.
Here we briefly compare these two complexities.
When we consider a two-side black hole, it can be mapped to a maximal entangle state, where $N$ qubits on the left maximally entangle with $N$ qubits on the right.
Then there may be some unitary evolution for both sides.
The difference between restricted and unrestricted complexity is if we require the corresponding $U$ which maps it back to a maximal entangle state can only apply on one side.
For example, we apply $U$ only on the right $N$ qubits for the restricted case.
However, for the case we are interested in, which is a one-side black hole, these two definitions are the same, and it is the post-selection that will increase the complexity exponentially.

\begin{figure}[tbp]
    \centering 
    {\includegraphics[width=.5\textwidth]{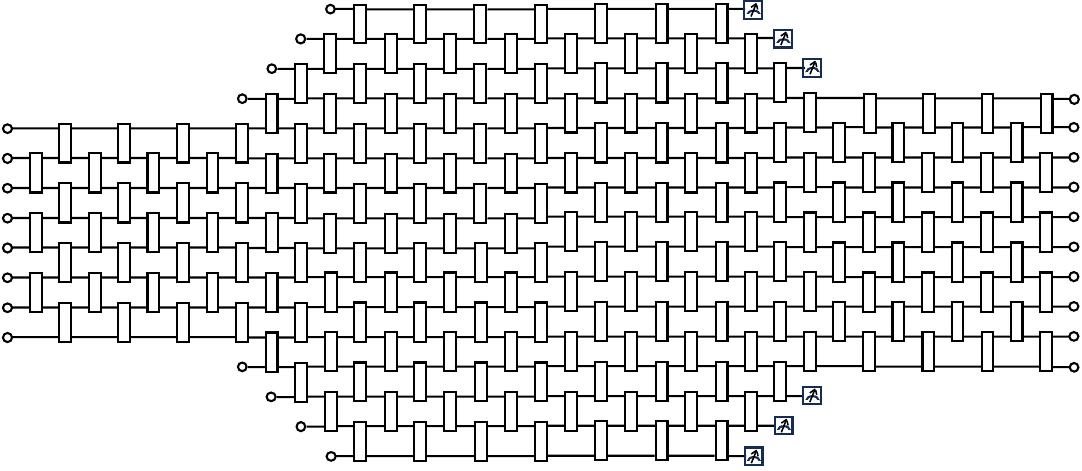}}
    \caption{\label{afig:diagram_of_Python's_lunch} Python's Lunch geometry with quantum circuit realization. Here we add some degrees of freedom after unitary evolution on initial system. Then we apply unitary transformation for the whole system and projective measurement in a subregion.}
\end{figure}

Now we discuss the Python's Lunch geometry and introduce the conjecture of the complexity of Python's Lunch in reference~\cite{brown2019pythons}.
A simple example of Python's Lunch is shown in figure~\ref{afig:diagram_of_Python's_lunch}.
There are two minimal areas and one maximal area, and the corresponding numbers of qubits from left to right are $N$, $(1+\alpha) N$ and $(1+\gamma) N$ with $\alpha>\gamma>1$.
Therefore, using the language of state complexity, we want to calculate $\mathcal{C}(\left|I\right>\left|0\right>^{\otimes m_L},\left|\psi\right>\left|0\right>^{\otimes m_R})$, where $m_L=\alpha N$ and $m_R=(\alpha-\gamma)N$.
In the reference, authors use the Grover algorithm to construct a decomposition of $U$ with $\left|\psi\right>\left|0\right>^{\otimes m_R}=U\left|I\right>\left|0\right>^{\otimes m_L}$.
And the total number of gates that need to construct $U$ and apply on the system is $2^{\frac{m_R}{2}}\mathcal{C}_{\rm TN}$, where $\mathcal{C}_{\rm TN}$ is the number of nodes in the tensor network.
So the complexity is $\mathcal{C}(U)=2^{\frac{m_R}{2}}\mathcal{C}_{\rm TN}$.
Besides, the authors also argue that for general state $\left|I\right>$, the corresponding complexity is invariant.
They also give a naive argument about the complexity of unitary transformation $U$, which will lead to a larger number of gates, but the main conclusion is the same.
Starting from the initial state $\left|I\right>$, we apply the unitary evolution on it and form the left side of figure~\ref{afig:diagram_of_Python's_lunch}.
Then by adding ancilla to the system, we can apply more unitary evolution and form the middle part of the diagram.
However, for the right-hand side, we have a smaller system, which means we must apply some measurement and project out some qubits.
Then after applying more unitary evolution, we will get the final state $\left|\psi\right>$.
Totally, we apply $\mathcal{C}_{\rm TN}$ number of gates in this process, but it needs some measurement which is non-unitary.
To project $m_R$ additional qubits on $\left|0\right>^{\otimes m_R}$, we must consider the post-selection effect, which means there are $2^{m_R}$ possible outcomes and only one is what we want.
Therefore, if we want to successfully construct a state $\left|\psi\right>$ from $\left|I\right>$, on average, we must repeat this process $2^{m_R}$ times.
Finally, we find that the total number of unitary gates we need is $2^{m_R}\mathcal{C}_{\rm TN}$, which is almost the same as the Grover algorithm, and the only difference is the prefactor on exponent.

According to the discussion above, they propose a conjecture about the complexity of Python's Lunch.
For a Python's Lunch geometry with min-max-min areas $\mathcal{A}_L, \mathcal{A}_{\rm max}$ and $\mathcal{A}_R$, and with $\mathcal{A}_L<\mathcal{A}_R$, the restricted complexity on the right system is
\begin{equation}
\label{aeq:conjecture of Python's Lunch complexity}
    \mathcal{C}_R(U)={\rm const}\times\mathcal{C}_{\rm TN} \exp{\left[\frac{1}{2}\frac{\mathcal{A}_{\rm max}-\mathcal{A}_R}{4G\hbar}\right]},
\end{equation}
where $\mathcal{C}_{\rm TN}=\frac{V}{G\hbar l_{\rm AdS}}$ is the volume of wormhole.
If we apply this conjecture to the case in figure~\ref{afig:diagram_of_Python's_lunch}, then $\mathcal{A}_L\approx N\cdot4G\hbar$, $\mathcal{A}_{\rm max}\approx(1+\alpha)N\cdot4G\hbar$ and $\mathcal{A}_R\approx(1+\gamma)N\cdot4G\hbar$, and $\mathcal{C}_R(U)={\rm const}\times\mathcal{C}_{\rm TN}\ e^{[(\alpha-\gamma)N]/2}$.
And $\mathcal{C}_{\rm TN}\geq N\log{N}$.
Besides, to be more concrete, two minimums are well-defined by local minimums, but the maximum is more complicated.
We should first choose one foliation of the geometry which is known as ``sweepout'', and find the global maximum.
Then we minimize all maximums for all foliations, and the minimum of maximum is $\mathcal{A}_{\rm max}$.

Correspondingly, they also give a covariant version of the conjecture above.
For a covariant Python's Lunch geometry with min-max-min generalized entropy $S^{\rm (gen)}_{L}$, $S^{\rm (gen)}_{\rm max}$ and $S^{\rm (gen)}_{R}$, and $S^{\rm (gen)}_{L}<S^{\rm (gen)}_{R}$, the restricted complexity on the right system is
\begin{equation}
\label{aeq:conjecture of Python's Lunch complexity cobariant version}
    \mathcal{C}_R(U)={\rm const.}\times\mathcal{C}_{\rm TN} \exp{\left[\frac{1}{2}\left(S^{\rm (gen)}_{\rm max}-S^{\rm (gen)}_{R}\right)\right]},
\end{equation}
where $\mathcal{C}_{\rm TN}$ is the size of the tensor network.

\subsection{Python's Lunch realization}
\begin{figure}[tbp]
    \centering 
    {\includegraphics[width=.5\textwidth]{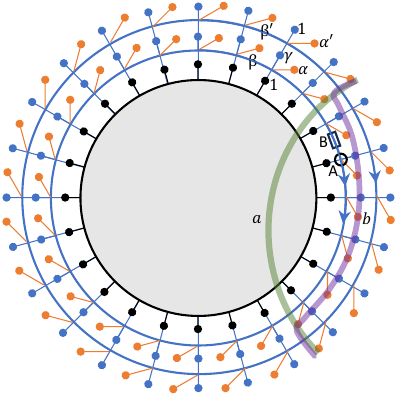}}\caption{\label{afig:Python's_lunch_realization} Python's lunch realization with MERA and additional layers of tensor network. Here the gray disk corresponds to MERA with the black dots denoting degrees of freedom on the boundary. 
    The blue dots mean degrees of freedom after measurements. 
    The orange dots denote ancilla qudits. 
    The green and purple curves indicate two geodesics.}
\end{figure}

In this section, we try to construct a tensor network to realize Python's lunch, which is related to the bubble-inside-horizon phase.
We consider a state dual to such a geometry.
The inner AdS space can be considered as the initial state before measurements, and the black hole geometry can be considered as the effect of general measurements.
Therefore, we can first construct the inner AdS space with a MERA, and then construct the additional black hole region with a tensor network which consists of both unitary evolution and post-selection.
We illustrate our construction in figure~\ref{afig:Python's_lunch_realization}.

The inner tensor network of MERA is shown in gray, and the external tensor network of measurements is shown in blue and orange.
We can construct MERA with unitary circuits and ancilla qudits with the method, e.g., in reference~\cite{Haghshenas_2022}, which will be easier to compute the complexity.
The key is how to design the tensor network of quantum quench, and show that for some parameters it corresponds to the bubble-inside-horizon phase and has exponentially big complexity, and for other parameters it corresponds to the bubble-outside-horizon phase with much smaller complexity. 
A natural construction is shown in figure~\ref{afig:Python's_lunch_realization}.
There are two layers of tensors for measurements (from interior to outside, we denote them as the first and second layers).
The bond dimension of each tensor is to be determined in the following.

We define the bond dimension $D^{(\cdot)}$ and simply use $(\cdot)$ to denote the bond dimension.
Because we hope that, after the weak measurements, the dimension of the Hilbert space of boundary qudits is invariant, i.e., the dimensions of the MERA state before and after weak measurements are the same.
For simplicity, we set the bond dimensions of legs in the MERA and of the legs in the outermost layer of the tensor network to be $1$.
Besides, we expect that there is a min-max-min structure in tensor network for Python's lunch. 
Here the first minimum can be seen as the center of MERA.
Then for the circle with the same center of MERA, the corresponding perimeter is larger for larger radius, and the maximum may locate near the boundary of MERA.
Now if we increase the radius, then the area will rely on the bond dimension of the legs between two external layers of tensors, which we define as $\gamma$.
Therefore, $\gamma<1$ will give another minimal area, and the corresponding geometry can be considered as Python's lunch.
However, if $\gamma=1$, there will be no maxima and second minimum, which means there doesn't exist Python's lunch. (Here we won't consider $\gamma>1$ because it may induce additional post-selection on the second layer.)
Now comparing the area of the circle and the corresponding area in the bubble phase in section~\ref{sec:phase transition for measurement and geodesic for two phases with black hole}, we can find that the bubble-inside-horizon phase corresponds to $\gamma<1$, while the bubble-outside-horizon phase corresponds to $\gamma=1$.
And we require tensors in the weak measurement should also be perfect tensors~\cite{Pastawski_2015}, which will be helpful to compute the geodesic and minimal surface.
Now we have constructed a tensor network to realize two phases and Python's lunch.

Intuitively, from the discussion before, we expect that for $\gamma<1$ the complexity will be exponentially large, but for $\gamma=1$ it cannot be larger than a power law.
Now we discuss the complexity of figure~\ref{afig:Python's_lunch_realization} in details.

\begin{figure}[tbp]
    \centering 
    {\includegraphics[width=.8\textwidth]{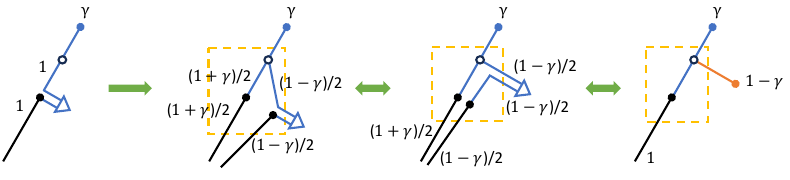}}
    \caption{\label{afig:post-selection_times} Diagram of equivalence of post-selection on maximally mixed state and general post-selection on ancilla with dimension $D^{1-\gamma}$.}
\end{figure}

Firstly, we discuss the ways of contracting tensors in the tensor network.
As the definition of complexity, we must decompose the tensor network to many small unitary tensors and apply them to initial qudits one by one.
For each tensor, there are two ways to contract them.
One is applying post-selection, and we choose the maximally entangled state, which can be considered as a straight line in tensor network and shown in figure~\ref{afig:post-selection_times} (first panel).
This method doesn't require other condition.
Although this method is general, we may have more post-selection, which leads to larger complexity.
There is another way of contracting the tensor by applying it as a unitary transformation.
Actually, we can always divide the legs of tensor into two parts to consider it as a unitary transformation.
For example, when we have a tensor with four bonds and their dimensions are $(1,\beta,\beta,\gamma)$, we can still consider it as a unitary tensor by dividing them into two equal parts with bond dimension $\beta+\frac{1+\gamma}{2}$.
However, although a tensor can be seen as a unitary tensor, we may have additional constraints.
Considering one tensor in the measurement part, because of the rotation symmetry, we require that two bonds on the tangent direction (angular direction) will be one ``in'' and one ``out'' with the same bond dimension.
Besides, tensors in the first layer will be connected to the boundary of MERA, and the corresponding bond should also be ``in''.
Therefore, if we require $\gamma\neq1$, then the tensor is not a unitary transformation because the total bond dimension of in and out are different.
It is the reason why we add additional dangling bonds on each site, and their bond dimensions are to be determined to minimize the post-selection.
Because additional bonds will increase the post-selection times, it is a trade-off for the complexity.
To summarize, for the first method we don't need additional bonds but post-selection to get maximally entangled state, while in the second method we need add some additional dangling bonds but may have less post-selection.
So we need to compare them and choose the smaller one.

As shown in figure~\ref{afig:post-selection_times}, we first consider the tensors with bond dimension $(1,\beta,\beta,\gamma)$ in the first layer that connect to the boundary of MERA.
Because of the rotation symmetry, we don't need to consider two tangent bonds with dimension $\beta$.
If we naively apply post-selection on the bond with dimension $1$ (to connect to the boundary of MERA), the Hilbert space of post-selection is $D^1\cdot D^1=D^2$.
However, we can use another way with the property of perfect tensors to reduce the dimension of post-selection.
We can divide the total bond into two equal parts with dimension $\frac{1+\gamma}{2}<1$.
Then we separate the bond with dimension $1$ into two groups with dimension $\frac{1+\gamma}{2}$ and $\frac{1-\gamma}{2}$, and apply the tensor as a unitary transformation on the bond with dimension $\frac{1+\gamma}{2}$.
Now what we need to do is applying post-selection on the left bond with dimension $\frac{1-\gamma}{2}$, which corresponds to the Hilbert space with dimension $D^{\frac{1-\gamma}{2}}\cdot D^{\frac{1-\gamma}{2}}=D^{1-\gamma}<D^2$.
It means that with the first method we must at least apply post-selection on Hilbert space with dimension $D^{1-\gamma}$, which is shown in figure~\ref{afig:post-selection_times} (second panel).
Now we consider the second method, where we add a dangling bond with dimension $\alpha$.
Because we want to apply tensor as a unitary transformation on the boundary of MERA, it requires $\alpha \geq 1-\gamma$.
Therefore, with a dangling bond with dimension $\alpha=1-\gamma$, we must apply post-selection on this bond after connecting the tensor to MERA boundary as a unitary transformation.
It means the dimension of Hilbert space under post-selection is $D^\alpha=D^{1-\gamma}$, which is shown in figure~\ref{afig:post-selection_times} (fourth panel).
Therefore, we find that for both methods, we must at least apply post-selection on a Hilbert space with dimension $D^{1-\gamma}$.
This equivalence is shown in figure~\ref{afig:post-selection_times}.
We can bind two bonds with dimension $\frac{1-\gamma}{2}+\frac{1-\gamma}{2}$ under post-selection to maximally entangled state in the first method, and consider them as the additional dangling bond we add in the second method.
This result is also consistent with the naive argument before, where we consider the times of post-selection as $(D^{1-\gamma})^N$ and $N$ is the number of dangling bonds on one layer.

Besides, the first and last tensors we add for each layer are special, where the bond with dimension $\beta$ is vital.
It is shown in figure~\ref{afig:Python's_lunch_realization} that, for the first layer, we start to construct the network from the tensor located at A.
For this tensor, the bond with dimension $1$ is ``in'', and the bonds with dimension $\beta,\beta$ and $\gamma$ are ``out''.
It means there are two additional out bonds.
So we may use the dangling bonds to decrease the dimension of post-selection.
For example, if $1+\alpha>\gamma+2\beta$, we can divide the dangling bond into two parts with dimension $(-1+\alpha+\gamma+2\beta)/2$ and $(1+\alpha-\gamma-2\beta)/2$ as ``in'' and ``out'' bonds.
Then the total dimension of ``in'' and ``out'' bond are the same.
Therefore, the first tensor needs post-selection of Hilbert space with dimension $D^\frac{(1+\alpha-\gamma-2\beta)}{2}$.
For $1+\alpha<\gamma+2\beta$, a naive method is directly applying post-selection on the bond with dimension $1$ to maximally entangled state (without introducing additional ancilla).
For the last tensor located on $B$, we will find that after connecting it to the whole network, there are two bonds with dimension $\beta$ left.
To connect them, we need to apply post-selection on them.
Therefore, for two layers there are additional post-selections, nevertheless, the complexity resulted from these post-selections won't scale with the total number of qudits in one layer.

The discussion before actually is only valid for the first layer with $\gamma<1$.
For the second layer, the method of connecting tensors is similar.
With rotation symmetry, we focus on the tensor except the first and last one, and do not need to consider two tangent bonds with dimension $\beta$.
Similar to the first layer, we can set $\alpha'=\alpha=1-\gamma$.
Then we can consider two bonds with dimension $\gamma$ and $\alpha'$ as ``in'' bond and the bond with dimension $1$ as ``out'' bond.
Therefore, we can connect the tensor by regarding it as a unitary transformation.
The difference between the first and second layers is that, for the first layer, the dangling bonds act as the Hilbert space under post-selection, while for the second layer, the dangling bonds act as ancilla qudits. 
These ancilla qudits are coupled to the system without the need of post-selection.
Besides, the post-selection is also needed for the first and last tensors, which do not scale with the system size.

To summarize, for $\gamma<1$, which corresponds to the Python's lunch geometry, the complexity of the tensor network is about $\mathcal{C}(U)={(\rm const)}\cdot D^{(1-\gamma)N}$, where $N$ is the total system size (the number of qudits on the boundary), and ${(\rm const)}$ includes the contribution of complexity from MERA (and the first and last tensors in two external layers that do not scale with the system size).
While, for $\gamma=1$, which corresponds to the bubble-outside-horizon phase without Python's lunch, the complexity is $\mathcal{C}(U)={(\rm const)}'$ which only comes from the contribution of MERA (and connecting the first and last tensors in the angular direction in two external layers).

There are some remarks about the discussion above.
(i) Here, we require the tensors in the tensor network realization of Python's lunch to be perfect tensors.
It is because this requirement will be useful when we compute the geodesic length and regard tensors as unitary transformations.
For example, with Python's lunch geometry, when we consider the geodesic bounded by two endpoints on the boundary, there is a local minimal which is shown in figure~\ref{afig:Python's_lunch_realization} with label (b).
The length of this geodesic is $d=(\log{D^\gamma})\cdot L+{(\rm const)}$ where $L$ is the size of the subsystem and ${(\rm const)}$ is a constant, and it corresponds to volume law entanglement.
But the minimal geodesic is labeled in figure~\ref{afig:Python's_lunch_realization} by (a) with length $d=\log{D}\cdot \log{L}+{\rm (const)'}$ where $\log{L}$ and ${\rm (const)'}$ terms represent the contributions from the AdS region and the boundary region associated with the measurements, respectively.
And it leads to the $\log$-law entanglement.
These geodesics can be gotten with a greedy algorithm for perfect tensor in reference~\cite{Pastawski_2015}.
For $\gamma=1$, there is no local minimal geodesics, and the only one has the length $d=\log{D}\cdot \log{L}+{\rm (const)'}$.
These properties for $\gamma<1$ and $\gamma=1$ are consistent with the results in section~\ref{sec:phase transition for measurement and geodesic for two phases with black hole}.
(ii) Here we use MERA in reference~\cite{Haghshenas_2022} figure 2 (b), where we can consider gray circles as ancilla qudits, and each tensor has two ``in'' legs with arrows pointing outside and two ``out'' legs with arrows pointing inside.

\bibliographystyle{JHEP}
\bibliography{reference.bib}

\end{document}